\documentclass[11pt]{article}
\usepackage{graphicx}
\usepackage{rotating}
\newcommand{\be}{\begin{equation}}
\newcommand{\ee}{\end{equation}}
\newcommand{\ben}{\begin{eqnarray}}
\newcommand{\een}{\end{eqnarray}}
\newcommand{\bc}{\begin{center}}
\newcommand{\ec}{\end{center}}
\begin{document}

\title{Supernova remnants and $\gamma$-ray sources\footnote{
Submitted to the Physics Reports, astro-ph/0209565}}

\author{Diego F.
Torres$^{1}$\thanks{e-mail: dtorres@igpp.ucllnl.org}, Gustavo E.
Romero$^2$, Thomas M. Dame$^3$, Jorge A. Combi$^4$, and Yousaf M.
Butt$^3$}

\maketitle

\noindent
\begin{center}
{\small $^1$Lawrence Livermore Laboratory, 7000 East
Ave. L-413, Livermore, CA 94550, USA }\\
{\small $^2$Instituto Argentino de Radioastronom\'{\i}a, C.C.5,
1894 Villa Elisa, Buenos Aires, Argentina}\\ {\small $^3$
Harvard-Smithsonian Center for
Astrophysics, 60 Garden Street, Cambridge, MA 02138, USA}\\
\end{center}

\begin{abstract}
A review of the possible relationship
between $\gamma$-ray sources and supernova remnants (SNRs) is presented.
Particular
emphasis is given to the analysis of the observational status of the problem of
cosmic ray acceleration at SNR shock fronts.
All positional coincidences between SNRs and unidentified $\gamma$-ray
sources listed in the Third EGRET Catalog at low Galactic latitudes
are discussed on a case by case
basis.  For several coincidences of particular interest,
new CO(J=1-0) and radio continuum maps are shown,
and the mass content
of the SNR surroundings is determined.
The contribution to the $\gamma$-ray flux
observed that might come from cosmic
ray particles (particularly nuclei) locally accelerated at the SNR
shock fronts is evaluated.
We discuss the prospects for future research in this field and remark on the
possibilities for observations with forthcoming $\gamma$-ray instruments.\\

{\bf Keywords:} gamma-rays: observations, gamma-rays: theory, ISM:
supernova remnants, ISM: clouds, cosmic rays.
\end{abstract}

\newpage

{\small \tableofcontents }
\section{ Introduction}

Gamma-ray astronomy has unveiled some of the most exotic and
energetic objects in the universe: from supermassive black-holes
in distant radio galaxies to radio-quiet pulsars and the still
enigmatic gamma-ray bursts. However, it has been conspicuously
less successful in achieving one of its original goals of shedding
light on the sources of Galactic cosmic ray nuclei. In this report
we focus on the remnants of galactic supernovae, and their
possible association with discrete sources of ($> 70$ MeV)
$\gamma$-rays, as seen by the Energetic Gamma-ray Experiment
Telescope (EGRET). In doing so, we attempt to lay a framework in
which the long-standing question of the supernova remnant origin
of Galactic cosmic rays may be addressed. \\

The first firm detection of celestial high-energy $\gamma$-rays
was achieved by Clark, Garmire and Kraushaar using the Orbiting
Solar Observatory (OSO-3), when they discovered that the plane of
the Galaxy was a source of photons  with $E> 70$ MeV (Clark,
Garmire \& Kraushaar 1968; Kraushaar et al.  1972). Higher spatial
resolution studies made with the SAS-2 satellite, launched in
1972, revealed individual sources of $\gamma$-rays from the Vela
pulsar (Thompson et al. 1975), and confirmed the high-energy
emission from the Crab (Kniffen et al. 1974). The long life of
ESA's COS-B satellite (1975-1982) produced another major
breakthrough in $\gamma$-ray astronomy: for the first time a
significant number of  $\gamma$-ray sources were seen which could
not be identified with objects known at other wavelengths (see
Bignami \& Hermsen 1983, for a review of COS-B results). Figure
\ref{cosb} shows the region surveyed by COS-B and the point
sources discovered, as reported in the second COS-B Catalog
(Hermsen 1981, Swanenburg et al. 1981).\\

In 1991, the EGRET telescope was launched onboard the Compton
Gamma-Ray Observatory (see Gehrels \& Shrader 2001, for a recent
review). The Compton satellite (1991-2000), the heaviest orbital
scientific payload at the time of its launch, had three other
experiments apart from EGRET. All of them have contributed to our
understanding of the $\gamma$-ray sky, although we shall
particularly focus on EGRET results in this report. The Third
EGRET (3EG) Catalog, whose point-like detections are shown in
Figure~\ref{egret}, is now the latest and most complete source of
information on high-energy $\gamma$-ray sources. It contains 271
detections with high significance, including 5 pulsars, 1 solar
flare, 66 blazar identifications, 1 radio galaxy (Cen A), 1 normal
galaxy (LMC), and almost two hundred unidentified sources, $\sim
80$ of them located at low galactic latitudes (see Grenier 2001
and Romero 2001 for recent reviews). \\

\begin{figure}
\begin{center}
\caption{Plot of source locations (galactic coordinates) in the
COS-B catalog. The undashed region of the sky was surveyed by the
satellite. Most of the sources had fluxes higher than $1.3 \times
10^{-6}$ photons cm$^{-2}$ s$^{-1}$ above 100 MeV.
  From Bignami \& Hermsen (1983).} \label{cosb}
\end{center}
\end{figure}

The detection of pulsed high-energy emission from some $\gamma$-ray sources,
on one hand, and the identification of Geminga as a radio quiet
pulsar, on the other, have prompted several authors to explore the
possibility that all unidentified low-latitude sources contained
in earlier versions of the EGRET Catalog (i.e. the Second EGRET --2EG--
Catalog, Thompson et al. 1995, 1996) could be pulsars as well
(excepting a small
extragalactic and isotropic
component which should be seen through the disc of the Galaxy).
In
particular, Kaaret \& Cottam (1996) used OB associations as
pulsar tracers, finding a significant positional correlation
with 2EG unidentified sources. A similar study, including SNRs and
HII regions (considered as tracers of star forming
regions and, hence, of possible pulsar concentrations) has
been carried out by Yadigaroglu \& Romani (1997), who also
concluded that the pulsar hypothesis for the unidentified 2EG sources was
consistent with the available information.\\

However, spectral analysis by Merck et al. (1996) and Zhang \&
Cheng (1998) showed that several 2EG sources were at odds with the
pulsar explanation; the spectra of many sources are too different
from what is expected from outer or polar gap models of pulsar
emission. Time variability in the $\gamma$-ray flux of many sources
(discussed below) also argued against a unique population behind
the unidentified galactic $\gamma$-ray sources.\\

\begin{figure}[t]
\begin{center}
\caption{Plot of source locations  (galactic coordinates) in the
Third EGRET Catalog. Different populations are marked in different
grades of light colours.
The size of the dots gives a qualitative idea of the detected flux. From
Hartman et al. (1999).} \label{egret}
\end{center}
\end{figure}

Most likely, the unidentified $\gamma$-ray sources at low galactic
latitudes are related to several different galactic populations
(e.g. Grenier 1995, 2000; Gehrels et al. 2000; Romero 2001). Among
them there are surely several new $\gamma$-ray pulsars (e.g. Kaspi
et al. 2000, Zhang, Zhang \& Cheng 2000, Torres, Butt \& Camilo
2001, Camilo et al. 2001, D'Amico et al. 2001, Mirabal et al.
2000, Mirabal \& Halpern 2001, Halpern et al. 2002). Pulsars
remain as the only confirmed low-latitude population, since pulsed
$\gamma$-ray radiation has been already detected for at least six
different sources (Thompson et al. 1999, Thompson 2001). Other
populations might include X-ray transients (Romero et al. 2001),
persistent microquasars (Paredes et al. 2000, Grenier 2001,
Kaufman-Bernad\'o et al. 2002), massive stars with strong stellar
winds (Benaglia et al. 2001, Benaglia \& Romero 2003), isolated
and magnetized stellar-size black holes (Punsly 1998a,b; Punsly et
al. 2000), and middle-mass black holes (Dermer 1997). Finally,
there is a possibility that some $\gamma$-ray sources could be
generated by supernova remnants (SNRs), especially those
interacting with, or located close to, molecular clouds (e.g.
Montmerle 1979; Dorfi 1991, 2000; Aharonian, Drury \& V\"olk 1994;
Naito \& Takahara 1994, Combi \& Romero 1995; Aharonian \& Atoyan
1996; Sturner, Dermer \& Mattox 1996; Esposito et al. 1996; Combi
et al. 1998, 2001; Butt et al. 2001). This review is devoted to
discuss this latter possibility in light
of recent observations. \\

SNRs are thought to be the main source of both cosmic ray (CR)
ions and electrons with energies below the knee in the galactic CR
spectrum, at $\sim10^{15}$ eV -- however, see Plaga (2002) for
alternate theories. The particle
acceleration mechanism in individual SNRs is usually assumed to be
diffusive shock acceleration, which naturally leads to a power-law
population of relativistic particles. In the standard version of
this mechanism (e.g. Bell 1978), particles are scattered by
magnetohydrodynamic waves repeatedly through the shock front.
Electrons suffer synchrotron losses, producing the non-thermal
emission from radio to
X-rays usually seen in shell-type SNRs. The maximum energy achieved
depends on the shock speed and age as well as on any
competing loss processes. In young SNRs, electrons can easily reach
energies in excess of 1 TeV, where they produce X-rays by
synchrotron mechanism (see, for example, Reynolds 1996, 1998).
Non-thermal X-ray
emission associated with shock acceleration has been clearly
observed in at least 11 SNRs, and this number seems to be steadily
increasing with time. In the case of the very nearby
remnant RX J0852.0-4622 (also known as Vela Jr.) the discovery was
originally made at X-rays (Aschenbach 1998) and only then the
source was detected at radio wavelengths (Combi, Romero \& Benaglia 1999).\\

\begin{figure}
\begin{center}
\caption{Distribution of the Green's SNRs (circles) together
with EGRET unidentified sources (diamonds), shown in galactic
coordinates. Some of the coincident pairs that are studied in this
report are marked. The top marks are SNRs for which TeV radiation
has been detected. From Mori (2001).} \label{snr-egret}
\end{center}
\end{figure}

As early as 1979, Montmerle suggested that SNRs within OB stellar
associations, i.e. star forming regions with plenty of molecular
gas, could generate observable $\gamma$-ray sources. Montmerle
himself provided statistical evidence for a correlation between
COS-B sources and OB associations. Pollock (1985) presented
further analysis of some COS-B sources in the same vein.
Statistical correlation studies of EGRET sources and SNRs have
been presented by Sturner \& Dermer (1995), Sturner et al. (1996),
Yadigaroglu \& Romani (1997), and Romero et al. (1999a). These
studies show that there is a high-confidence correlation between
remnants and $\gamma$-ray sources. Figure~\ref{snr-egret} shows
the distribution of the SNRs in Green's Catalog (2000) along with
the 3EG unidentified sources. Some of the coincident pairs that
are studied in this report are marked. It is worth noticing,
however, that the EGRET observations may contain important
information about the hadronic component of cosmic rays concerning
only the low-energy domain (typically less than several tens of
GeV). Hence, these observations alone cannot solve the problem of
galactic cosmic rays. The latter effort require TeV observations,
and that is why we also discuss future TeV observations of EGRET
sources in this review. Note, in passing, that whereas an all-sky EGRET map
is available, only pointed TeV observations are possible.\\

SNRs can produce high-energy $\gamma$-rays through nucleus-nucleus
interactions leading to $\pi^0$-production and subsequent
$\gamma$-decays. The resulting $\gamma$-ray luminosity will depend
on the local enhancement of the CR energy density as well as on
the density of the ambient media. However, as shown below in Eq.
(\ref{ah1}), the expected fluxes of $\pi^0$-decay gamma-rays from
a SNR are generally well below the EGRET sensitivity (quite
importantly, the estimate given by Eq. (\ref{ah1}) is almost
independent of the proton spectrum). Thus, any detection of
gamma-ray flux from a supernova remnant neighborhood would imply a
significant enhancement of gamma-ray production. Such an
amplification would be possible only through the assumption of the
interaction of CRs accelerated by the shell of the SNR with a
nearby high density environment - e.g. a dense molecular cloud.
GeV $\gamma$-rays (and also TeV photons, see e.g. Pohl 1996) can
be produced also by electrons through relativistic Bremsstrahlung
and inverse Compton upscattering of cosmic microwave background
photons, diffuse Galactic infrared/optical radiation, and/or the
radiation field of the remnant itself (e.g. Mastichiadis 1996, de
Jager \& Mastichiadis 1997). Gaisser et al. (1998) modeled these
processes in detail in order to fit the observational data for the
SNRs IC 443 and $\gamma$-Cygni. Sturner et al. (1997) and Baring
et al. (1999) also modeled IC 433 with synchrotron emission in the
radio band and relativistic Bremsstrahlung in $\gamma$-rays. De
Jager \& Mastichiadis (1997) included inverse Compton scattering
in their model of SNR W44. Bykov et al. (2000) have recently
analyzed the non-thermal emission from a SNR interacting with a
molecular cloud, modeling it as a highly inhomogeneous structure
consisting of a forward shock of moderate Mach number, a cooling
layer, a dense radiative shell, and an interior region filled with
hot tenuous plasma. Particularly for SNRs with mixed morphology
(remnants which are shell-like in radio and dominated by central
emission in X-rays, Rho \& Petre 1998), they found that
Bremsstrahlung, synchrotron, and inverse Compton radiation of the
relativistic electron population produce multiwavelength photon
spectra in quantitative agreement with radio and high-energy
observations. These are only some of the works devoted to
high-energy emission from SNRs published in recent years.
Differentiating the $\gamma$-ray emission produced by ions from
that originating in leptons is crucial for determining the origin
of cosmic-ray nuclei (for some recent reviews and more references
the reader is referred to V\"olk 2001, 2002; Drury et al. 2001;
Kirk \& Dendy 2001).\\

After introducing a simple theoretical model for evaluating the
possible hadronic $\gamma$-ray emission from SNRs and nearby
clouds, we characterize the sample to be investigated, discuss the
$\gamma$-ray flux variability of the sources, and study the
possibility that pulsars might be possible counterparts. For
each SNR-EGRET source pair we review and analyze the different
scenarios proposed as an explanation of the $\gamma$-ray emission.
We present CO(J=1-0) mm wavelength observations to evaluate
whether there are molecular clouds in the vicinity of the SNRs and
estimate the $\gamma$-ray flux that would be produced in each case
via $\pi^0$-decays. Some new radio continuum maps are also
presented. These latter maps have been processed to eliminate, as
far as possible, the galactic contaminating diffuse emission.\\

\begin{figure}
\begin{center}
\caption{Expected distribution of source locations  (Galactic
coordinates) for one year survey of the LAT experiment onboard
GLAST. Courtesy of the GLAST Science Team and NASA.} \label{lat}
\end{center}
\end{figure}


Our aim with this review is to provide a quantitative basis to
analyze the possible $\gamma$-ray production in SNRs, providing the
reader with useful
information to guide future studies. Specifically, the role of INTEGRAL,
AGILE, and GLAST satellites, and the \v{C}erenkov telescopes HESS, VERITAS,
MAGIC, and CANGAROO III is discussed. Several target candidates for
observations with all these telescopes and satellites are
mentioned. As an example, in the GeV band, the Gamma-ray Large
Area Telescope (GLAST), which will be launched in a few years, is
expected to detect $\sim 10^4$ high-energy $\gamma$-ray sources,
thousands of them belonging to our Galaxy (see
Figure~\ref{lat}).  A technical Appendix quotes the main features of GLAST,
as well as its predecessors, AGILE and INTEGRAL, for quick reference.
It is also important to clarify
what this review is not about. Many authors have studied the
evolution of SNRs or their
emission properties from very sophisticated numerical modeling
points of view, during the last years. We shall not particularly
focus on those, except briefly when dealing
with the case by case analysis for specific SNR-EGRET source pairs.\\

The rest of this work is organized as follows. In the next Section we
introduce a simple model
to account for the hadronic $\gamma$-ray emission in SNRs and their
neighborhoods. Section 3 refers to
Relativistic Bremsstrahlung as a competing process. Section 4
analyzes the spectral changes that
difussion could produce on the observed $\gamma$-ray spectrum. The
general characteristics
of the SNR sample that we shall analize and the possible pulsar
counterparts are presented in Sections
5, 6, and 7. The variability in the $\gamma$-ray emission for the
Third EGRET sources
under study is assesed in Section 8. Section 9 gives a brief account
of the observations and data extraction
techniques used in this review. A case by case analysis of all
coincident pairs between SNRs and
$\gamma$-ray sources is given in Section 10.  Some particular cases
in which SNRs were discovered
by their high energy emission are discussed in Section 11. The
TeV-emission properties and the
prospects for new observations using new TeV-telescopes are discussed
in Section 12. Finally,
Section 13 presents a very brief overview and some concluding remarks.

\section{ Phenomenological model for the hadronic $\gamma$-ray emission in SNRs
and their environs}

We first present a simple model for the hadronic $\gamma$-ray
emission from ``bare" SNRs, and those interacting with molecular
clouds. Further details can be obtained from Dorfi (1991, 2000),
Drury et al. (1994), Aharonian et al. (1994), and Aharonian \&
Atoyan (1996). We shall partially follow Morfill et al. (1984) and
Combi \& Romero (1995) in our presentation. Our intention is not
to arrive at the most precise theoretical model for an individual
SNR, but rather to have a simple, straightforward and robust,
albeit crude, method of obtaining and inter-comparing
$\gamma$-ray fluxes due to nucleus-nucleus interactions in
interacting SNRs.\\

Let us consider the expansion of a SNR in a homogeneous medium.
If this expansion is adiabatic, we can use the Sedov's solutions
(1959), which give the time since the explosion and the velocity
of the shock front, respectively, as \be t \sim 1.5\; \times 10^3
n_{-1}^{1/2} E_{51}^{-1/2} R_1^{5/2} {\rm yr}, \ee \be v_{\rm s}
\sim 21.6 \; \times 10^2 n_{-1}^{-1/2} E_{51}^{1/2} R_1^{-3/2}
{\rm km\; s}^{-1}.\ee Here, $E_{51}$ is the energy of the SN
explosion in units of $10^{51}$ erg, $R_1$ is the SNR radius in
units of 10 pc, and $n_{-1}$ is the medium density in units of 0.1
cm$^{-3}$. The CR energy per time unit incoming to the SNR is \be
\dot E(t) = k_{\rm s}\; \epsilon_{\rm CR} \;4\pi R_{\rm s}(t)^2\;
v_{\rm s}(t), \ee where the dot means derivative with respect to
time, $R_{\rm s}$ is the SNR time-dependent radius, $\epsilon_{\rm
CR}$ is the background CR ambient density ($\sim$ 1 eV cm$^{-3}$
in the solar neighborhood), and $k_{\rm s}$ is the enhancement
factor due to re-acceleration by Fermi mechanism at the shock
front (see Jones 2001 for a recent review and references on
acceleration details). If we assume equipartition, i.e. that the
energy flux from the unshocked medium is converted in equal parts
into electromagnetic energy, thermal energy, and CR enhancement
(Morfill et al. 1984), we can write the previous expression as \be
\dot E \sim \frac{4\pi}{3} \frac{1-\xi}{1-2\mu n v_{\rm s}^3}
R_{\rm s}^2,\ee where $\xi$ is the downstream to upstream ratio of
kinetic energy flux in the shock frame ($\xi \sim 0.06 $ for
strong shocks), $\mu$ is the mean molecular weight, and $n$ is the
unshocked particle density. This expression states that the power
available for accelerating CRs is 1/3 of the mechanical energy
flux across the shock.\\

When the SNR expands from a radius comprising a volume $V(t_1)$ to
a volume $V(t_2)$, the energy decreases accordingly as \be \frac
{E(t_2)}{E(t_1)}= \left( \frac {V(t_1)}{V(t_2)} \right)^{\gamma -1
},\ee with the adiabatic index being $\gamma=4/3$, Using the
previous equations, the CR energy in the SNR between times $t_1$
and $t_2$ is
  \be E_{\rm CR}(t_1,t_2)= \frac {2\pi \mu n
(1-\xi)}{3 R_{\rm s}(t_2)^{3(\gamma-1)}} \int_{t_1}^{t_2} dt \; v_{\rm s}^3\;
R_{\rm s}(t)^{3(\gamma-1)}. \label{g}\ee
Through the Sedov solutions, this
leads to a ratio
\be \theta = \frac{E_{\rm CR}}{E_{\rm SN}}\sim
\frac{\pi}{5} \left( 1 - \left( \frac{t_1}{t_2} \right)^{2/5}
\right).\label{mor}\ee
We shall adopt $t_2$ as the actual age of the SNR,
estimated from observational data and the
Sedov solutions, and we shall assume the initial time $t_1$ as that
obtained when the SNR has swept about $5M_\odot$ of interstellar
material, starting then the Sedov phase (Lozinskaya
1992, pp. 205ff). The radius at which this happens is
$R_{\rm s}=(3 \cdot 5M_\odot/4\pi\mu m_H n_0)^{1/3}$, and
typical values for $t_1$ are in the range 200-2000 years.\\

In simplified models of SNRs, the remnant is divided into three
regions:  an interior region filled with hot gas and accelerated
particles but very little mass, an immediate post shock region
where most of the matter is concentrated, and a shock precursor
region where the accelerated particles diffusing ahead of the
shock affect the ambient medium. Following Drury et al. (1994) the
production rate of $\gamma$-rays per unit volume can be written as
\begin{equation}
Q_\gamma = {\cal E}_\gamma n = q_\gamma n E_{\rm CR},
\end{equation}
where $n$ is the number density of the gas, $E_{\rm CR}$ is the CR
energy density, and $q_\gamma$ is the $\gamma$-ray emissivity
normalized to the CR energy density, $q_\gamma={\cal
E}_\gamma/E_{\rm CR}$. The total gamma-ray luminosity is given by
$\int q_{\gamma} n E_{\rm CR}\,d^3r$, which can be written as
$q_\gamma\left(M_1E_{{\rm CR}\,1}+M_2E_{\rm CR\,2}\right)$, where
$M_1$ is the total mass in the precursor region, $M_2$ that in the
immediate post-shock region, and $E_{{\rm CR}\,1,2}$ are the
corresponding CR energy densities. Since particle diffusion occurs
across the shock front, we have $E_{{\rm CR}\,1}=E_{{\rm CR}\,2}$.
This value is also probably not very different from $E_{{\rm
CR},3}$, the energy density in the interior of the remnant,
because of two reasons. First, there is diffusive coupling between
the acceleration region around the shock and the interior of the
remnant (Drury et al. 1994). Second, if the acceleration is
efficient, CRs provide a substantial, if not the dominant, part of
the interior pressure and the interior of the remnant has, for
dynamical reasons, to be in pressure equilibrium. It follows that,
to order of magnitude, the CR energy density throughout the
remnant and in the shock precursor can be taken as $E_{{\rm
CR}\,1}=E_{{\rm CR}\,2}\approx E_{{\rm CR}\,3}\approx 3 \theta
E_{\rm SN}/4\pi R^3$, where $\theta$ is, again, the fraction of
the total supernova explosion energy, $E_{\rm SN}$, converted to
CR energy and $R$ is the remnant radius. Thus, the $\gamma$-ray
luminosity results
\begin{equation} L_\gamma = q_\gamma (M_1+M_2) {3 \theta E_{\rm
SN}\over 4\pi R^3}
\approx \theta q_\gamma E_{\rm SN} n
\approx  10^{38} \theta \left( \frac{E_{\rm SN}}{10^{51}{\,\rm erg}}\right)
\left(\frac{n}{1\,\rm cm^{-3}}\right){\rm ph\;  s^{-1}},
\end{equation}
where $n$ is the ambient density. The exact value of $\theta$
depends on the details of the model, for which Eq. (\ref{mor})
gives an example. For different plausible injection models,
$\theta$ is roughly constant throughout the Sedov phase with only
a moderate dependence on external parameters such as the ambient
density (Markiewicz et al. 1990). If the SNR is located at a
distance $d$, the hadronic $\gamma$-ray flux is \be \label{ah1}
F(>100 {\rm MeV})_{{\rm SNR}} \sim 4.4 \; \times 10^{-7} \theta
\left( \frac{E_{\rm SN}}{10^{51} {\rm erg}} \right) \left( \frac
{d}{{\rm kpc}} \right)^{-2} \left( \frac {n}{{\rm cm}^3} \right)
{\rm ph\; cm^{-2} s^{-1}},\ee where $d$ is the distance to the
remnant. Only for very high densities can the usually observed
$\gamma$-ray sources can be due to the remnant  itself. In
general, the flux provided by the previous equation is far too low
to produce a detectable EGRET source
(Drury et al. 1994), at typical galactic distances.\\

However, stronger emission can be produced if there are molecular
clouds in the vicinity of the SNR where the locally accelerated
protons interact with target ions, producing pions and hence
enhancing the $\gamma$-ray flux (eg. Montmerle 1979; Dorfi 1991,
2000; Aharonian et al. 1994). The expected total flux is
\be
F_\gamma= \frac{1}{4\pi d^2} \int_{V_0} n(\bar r) q_\gamma ({\bar r}) d^3r.
\ee
Neglecting all possible gradients within the cloud, this equation reduces to
\be
F_\gamma= \frac{M_{{\rm cl}}}{m_p} \frac{q_\gamma}{4\pi d^2},
\ee
where $M_{{\rm cl}}$ is the mass of the cloud. In particular, we may write
\be F(>100 {\rm
MeV})^{{\rm cloud}} \sim 10^{-9} M_3 \left( \frac {d}{{\rm kpc}}
\right)^{-2} q_\gamma(>100 {\rm
MeV}) \; {\rm ph\; cm^{-2} s^{-1}}, \label{hadro}
\ee
where $M_3$ is the mass of the target cloud in units of $10^3
M_\odot$, and $q_\gamma$ is the $\gamma$-emissivity in units of
$10^{-25}$ s$^{-1}$ (H -- atom)$^{-1}$. The factor $q_\gamma$ will
be enhanced in comparison with its normal value because of the local
CR source. In a
passive giant molecular cloud exposed to the same
proton flux measured at the Earth,
the $\gamma$-ray emissivity above 100 MeV is equal to
$1.53  \ \eta \  q_{-25}(\geq 100 \, \rm MeV) \rm (H-atom)^{-1}
s^{-1}$, where the parameter $\eta  \simeq 1.5$ takes into account
the contribution of nuclei both in CRs and in the
interstellar medium (Dermer et al. 1986; Aharonian 2001). In clouds near
CR accelerators, it may be much higher than this value. If the
shape of the CR spectrum in the cloud does not differ much from
that existing near the Earth, we can approximate
\be
\frac {q_\gamma}{q_{\gamma,0}} \sim \frac{\epsilon_{\rm
CR}}{\epsilon_{CR,0}} \sim k_{\rm s} .\ee Following Morfill \&
Tenorio Tagle (1983), we can use Eq. (\ref{g}) to obtain \be
k_{\rm s}=\frac {3}{20} \frac{E_{\rm SN}}{R_{\rm s}(t_2)^3} \left(
1- \frac{R_{\rm s}(t_1)}{R_{\rm s}(t_2)} \right) \frac
1{\epsilon_{\rm CR}}. \label{kk}\ee The Sedov solutions for each
of the SNRs considered below should be used in this latter
expression to obtain $k_{\rm s}$, and thus $F(E>100 {\rm
MeV})^{{\rm cloud}}$. One immediate test of energetic consistency
is to check that $E_{\rm CR}=k_{\rm s} \epsilon_{\rm CR} / (4/3)
\pi R^3 < E_{\rm SN}$, for the obtained value of $k_{\rm s}$ and
the assumed value of $E_{\rm SN}$. If the previous inequality is
not valid, one or more of the simplifying assumptions of the model
are not correct for the particular case under analysis.\\

The expected $\gamma$-ray flux in the TeV region
by a SNR is (Aharonian et al. 1994)
\be F_\gamma (>E) = f_\Gamma \; 10^{-10} \; \left(
\frac{E}{{\rm TeV}} \right)^{-\Gamma+1} \; A\; {\rm cm^{-2}
s^{-1}},\ee
where the factor $A$ is \be A= \theta \left(
\frac{E_{\rm SN}}{10^{51} {\rm erg}} \right) \left( \frac {d}{{\rm
kpc}} \right)^{-2} \left( \frac {n}{{\rm cm}^3} \right) {\rm ph\;
cm^{-2} s^{-1}},\ee $n$ is the medium density, and $f_\Gamma$
is a function of the index in the differential power-law proton
spectrum ($\Gamma$),
equal to 0.9, 0.43, and
0.19 for $\Gamma$=2.1, 2.2, and 2.3, respectively. This
estimate, however, usually exceeds that obtained when the GeV
spectral index is extrapolated up to TeV energies. When that is
the case, the extrapolated flux (with the same spectral index)
will be considered a safer estimate. To extrapolate the GeV
flux up to TeV energies we assume (Thompson et al. 1996)
\be \frac{dN}{dE} = K
\left(\frac{E}{E_0}\right)^{-\Gamma},\ee
where $K$ is a constant, $E$ is given in MeV, and $E_0$ is a reference energy.
This
constant can be obtained simply by integrating the flux,
\be
K \equiv F_{\rm ph}\; /\; \int_{100 {\rm MeV}}^{10 {\rm GeV}}
\left(\frac{E}{E_0}\right)^{-\Gamma}\; dE, \ee where $F_{\rm ph}$
is the observed total flux (that quoted in the 3EG Catalog, for
instance). Once $K$ is known, the flux in any given energy
interval $E_1$ -- $E_2$ is just, \be F(E_1,E_2)=
K\;\int_{E_1}^{E_2} \left(\frac{E}{E_0}\right)^{(-\Gamma + 1)}\;
dE .\ee We are extrapolating, then, the measured spectral index at
MeV-GeV energies assuming that there is no spectral change at
higher energies. Actually, this assumption is a simplification not
compatible with TeV observations of several sources (see below).
In any case, this extrapolation will always provide an upper
bound to the high-energy photon flux.

\section{ Relativistic Bremsstrahlung}

Since relativistic electron Bremsstrahlung and nucleus-nucleus
induced pion-decay are competing processes in the generation of
$\gamma$-rays from molecular clouds exposed to a nearby CR
accelerator, it is necessary to assess the relative weight of each
contribution if we are to quantitatively address the question of the
possible SNR origin of nucleonic CRs.\\

The $\gamma$-ray emissivity at a given energy $E$ from
relativistic Bremsstrahlung is (e.g. Longair 1994, p.267-269) \be
q_B(E)=\frac{10^{-21}}{p-1} n_{{\rm m}^{-3}}\; K E^{-p}\; {\rm
m}^{-3} \;{\rm s}^{-1} \;{\rm GeV}^{-1} ,\ee where it is assumed
an electron power-law distribution, $N_e(E)=KE^{-p}$. We are
interested in the $\gamma$-ray radiation above 100 MeV, so we
integrate the previous expression to obtain \be q_B(E>0.1 {\rm
GeV})=\frac{10^{-21}}{p-1} n_{{\rm m}^{-3}}\; K \int_{0.1 {\rm
GeV}}^\infty E^{-p} dE\;\; {\rm m}^{-3} \;{\rm
s}^{-1}=\frac{10^{-21}}{(p-1)^2} n_{{\rm m}^{-3}}\; K\;
0.1^{-(p-1)}\;\; {\rm m}^{-3} \;{\rm s}^{-1}.\ee This same
population of relativistic electrons will also radiate at radio
wavelengths, via the synchrotron mechanism. The synchrotron
spectrum of a power-law electron energy distribution is (Longair
1994, p. 261) \be J(\nu)=
23.44 \; a(p)\; B^{(p+1)/2} \; K \; \left( \frac{1.253 \; \times
10^{37}}{\nu} \right)^{(p-1)/2} {\rm Jy\; m^{-1} } ,\ee where $B$
is the magnetic field measured in Tesla, and \be a(p) =
\frac{\sqrt{\pi}}{2} \frac{\Gamma(\frac p4+\frac{19}{12})
\Gamma(\frac p4-\frac1{12})\Gamma(\frac p4+\frac54)}{(p+1)
\Gamma(\frac p4+\frac74)},\ee is a numerical coefficient depending
on the spectral index. Then, the ratio between the $\gamma$-ray
flux emitted by relativistic Bremsstrahlung and the synchrotron
emission results \be \label{ratio} R= \frac{q_B(E>100{\rm
MeV})}{J(\nu)}=\frac{F(E>100{\rm MeV})}{F_\nu [{\rm Jy}]}=
\frac{4.3 \; \times 10^{-21}}{b(p)}\;
  n_{\rm cm^{-3}} B_{\mu {\rm G}}^{-(1+p)/2} \;
\nu_{{\rm Hz}}^{(p-1)/2}\; {\rm Jy}^{-1}\; {\rm cm}^{-2}\;{\rm
s}^{-1},\ee
where we have defined
\be b(p)= 10^{-5(1+p)} (3.2 \;
\times 10^{15})^{(p-1)/2}\;(p-1)^2\;a(p),\ee
and converted units to the cgs system.\\

If $F(E>100\; {\rm MeV})$ is known,  estimating the right hand
side of Eq. (\ref{ratio}) for the measured spectral photon index
and the derived density and magnetic field, the expected value of
$F_\nu [{\rm Jy}]$ can be obtained. This is the radio emission
that {\sl should be observed if the $\gamma$-rays are from
relativistic Bremsstrahlung}. If the $\gamma$-ray source is not
superposed with the bulk of the synchrotron radio/X-ray emission
from the SNR, this tends to favor a nucleus-nucleus origin of the
high-energy flux, rather than a electron Bremsstrahlung scenario.
In general, at the high densities found in molecular clouds,
inverse Compton scattering can be ruled out as the main mechanism
contributing to the $\gamma$-ray emission (e.g. Gaisser et al.
1998). Above 100 MeV, the relevant cross sections and estimates of
the electron-nucleon density ratio show that relativistic
Bremsstrahlung dominates over inverse Compton processes (see, for
instance, Stecker's 1977, Figure 1 and 2). In particular, de Jager
\& Mastichiadis (1997) have shown that for molecular densities
above 10 cm$^{-3}$ $\gamma$-ray fluxes above 70 MeV are dominated
by Bremsstrahlung when electrons are considered (see their Figure
4). In what follows, since we shall mainly consider high-density
scenarios, relativistic Bremsstrahlung will be the main
alternative to nucleonic interactions in evaluating the origin of
the observed $\gamma$-ray flux in the MeV-TeV energy range.\\

\section{ Diffusion of CRs and $\gamma$-ray spectral evolution \label{difff}}

The spectrum of $\gamma$-rays generated through $\pi^0$-decay at
a source of proton density $n_{p}$ is
\begin{equation}\label{Fgam}
   F_{\gamma}(E_{\gamma})=2\int^{\infty}_{E_{\pi}^{\rm min}}
   \frac{F_{\pi}(E_{\pi})}{\sqrt{E_{\pi}^{2}-m_{\pi}^{2}}}
   \;dE_{\pi},
\end{equation}
where
\begin{equation}\label{Epi}
E_{\pi}^{\rm
min}(E_{\gamma})=E_{\gamma}+\frac{m_{\pi}^{2}}{4E_{\gamma}},
\end{equation}
and
\begin{equation}\label{Fpi}
F_{\pi}(E_{\pi})=4\pi n_{p}\int^{E^{\rm max}_{p}}_{E^{\rm
min}_{p}} J_p(E_p) \frac{d\sigma_{\pi}(E_{\pi},\;E_{p})}{dE_{\pi}}
\;dE_{p}.
\end{equation}
Here, $d\sigma_{\pi}(E_{\pi},\;E_{p})/dE_{\pi}$ is the
differential cross-section for the production of $\pi^0$-mesons of
energy $E_{\pi}$ by a proton of energy $E_{p}$ in a $p-p$
collision. If the proton spectrum $J_p(E_p)$ at the $\gamma$-ray
production site is
\begin{equation}\label{powerlawp}
   J_p(E_p)=K E_p^{-\Gamma},
\end{equation}
we can also expect a power-law spectrum at $\gamma$-rays:
\begin{equation}\label{powerlawg}
   F_{\gamma}(E_{\gamma})\propto E_{\gamma}^{-\Gamma}.
\end{equation}
However, the spectrum given by Eq. (\ref{powerlawp}) is not
necessarily the same proton spectrum at the acceleration site. If
there is diffusion, we shall have, instead
\begin{equation}\label{f}
   J_p(E_p,\;r,\;t)=\frac{c}{4\pi} f,
\end{equation}
where $f(E_p,\;r,\;t)$ is the distribution function of protons at
an instant $t$ and distance $r$ from the source. The distribution
function satisfies the well-known diffusion equation (Ginzburg \&
Syrovatskii 1964):
\begin{equation}\label{difeq}
   \frac{\partial f}{\partial t}=\frac{D(E_p)}{r^2} \frac{\partial}{\partial
   r} r^2 \frac{\partial f}{\partial r} + \frac{\partial}{\partial
   E_p}(Pf)+Q,
\end{equation}
where $P=-dE_p/dt$ is the continuous energy loss rate of the
particles, $Q=Q(E_p,\;r,\;t)$ is the source function, and $D(E_p)$
is the diffusion coefficient, for which we assume here no
dependency on $r$ or $t$, i.e. the particles diffuse through an
homogeneous, quasi-stationary medium.\\

We assume that $D(E_p)\propto E_p^\delta$ and $f\propto
E_p^{-\Gamma}$ with continuous injection given by $Q(E_p, \;
t)=Q_0 E_p^{-\Gamma} q(t)$, which is appropriate for a supernova
remnant (Aharonian \& Atoyan 1996). Further simplicity can be
achieved assuming that the source is constant after turning on at
some instant, i.e. $q(t)=0$ for $t<0$ and $q(t)=1$ for $t\geq 0$.
Atoyan et al. (1995) have found a general solution for Eq.
(\ref{difeq}) with arbitrary injection spectrum, which with the
listed assumptions leads to:
\begin{equation}\label{solf}
f(E_p,\;r,\;t)=\frac{Q_0 E_p^{-\Gamma}}{4\pi D(E_p) r}\left(
\frac{2}{\sqrt{\pi}}\right)\int^{\infty}_{r/R_{\rm diff}} e^{-x^2}
dx.
\end{equation}
In this expression, $R_{\rm diff}=R_{\rm diff}(E_p,\; t)$ is the
diffusion radius which corresponds to the radius of the sphere up
to which the particles of energy $E_p$ propagate during the time
$t$ after the injection. Now, for $D(E_p)=a D_{28} E_p^{\delta}$,
where $D_{28}=D/10^{28}$ cm$^{2}$ s$^{-1}$, and for $R_{\rm
diff}\gg r$, i.e. when the target is well immerse into the cosmic
ray flux, Eq. (\ref{solf}) reduces to
\begin{equation}\label{solfred}
f(E_p,\;r)=\frac{Q_0 E_p^{-(\Gamma+\delta)}}{4\pi a D_{28} r},
\end{equation}
and then, from Eq. (\ref{f}), we get:
\begin{equation}\label{Jmod}
J_p(E_p,\;r)=\frac{c Q_0 E_p^{-(\Gamma+\delta)}}{(4\pi)^{2} a
D_{28} r}.
\end{equation}
Hence, as has been emphasized by Aharonian \& Atoyan (1996), the
observed $\gamma$-ray flux $F_{\gamma}(E_{\gamma})\propto
E_{\gamma}^{-(\Gamma+\delta)}$ can have a significantly different
spectrum from that expected from the particle population at the
source (the SNR). Standard diffusion coefficients $\delta\sim
0.3-0.6$ can explain $\gamma$-ray spectra as steep as $\Gamma\sim
2.3-2.6$ in sources with particles accelerated to a power-law
$J_p(E_p)\propto E^{-2}$ if the target is illuminated by the
$\pi^0$-decays are at sufficient distance from the accelerator.
This can explain observed discrepancies in the particle spectral
indices inferred from SNR at different frequencies, {\em even if
all particles, leptons and hadrons, are accelerated to the same
power-law in the source}.\\

\begin{figure}
\begin{center}
\caption{Temporal and spectral evolution of CR fluxes at different
(10 pc, 30 pc, and 100 pc) distances from an {\it impulsive}
proton accelerator. A power-law proton spectrum with $\Gamma =2.2$
and total energy $W_{\rm p}=10^{50} \, \rm erg$ are assumed.
Curves 1, 2, 3, and 4  correspond to an age of the source of
$t=10^3\,\rm yr$, $10^4 \,\rm yr$, $10^5 \,\rm yr$, and $10^6\,\rm
yr\,$, respectively. An energy-dependent diffusion coefficient
$D(E)$ with power-law index $\delta =0.5$ is adopted. The left
panel presents results for $D=D_{28} \,\rm cm^2 s^{-1}$, whereas
the right panel corresponds to $D=10^{-2}D_{28} \,\rm cm^2
s^{-1}$. The hatched curve shows the local (directly measured)
flux of CR protons. More details are given in Aharonian \& Atoyan
(1996).} \label{dif}
\end{center}
\end{figure}

CRs with total energy $W_{\rm p}$ and injected in the interstellar
medium by some local accelerator reach a radius $R(t)$ at instant
$t$. Their mean energy density is $w_{\rm p} \approx 0.5\, (W_{\rm
p}/10^{50} \, \rm erg)(R/100 \, \rm pc)^{-3} \; \rm eV cm^{-3}$.
Thus, in regions up to 100 pc around a CR accelerator with $W_{\rm
p} \sim 10^{50} \, \rm erg$, the density of relativistic particles
may significantly exceed the average level of the ``sea'' of
galactic CRs, $w_{\rm GCR} \sim 1 \, \rm eV/cm^3$. In Figure
\ref{dif}, the differential flux of protons at distances $R=$10,
30, and 100 pc from an ``impulsive'' accelerator,  with total
energy $W_{\rm p}=10^{50} \, \rm erg$ are shown. The spectrum of
CRs at the given time and spatial location can
differ significantly from the source spectrum. The diffusion
coefficient in this
figure is assumed in a power-law form, $D(E) \propto E^{0.5}$
above 10 GeV, and constant below 10 GeV. The commonly adopted
value at 10 GeV is about $10^{28} \, \rm cm^2 s^{-1}$, however smaller
values,  e.g. $10^{26} \, \rm cm^2 s^{-1}$, cannot be excluded,
especially in active star forming regions (Aharonian 2001). The
existence of massive gas targets like molecular clouds in these
regions may result in $\gamma$-ray fluxes detectable by EGRET, if
$(W_{50} \cdot M_5)/d_{\rm kpc}^2 \geq \sim 0.1$, where $M_5$ is
the mass of the cloud in units of $10^5 M_\odot$ (Aharonian 2001).
In the case of energy-dependent propagation of CRs, large variety
of $\gamma$-ray spectra is then expected, depending on the age of
the accelerator, duration of injection, the diffusion coefficient,
and the location of the cloud with respect to the accelerator. \\

\begin{figure}[t]
\begin{center}
\caption{Gamma-ray  emissivities in terms of $E_{\gamma}^2 \times
q_\gamma(E_{\gamma})$ at different times $t$ and distances $R$
from a proton accelerator. The right hand side axes shows the
$\gamma$-ray fluxes, $E_{\gamma}^2 \times F_\gamma(E_{\gamma})$,
which are expected from a cloud  with parameter $M_5/d_{\rm kpc}^2
=1$. The thin and bold curves correspond to times $t=10^3$ and
$10^5\,\rm yr$, respectively. Fluxes at distances $R=10$ and
$30\,\rm pc$ are shown by solid and dashed lines. The power-law
index, the total energy of protons, and the diffusion coefficient
are the same as in the right panel of Figure \ref{dif}. The curve
shown by full dots corresponds to  the $\gamma$-ray emissivity
(and flux) calculated for local CR protons. In order to take into
account the contribution of nuclei, all curves should be increased
by a factor of $\eta \approx 1.5$. More details can be found in
Aharonian \& Atoyan (1996).} \label{difc}
\end{center}
\end{figure}

The comparison of $\gamma$-ray fluxes from clouds located at
different distances from an accelerator may provide unique
information about the CR diffusion coefficient $D(E)$. Similar
information may be obtained from a {\it single} $\gamma$-ray
emitting cloud, but in different energy domains. For the
energy-dependent propagation of CRs  the probability for
simultaneous detection of a cloud in GeV and TeV $\gamma$-rays is
not very high, because the maximum fluxes at these energies are
reached at different epochs (see Figure \ref{difc}). The higher
energy particles propagate faster and reach the cloud earlier,
therefore the maximum of GeV $\gamma$-radiation appears at the
epoch when the maximum of the TeV $\gamma$-ray flux is already
over. In the case of energy-independent propagation (e.g. due to
strong convection) the ratio $F_\gamma(\geq 100 \, \rm
MeV)/F_\gamma(\geq 100 \, \rm GeV)$ is independent of time,
therefore the clouds that are visible for EGRET at GeV energies
would be detectable also at higher energies, provided that  the
spectral index of the accelerated protons be  $\Gamma \leq 2.3$.\\

Summing up, special care must be taken in analyzing the a priori
expectations for the SNR-cloudy medium scenario: when particles
diffuse into the ISM before reaching the cloud, a quite different
spectrum from that of the injecting source can finally emerge,
depending on the distance to the cloud and the diffusion index.
Even when molecular clouds are overtaken by the expanding shell
of the SNR and  there is clear evidence of interaction, a strong
magnetic field could produce differences in the spectra.

\section{ Sample and correlation analysis}

\begin{sidewaystable}
\begin{center}
\caption{Positional coincidences between supernova remnants quoted
in Green's Catalog (2000), and unidentified 3EG EGRET sources.
See
text for the meaning of the different columns. }\vspace{0.2cm}
{\small
\begin{tabular}{llllllllllll}
\hline $\gamma$-source & $F_{\gamma} $ & $\Gamma$ & Class  &$I$&
$\tau $& P? &
${\rm SNR}$ & Other name & $\Delta\theta$ & Size & T  \\
\noalign{\smallskip} \hline \noalign{\smallskip}
0542$+$2610 & 14.7$\pm$3.2 & 2.67$\pm$0.22 & em C& 3.16 &
$0.70_{0.34}^{1.40}$& &G180.0$-$1.7& &
2.04& 180 & S  \\
0617$+$2238$^{1,2}$ & 51.4$\pm$3.5 & 2.01$\pm$0.06 &  C& 1.68&
$0.26_{0.15}^{0.38}$& &G189.1$+$3.0& IC443 &
0.11& 45 & S  \\
0631$+$0642$^{1,3}$  & 14.3$\pm$3.4 &
2.06$\pm$0.15 & C& 1.52 &$75.8_{7.89}^{\infty}$& &G205.5$+$0.5&
Monoceros & 1.97& 220 &  S   \\
0634$+$0521 & 15.0$\pm$3.5 & 2.03$\pm$0.26 & em C& 1.02
&$72.0_{5.15}^{\infty}$& &G205.5$+$0.5 & Monoceros& 2.03& 220 & S \\
1013$-$5915 & 33.4$\pm$6.0 & 2.32$\pm$0.13 & em C& 1.63
&$0.22_{0.00}^{0.46}$& y &G284.3$-$1.8&MSH 10-53 & 0.65& 24 &   S\\
1102$-$6103 & 32.5$\pm$6.2 & 2.47$\pm$0.21 & C & 1.86
&$0.00_{0.00}^{0.90}$&&G290.1$-$0.8 &MSH 11-61A&
0.12& 19 & S   \\
& & &&  && &G289.7$-$0.3 & &0.75& 18 &   S \\
1410$-$6147$^4$  & 64.2$\pm$8.8 & 2.12$\pm$0.14 & C& 1.22
&$0.33_{0.16}^{0.55}$&y&G312.4$-$0.4&
          & 0.23& 38 &  S\\
1639$-$4702 & 53.2$\pm$8.7 & 2.50$\pm$0.18  & em C& 1.95
&$0.00_{0.00}^{0.38}$&y&G337.8$-$0.1 &Kes 41&
0.07& 9  & S   \\
       &  &  &            &   &           &   & G338.1$+$0.4  &       & 0.65& 15
&  S  \\
        &  & &             &  &            &  & G338.3$+$0.0   &      & 0.57& 8
&  S  \\
1714$-$3857 & 43.6$\pm$6.5 & 2.30$\pm$0.20  & em C& 2.17 &
$0.15_{0.00}^{0.38}$&y&G348.5$+$0.0 && 0.47& 10 &  S  \\
         &  &&          &     &             & & G348.5$+$0.1    & CTB
37A    & 0.50&  15
&  S \\
         &&  &         &    &             & & G347.3$-$0.5       &  & 0.85&  65
&  S\\
1734$-$3232$^5$  & 40.3$\pm$6.7 &     --      & C& 2.90&
$0.00_{0.00}^{0.24}$&&G355.6$+$0.0&
& 0.16& 8  & S  \\
1744$-$3011 & 63.9$\pm$7.1 & 2.17$\pm$0.08 & C& 1.80 &
$0.38_{0.20}^{0.62}$&&G359.0$-$0.9& & 0.41&
23 & S \\
           && &        &     &               & & G359.1$-$0.5 &
& 0.25&  24
&  S  \\
1746$-$2851$^6$ &119.9$\pm$7.4 & 1.70$\pm$0.07 & em C& 2.00 &
$0.50_{0.36}^{0.69}$&&G0.0$+$0.0 && 0.12& 3.5   & S \\
           &&  &       &     &              &  & G0.3$+$0.0&           & 0.19&
           16
&  S  \\
1800$-$2338$^{1,7}$  & 61.3$\pm$6.7 & 2.10$\pm$0.10 & C& 1.60 &
$0.03_{0.00}^{0.32}$&y&G6.4$-$0.1 & W28& 0.17&  42 & C   \\
1824$-$1514 & 35.2$\pm$6.5 & 2.19$\pm$0.18 & C& 3.00 &
$0.00_{0.00}^{0.51}$&y&G16.8$-$1.1 && 0.43&  30 & -- \\
1837$-$0423 & $<$19.1         & 2.71$\pm$0.44 & C& 5.41
&$12.0_{2.17}^{\infty}$&y&G27.8$+$0.6 &&
0.58& 50    & F    \\
1856$+$0114$^8$  & 67.5$\pm$8.6 & 1.93$\pm$0.10 & em C& 2.92
&$0.80_{0.50}^{1.51}$&y&G34.7$-$0.4 &W44&
0.17& 35 & S  \\
1903$+$0550$^4$  & 62.1$\pm$8.9 & 2.38$\pm$0.17 &  em C&2.28 &
$0.35_{0.18}^{0.60}$&y&G39.2$-$0.3 &3C396, HC24& 0.41& 8    &  S\\
2016$+$3657  & 34.7$\pm$5.7 & 2.09$\pm$0.11  & C& 2.06 &
$0.37_{0.08}^{0.75}$&&G74.9$+$1.2  &CTB 87        & 0.26& 8   &  F \\
2020$+$4017$^{1,9}$   &123.7$\pm$6.7 & 2.08$\pm$0.04 & C& 1.12
&$0.07_{0.00}^{0.18}$&?&G78.2$+$2.1& W66, $\gamma$-Cygni
   & 0.15&  60 &  S  \\
\noalign{\smallskip} \hline \noalign{\smallskip}
\end{tabular}
} \label{green} \mbox{}\\ {\small $^1$ Association proposed by
Sturner \& Dermer (1995) and Esposito et al. (1996). $^2$ GeV
J0617$+$2237 $^3$ GeV J0633$+$0645. $^4$ Association proposed by
Sturner \& Dermer (1995). $^5$ GeV J1732$-$3130. $^6$ GeV
J1746$-$2854. $^7$ GeV J1800$-$2328. $^8$ GeV J1856$-$0115. $^9$
GeV J2020$+$4023. GeV sources compiled in the GeV ASCA Catalog
(Roberts et al. 2001a).}
\end{center}
\end{sidewaystable}

Table \ref{green} shows those 3EG sources that are positionally
coincident with SNRs listed in the latest version of Green's
Catalog (2000). From left to right, columns are for the $\gamma$-ray
source name, the measured flux in the summed EGRET phases P1234
(in units of $10^{-8}$ ph cm$^{-2}$ s$^{-1}$), the photon spectral
index $\Gamma$, the EGRET class of source (em for possibly
extended and C for confused), the variability indices $I$ (as in Torres et al.
2001a,c) and $\tau_{{\rm lower\; limit}}^{{\rm upper\; limit}}$
(as in Tompkins 1999), information about coincidences with radio pulsars (``y''
  stands for a pulsar within the error box), the
SNR identification (including other usual names when available),
the angular distance between the center of the $\gamma$-ray source position
and the center of the remnant (in degrees), the size of the
remnant (in arcmin), and finally the SNR type T (S for shell-like emission,
F for
filled-centre or plerionic remnant, and C for composite). A separate
section below analyzes the
possible pulsar associations. All remnants were considered as
circles with a radius equal to the major axis of the ellipse that
better fits their shape, when such is given in Green (2000). It is
interesting to see that many of the 3EG sources involved in the
associations are classified as extended, and all of them as
confused. Also, it was already noted (Romero et al. 1999a) that
in the 3EG catalog not all the positional coincidences with SNRs
are SNOBs (SNRs in OB associations), as it was the case in the studies by
Montmerle (1979) and Yadigaroglu \& Romani (1997) using previous samples.\\

The adoption of the 2000 edition\footnote{A
new edition --2001-- of the catalog has recently appeared.
The results presented herein are not changed
when the 2001 edition is considered.} of Green's Catalog (2000)
does not produce any substantial statistical difference with respect to
the previous edition of 1998. Only 5 SNRs were added. However,
there is a notable particular difference in the case of the EGRET
source 3EG J1714-3857, which now coincides with three supernova
remnants instead of two. One of these SNRs (the new one in Green's
catalog, G347.3-0.5) appears to be amongst the strongest cases for
SNR shock/$\gamma$-ray/nucleonic cosmic-ray source associations
known to date (Butt et al. 2001, see below). In addition, 2EG
J1801-2312 shifted its position half a degree when converting
into 3EG J1800-2338, also affecting previous positional
coincidences. The evolution in the number of coincidences between
SNRs and EGRET sources since the First EGRET Catalog until the
current situation is shown in Table \ref{to} (Torres et al.
2001b).\\

\begin{table}
\begin{center} \caption{Evolution of the number of positional coincidences
between SNRs and unidentified EGRET sources. In the last row, 6 likely
artifacts are disregarded in the 3EG Catalog. }
{\small
\begin{tabular}{lllll}
\noalign{\smallskip} \hline \noalign{\smallskip} Catalog  &
Unidentified & Real & Number of SNRs & Significance \\ EGRET &
detections & coincidences & in Green's catalog & (statistical) \\
\hline First EGRET Catalog$^{\rm a}$ & 37 & 13 (35\%) & 182 &
1.8$\sigma$ \\ 2EG$^{\rm b}$ & 32 & 7 (22\%)  & 194 & ? \\
2EG$^{\rm c}$ & 32 & 5 (16\%) & 14$^{\rm g}$ & ? \\
  2EG$^{\rm d}$ & 33 &
10 (30\%) & 194 &  h \\ 3EG$^{\rm e}$ & 81 & 22 (27\%) & 220 &
5.7$\sigma$ \\ 3EG$^{\rm f}$ & 75 & 19 (25\%) & 220 & 4.8$\sigma$
\\
\noalign{\smallskip} \hline \noalign{\smallskip}
\end{tabular}
}
\end{center}
{\small a: Sturner \& Dermer (1995). b: Sturner, Dermer \& Mattox
(1996).  c: Esposito, Hunter, Kanbach \& Sreekumar (1996). d:
Yadigaroglu \& Romani (1997). e: Romero, Benaglia \& Torres
(1999). f: Torres et al. (2001b). g: Only radio-bright SNRs, flux at 1 GHz
greater than 100 Jy, were considered. h: Computed for pairs.}
\label{to}
\end{table}
\begin{figure} [t]\vspace{-1cm}
\begin{center}
\end{center}
\vspace{-1cm} \caption{Left: Statistical results for the random
association between SNRs and EGRET sources at low-latitudes.
Right: Distribution of the $\gamma$-ray fluxes as a function of the
photon spectral index. Two sources seem to differentiate from the rest.
One of these sources (see below) has been recently identified with
a blazar. } \label{stat}
\end{figure}

The Poisson probability for the 19 coincidences to be a chance
effect is $1.05 \times 10^{-5}$, i.e., there is an a priori 0.99998
probability that at least one of the positional associations in
Table \ref{green} is physical. This expected chance association was
computed using thousands of simulated sets of EGRET sources, by
means of a numerical code described elsewhere (Romero et
al. 1999a,b, Sigl et al. 2001). Figure \ref{stat} shows the result of numerical
simulations for these random populations of $\gamma$-ray sources.
Figure \ref{gdis} shows the distribution of the $\gamma$-ray photon
spectral index
for the sample of 3EG sources coincident with SNRs. Some cases of
possible physical associations mentioned in the literature and
discussed in the next sections are indicated.\\

\begin{figure}[t]
\begin{center}
\caption{Distribution of the $\gamma$-ray photon spectral index.
Some SNR-3EG coincidences for which the physical
association has been suggested in the literature are indicated.
  \label{gdis}}\end{center}
\end{figure}
\begin{table}
\begin{center}\caption{Supernova remnants and nearby (but not coincident)
unidentified 3EG.} \vspace{0.2cm}
{\small
\begin{tabular}{lllllllllll}
\hline $\gamma$-source & $F_{\gamma} $ & $\Gamma$ & Class  & $I$&
$\tau $&
${\rm SNR}$  & $\Delta\theta$ & Size & T & Other \\
\noalign{\smallskip} \hline \noalign{\smallskip}
0229$+$6151 & 39.9$\pm$6.2 & 2.29$\pm$0.18 & C& 1.3
&$0.37_{0.16}^{0.74}$ & G132.7$+$1.3 &
1.50 & 82 & S & Of/OB  \\
1736$-$2908 & 51.5$\pm$9.1 & 2.18$\pm$0.12 & C& 2.4 &
$0.66_{0.40}^{1.09}$ & G359.1$+$0.9 &
0.73 & 12 & S &  -- \\
1741$-$2050 & 24.1$\pm$3.9 & 2.25$\pm$0.12 & C& 2.1 &
$0.41_{0.14}^{0.70}$ & G6.4$+$4.0 &
1.00 & 31 & S  &  -- \\
1823$-$1314$^{a,b}$ & 102.6$\pm$12.5 &2.69$\pm$0.19  & C&  2.9&
$0.72_{0.40}^{1.37}$ & G18.8$+$0.3 &
0.88 & 17 & S  & OB \\
1826$-$1302 & 66.7$\pm$10.1 & 2.00$\pm$0.11 & C& 2.6 &
$0.75_{0.49}^{1.28}$ & G18.8$+$0.3 &
0.88 & 17 & S &  OB \\
1928$+$1733 & 157.0$\pm$36.9 & 2.23$\pm$0.32 & em C& 3.9 &
$0.82_{0.43}^{2.01}$ & G54.1$+$0.3 &
1.21 & 1.5 & F? & -- \\
1958$+$2909$^c$ & 26.9$\pm$4.8 & 1.85$\pm$0.20 & em C& 1.6 &
$0.43_{0.15}^{0.98}$ & G65.1$+$0.6 &
1.36 & 90 & S &  -- \\
\noalign{\smallskip} \hline \noalign{\smallskip}
\end{tabular}
}
\end{center}
{\small $a:$ Association proposed by Sturner \& Dermer (1995), and
Esposito et al. (1996). $b:$ GeV J1825$-$1310. $c:$ GeV
J1957$-$8859.}\label{cerca}
\end{table}

If some SNRs interact, as expected, with nearby massive clouds
producing enhanced $\gamma$-ray emission through
hadronic/Bremsstrahlung interactions, cases in which there is just
a marginal coincidence between the center-points of the SNRs and
the centers of the EGRET sources should be also considered, since the peak
$\gamma$-ray emissivity will likely be biased towards the adjacent cloud. So we
have also looked at the positional coincidences between
unidentified EGRET sources and the region just around the SNRs.
We have done so by artificially enlarging
the size of the SNR by half a degree. We
found that there are 26 coincidences of this kind, including those
in Table \ref{green}. Then, there are 7 new cases perhaps worthy
of further study. These cases are shown in Table \ref{cerca}.
Interestingly, the expected chance coincidence in this case is at
the level of 14.8$\pm$3.14, still 4$\sigma$ lower than the real
result, implying a probability of $2.5 \times 10^{-3}$ for the
real result to be a random (Poisson) fluctuation. In order to
quantify the role played by the 7 new sources in this result
new simulations were carried out, now considering only these
sources. The chance result is 2.3$\pm$1.2, again 4$\sigma$ below
the real number of coincidences. In the present review, however, we
shall focus only on those SNRs that present positional
correlation with $\gamma$-ray sources, i.e. only the cases
listed in Table \ref{green}.\\

\section{ SNRs coincident with $\gamma$-ray sources}

We now analyze the sample of SNRs in Table \ref{green} in more
detail. Table \ref{snr} presents the radio fluxes of the
SNRs, together with their spectral index and known distance
estimates (with the corresponding references). Distances are only
approximate since several different values for the same SNR can be
found in the literature.
When no direct determination is available, estimates can be made
using the radio surface brightness-to-diameter relationship, known
as $\Sigma -D$ (Clark \& Caswell 1976, Milne 1979, Case \&
Bhattacharya 1998). For these cases, marked with a star in Table
\ref{snr}, the distances given by the new $\Sigma
-D$ introduced by Case \& Bhattacharya (1998) will -unless otherwise noted- be
adopted. A double star
symbol means that neither a distance determination nor an estimate
is available. There is only one such case in Table \ref{snr}, for
which the distance to a coincident OB association was assumed.
Thus, distances in Table \ref{snr} marked with one or two stars
are, respectively, more uncertain than the others.  \\

Using the estimated distance to each remnant in Table \ref{snr},
we have calculated the approximate intrinsic $\gamma$-ray
luminosity of the putative region producing the high-energy source
in the energy range 100 MeV--10 GeV using the observed EGRET flux
and photon indices (see Table \ref{green}), assuming isotropic
emission. We have also re-ordered Green's Catalog according to
descending radio flux, and the rank in this list is given for each
remnant. It is interesting that from the first 20 SNRs with the
highest radio fluxes, only 6 appear to be correlated with EGRET
sources.  Sturner \& Dermer (1996) have noted that SNRs not
correlated with 2EG sources were either more distant than
$\gamma$-Cygni (G78.2+2.1) and IC 433 (G189.1+3.0), or presenting
a far smaller radio flux. This trend is not observed now with the
larger 3EG sample.\\

\begin{table}
\begin{center}
\caption{Properties of the SNRs coincident with 3EG sources.
The distances quoted are reported or discussed in the cited references.
Radio fluxes and spectral indices ($\alpha$, such that
$S_{\nu}\propto\nu^{\alpha}$) are taken from
Green's (2000) Catalog. }\vspace{0.2cm}
{\small
\begin{tabular}{lllllllllll}
\hline $\gamma$-source & ${\rm SNR}$  & $d$ & Ref. & $F_{\rm
radio}^{1{\rm GHz}}$ & $\alpha$ & Rank & $L_\gamma$ \\ & & kpc & &
Jy &  &  &  erg s$^{-1}$
\\\noalign{\smallskip} \hline \noalign{\smallskip}
0542$+$2610 &  G180.0$-$1.7 & 0.8-1.6 & 1 & 65 & varies & 24 & 9.65 10$^{33}$\\
0617$+$2238 & G189.1$+$3.0 & 1.5 & 2& 160 & 0.36 & 11 & 1.01 10$^{35}$\\
0631$+$0642 &G205.5$+$0.5 & 0.8-1.6 &3 & 160 & 0.5 & 12 & 1.70 10$^{34}$\\
0634$+$0521 & G205.5$+$0.5 &  0.8-1.6 & 3 & 160 & 0.5 & 12 & 1.85 10$^{34}$\\
1013$-$5915 & G284.3$-$1.8 & 2.9 & 4 & 11 & 0.3? & 91 & 1.71 10$^{35}$ \\
1102$-$6103 & G290.1$-$0.8 & 7 & 5 & 42 & 0.4 & 34 &8.46 10$^{35}$  \\
&  G289.7$-$0.3 & 8.2 & * & 6.2 & 0.2? &123 & 1.11 10$^{36}$\\
1410$-$6147&G312.4$-$0.4    & 1.9-3.1& 6 & 45 & 0.26& 32 & 3.06 10$^{35}$\\
1639$-$4702 & G337.8$-$0.1 &12.3 & 7 & 18 & 0.5&67 &4.17 10$^{36}$\\
         & G338.1$+$0.4         & 9.9& * & 4?  &0.4 & 153 & 2.70 10$^{36}$\\
        & G338.3$+$0.0         & 8.6 & * & 7? & ? & 121 &2.04 10$^{36}$\\
1714$-$3857 & G348.5$+$0.0 & 11.3 & 8 &  9  &0.4? & 96 & 3.47 10$^{36}$\\
          & G348.5$+$0.1         & 11.3&  8 & 72 & 0.3&21 & 3.47 10$^{36}$\\
          & G347.3$-$0.5         & 6.3& 9 & ? & ? & ? & 1.07 10$^{36}$\\
1734$-$3232&G355.6$+$0.0 & 12.6 & *  & 3? & ? & 170 & --\\
1744$-$3011 & G359.0$-$0.9 & 6 & 10 & 23 & 0.5 & 54 & 1.65 10$^{36}$\\
           &G359.1$-$0.5    & 8.5-9.2 &  8-11 &  14  & 0.4? & 81 &
3.56 10$^{36}$\\
1746$-$2851&G0.0$+$0.0 & 8.5& 8   & 100? & 0.8? & 18 &1.20 10$^{37}$ \\
           & G0.3$+$0.0           & 8.5& 12 & 22 & 0.6 & 55 & 1.20 10$^{37}$\\
1800$-$2338&G6.4$-$0.1 & 1.6--4.2&  13 & 310   &varies &6& 4.04 10$^{35}$\\
1824$-$1514 &G16.8$-$1.1 & 1.48 & ** & 2? & ? &186& 5.42 10$^{34}$\\
1837$-$0423 & G27.8$+$0.6 & 2 & 14& 30 & varies & 44 & 3.40 10$^{34}$\\
1856$+$0114 & G34.7$-$0.4 & 2.5 & 15 & 230 & 0.30 &7 & 4.14 10$^{35}$\\
1903$+$0550  & G39.2$-$0.3 & 7.7-9.6 & 15    & 18  & 0.6 & 68 &2.12
10$^{36}$ \\
2016$+$3657  &G74.9$+$1.2          & 10& 15   &  9 & varies &104
&2.75 10$^{36}$\\
2020$+$4017 &G78.2$+$2.1 & 1.7 &  16 &  340  & 0.5 & 5 & 3.20 10$^{35}$\\
\noalign{\smallskip} \hline \noalign{\smallskip}
\end{tabular}
}
\label{snr}
\end{center}
{\small 1. Anderson et al. (1996) 2. Fesen (1984) 3. Jaffe et al.
(1997) and Hensberge et al. (2000) 4. Ruiz \& May (1986) 5. Kaspi
et al. (1997) 6. Caswell \& Barnes (1985), Case \& Bhattacharya
(1999) 7. Koralesky et al. (1998) 8. Green et al. (1997), see also
Reynoso \& Mangum (2000) 9. Slane et al. (1999) 10. Bamba et al.
(2000) 11. Uchida, Morris \& Yusef-Zadeh (1992) 12. Kassim \&
Frail (1996) 13. Frail et al. (1993) and Clark \& Caswell (1976)
14. Reich, Furst \& Sofue (1984) 15. Green (2000) and Caswell et
al. (1975) 16. Lozinskaya et al. (2000)  * From the $\Sigma-D$
relationship presented by Case \& Bhattacharya (1998) ** Distance
assumed equal to a coincident OB association, Romero, Benaglia \&
Torres (1999).}
\end{table}
\section{ Pulsars within the EGRET error boxes}

\begin{table}
\caption{Properties of the $\gamma$-ray pulsars
detected by EGRET. Pulsar parameters and distances are
taken from Kaspi et al. (2000), except PSR~B1055$-$52, for
which  a smaller value of distance was also considered (\"{O}gelman \&
Finley 1993, Combi et al. 1997), and Vela (Caraveo et al. 2001 and
references therein). $\tau= P/2\dot P$, and $\dot E=4\pi^2 I \dot
P /P^3$, with $I=10^{45}$\,g\,cm$^2$. The ``P1234'' $\gamma$-ray
fluxes and spectral indices are from the 3EG catalog (Hartman et
al. 1999). }
\vspace{0.2cm}
\centering
{\small
\begin{tabular}{lrrlllll}
\hline
{Pulsar/3EG source} & {$P$} & {$\tau$} &
{$\dot E$} & {$d$} & {$F^{\rm 3EG}_\gamma
[\times 10^{-8}]$} & {$\gamma^{3EG}$} & {$\eta $}\\
  & {(ms)}                              &
{(kyr)} & {(erg\,s$^{-1}$)}                  & {(kpc)} & {(ph
cm$^{-2}$ s$^{-1}$)}           &
  & {$^{(100{\rm MeV}-}_{-10{\rm GeV})}$} \\ \hline
Crab/0534$-$2200     &  33 &   1.2 & $5.0\times
10^{38}$ & 2.0  &
   226.2$\pm$4.7  & 2.19$\pm$0.02 & 0.01\% \\
Vela/0834$-$4511     &  89 &  12.5 & $6.3\times 10^{36}$ & 0.25  &
   834.3$\pm$11.2 & 1.69$\pm$0.01 & 0.08\%  \\
B1951+32/--     &  39 & 100.0 & $3.7\times 10^{36}$ & 2.4  &
--        & --      & 0.3\%  \\
B1706$-$44/1710$-$4439 & 102 &  15.8 & $3.1\times 10^{36}$ & 1.8 &
   111.2$\pm$6.2  & 1.86$\pm$0.04 & 1\%  \\
Geminga/0633+1751    & 237 & 316.2 & $3.1\times 10^{34}$ & 0.16 &
   352.9$\pm$5.7  & 1.66$\pm$0.01 & 3\%  \\
B1055$-$52/1058$-$5234 & 197 & 501.1 & $3.1\times 10^{34}$ &
0.5/1.5 &
   33.3$\pm$3.8   & 1.94$\pm$0.10 & 2/19\%  \\\hline
\end{tabular}
}
\label{eg-p}
\end{table}

Since both molecular clouds and pulsars can produce $\gamma$-rays,
and because both often lie close to SNRs, it is
important to explore the possible origin of the high-energy flux in neighboring
pulsars.  \\

In order to firmly identify a pulsar as the origin of the
$\gamma$-rays from an EGRET source, $\gamma$-ray pulsations must
be detected at the pulsar period. However, this is not always
possible because of the usually low photon counts observed in most
cases. There are only 6 high-energy $\gamma$-ray pulsars that are
already confirmed, 5 of them have associated $\gamma$-ray sources
in the 3EG Catalog. We give their properties in Table \ref{eg-p}.
Other candidates are usually judged by comparison with the
properties of the known EGRET pulsars.\\

The results of a correlation analysis
between the 3EG sources superposed to SNRs (Table \ref{green}) and
pulsars are presented in this Section. The latter were extracted from
the Princeton Catalog
(Taylor et al. 1993, available on line at
http://pulsar.princeton.edu/ftp/pub/catalog/) and from the
recently -partially- released Parkes Multibeam Survey (Manchester et al. 2001,
http://www.atnf.CSIRO.AU/research/pulsar/pmsurv/). Adding up both
surveys, there are more than 1000 known pulsars. \\

Table \ref{pulsar} presents the results of the correlative
3EG-radio pulsar spatial coincidence analysis: name of the 3EG
source, name of the pulsar found within the error box, their
angular separation, and the size of the 95\% confidence contour of
the 3EG source. Apart from the results quoted in Table
\ref{pulsar}, recently, Torres and Nuza (2003) discovered five new
coincidences between EGRET sources and pulsars in the 2003 version
of the Parkes Catalog, but they discarded their possible
association based on simple spin-down energetics. We refer the
reader to their paper for further details.\footnote{Discussion on
how gamma-ray observations of Parkes' pulsars can help
distinguishing the outer gap from other models for gamma-ray
emission can be found in Torres and Nuza (2003) and references
cited therein.} We provide also the available information about
the pulsar: its galactic coordinates, distance, characteristic
time $\tau= P/2\dot P$, with $P$ and $\dot P$ the period and
period derivative, respectively, spin-down energy release $\dot
E=4\pi^2 I \dot P /P^3$ (assuming a neutron star moment of inertia
$I=10^{45}$\,g\,cm$^2$), and the efficiency in converting the
spin-down luminosity into $\gamma$-rays, if the pulsar alone were
responsible for generating the 3EG source
flux.\\

The $\gamma$-ray efficiency was estimated as \be \eta \equiv
L_{\gamma}/\dot E = f 4 \pi d^2 F_{\gamma}/\dot E, \ee where
$F_{\gamma}$ is the observed $\gamma$-ray flux between 100 MeV and
10 GeV, and $f$ is the $\gamma$-ray beaming fraction ($0 < f \le
1$). This fraction is essentially unknown (see, e.g., Yadigaroglu
\& Romani 1995, Romero 1998), but it is common practice to assume
$f \equiv 1/4\pi$ (e.g. Thompson 2001, Kaspi et al. 2000,
D'Amico et al. 2001, Torres et al. 2001d). These efficiencies are
uncertain also because they suffer (quadratically) the imprecise
knowledge of the distance to the pulsar. \\

\mbox{}For the confirmed $\gamma$-ray pulsars, the observed
efficiencies are in the range $\eta \in (\sim 0.01\%, \sim
3$--19\%), where the interval of upper limits is caused by the
different estimates of the distance to PSR~B1055$-$52 (see
\"{O}gelman \& Finley 1993, Combi et al. 1997, Romero 1998). It
could be considered reasonable that a pulsar generates
$\gamma$-rays with efficiencies in the range $\eta \in (\sim
0.01\%, \sim 10\%)$. For higher values the pulsar would be too
close to the so-called death line, where the high-energy emission
is quenched (Usov 1994).\\

\begin{table}[t]\begin{center}
\caption{Positional coincidences between 3EG unidentified sources
superposed to SNRs and pulsars in the Princeton Catalog
and in the recently -partially- released Parkes Multibeam Survey. We show
measured and derived pulsar parameters as well. See text for
details. In the case of 3EG J1410-6147, second and third values of
efficiency are given for two different estimates of the distance to
G312.4-0.4, 1.9 kpc and 3.1 kpc, respectively. All efficiencies
are based on the new data of the 3EG Catalog (Hartman et al.
1999).
  }\vspace{0.2cm}
  {\small
\begin{tabular}{lllllllllll}
\hline 3EG J  &  PSR J & $\Delta\theta$& $\theta$& $(l,b)$ & $d$ &
$\tau$ & P & $\dot P$ &
$\dot E$ & $\eta$ (\%)  \\
  &  & deg& deg &deg & kpc & kyr & ms & $10^{-15}$ &
erg s$^{-1}$  &  (beamed) \\ \noalign{\smallskip} \hline
\noalign{\smallskip} \hline Princeton
\\ \hline
1013$-$5915& 1012$-$5857 &0.29 &0.72& 283.7,$-$2.1  & 10.1  &741 &
820 & 17.69 & 1.2 $\times 10^{33}$
  & $>100$\\
   1639$-$4702&1640$-$4715&0.29 &0.56& 337.7,$-$0.4  & 7.2  & 188    &
518    & 42.02    &
1.3 $\times 10^{34}$ &$\gg 100$\\
  1800$-$2338&1800$-$2343&0.12 &0.32&   6.1,$-$0.1  &4.8   &--     &
1030    & --    &
-- &--\\
  1824$-$1514&1825$-$1446 &0.46 &0.52&  16.8,$-$1.0  &5.4   &195     &
279    & 22.68    &
4.1 $\times 10^{34}$ &$>100$\\
  1837$-$0423&1836$-$0436&0.28 &0.52&  27.1,1.1   &4.6   & 33    &354
& 1.66    &
1.5 $\times 10^{33}$ &$\gg 100$\\
  1856$+$0114&1856$+$0113&0.05 &0.19&  34.5,$-$0.5  &3.3   &    10
&267      &  208.408   &
4.3 $\times 10^{35}$ &13\\
  1903$+$0550&1902$+$0556&0.26 &0.64&  39.5,0.2  &3.9   &912     &
746    & 12.896     &
1.2 $\times 10^{33}$ & $\gg 100$\\
  &1902$+$0615&0.48 &--&   39.8,0.3  &  10.1 & --    &673      &  --
  &-- & -- \\
\hline Parkes \\ \hline
1013$-$5915 &  1016$-$5857 & 0.16 & 0.72 & 284.1,$-$1.9 &3.0 & 21 &
107& 0.806& $2.6\times 10^{36}$&0.5\\
           &  1013$-$5934 & 0.31   & --      &  284.1,$-$2.6 &
11.3&12561  &442 & 0.557 &$2.5\times 10^{32}$& $\gg$100\\

1014$-$5705 & 1015$-$5719 &0.30 & 0.67 & 283.69,$-$0.58 &4.9 & 39 &140 &57.4 & 8.2 $\times 10^{35}$ & 5 \\

1410$-$6147 &  1412$-$6145 & 0.15   & 0.36&312.3,$-$0.3 & 9.3 & 50 &
315 &  98.7 & $1.2\times 10^{35}$ & $>100/12/30$\\

1420$-$6038 & 1420$-$6048 & 0.16 & 0.33 & 313.54,$+$0.23 & 8 & 13
& 68.2 & 82.85 & 1.0$\times 10^{37}$ & 2 \\

           &  1413$-$6141 & 0.28   &0.36 &312.4,$-$0.3 &11.0 & 13 &
286 & 333.4 & $5.7\times 10^{35}$ & 80/3/6\\
1639$-$4702 &  1637$-$4642 & 0.46   &0.56 &337.8,$+$0.3 & 5.8 & 41 &
154 &  59.2 & $6.4\times 10^{35}$&12\\
           &  1640$-$4648 & 0.37       & --  &338.1,$-$0.2 &6.1 &3501
& 178&0.806 & $5.6 \times 10^{33}$ & $\gg$100\\
           &  1637$-$4721 &  0.45      & -- &337.3,$-$0.1 &5.9 &4160
&1165 &4.44 &$1.1 \times 10^{32}$& $\gg 100$\\
1714$-$3857 &  1713$-$3844 & 0.30 &0.51 & 348.1,0.2 &6.5 &143
&1600
&177.41 &$1.7 \times 10^{33}$ &$\gg 100$\\
           &  1715$-$3903 & 0.23 & -- &348.1,$-$0.3 &4.8 &117 &278
&37.688 &$6.9 \times 10^{34}$ & 72\\
1837$-$0423 &  1838$-$0453 & 0.50 & 0.52& 27.1,$+$0.7 & 8.2 & 52 &
381 & 115.7 & $8.3\times 10^{34}$ &$<55$\\
1837$-$0606 & 1837$-$0559 &0.14 & 0.19 & 26.0,$+$0.38 & 5.0 &1045
& 201.0
&3.304 & 1.5$\times 10^{34}$ & $>100$ \\
      & 1837$-$0604 &0.16 & -- & 25.96,$+$0.27          & 6.2 &
34& 96.3 & 45.170 & 2.0$\times 10^{36}$ & 7 \\

\hline
\end{tabular}
}
\end{center}
\label{pulsar}
\end{table}

Two of the pulsars in Table (\ref{pulsar}), PSR J1800-2343 and PSR
J1902+0615, lack a confident determination of the period
Derivative; for these it is impossible to assess their expected
efficiencies. Based on the required efficiencies and derived
spin-down luminosities of the other members of the group shown in
the first (Princeton) panel, none of the positional associations
seem likely except that between 3EG 1856+0114 and PSR J1856+0113.
For this case, the pulsar should have an efficiency of 13\% in
converting rotational energy into $\gamma$-rays. PSR J1856+0113 was
already mentioned as possibly associated with the EGRET source by de
Jager \& Mastichiadis (1996).\\

The bottom panel of Table (\ref{pulsar}) contains data that were
analyzed elsewhere (Torres et al. 2001d, D'Amico et al. 2001,
Camilo et al. 2001); here are the main conclusions. 1. The
physical association between 3EG J1013-5915 and PSR J1016-5857 is
possible, as  first noted by Camilo et al. (2001). 2. The source
3EG J1410-6147 and either PSR J1412-6145 or J1413-6141 might be
physically associated if either of the latter lie closer than
their dispersion measure distances, say at the estimated distance
to G312.4-0.4 (2-3 kpc). 3. The pulsar PSR J1637-4642 could
contribute part of the high-energy budget of the source 3EG
J1639-4702. In addition to this, the pulsars J1015-5719,
J1420-6048 and J1837-0604 are also possible counterparts for their
respective EGRET sources.

\section{ Variability}

Ift SNRs, molecular clouds, or pulsars are responsible
for some EGRET sources, we would expect them to be
non-variable on the time scale of EGRET observations, i.e. from
weeks to a few years. Hence, variability analysis of the
$\gamma$-ray emission is an important tool to test the original
hypothesis, in the sense that variable sources could be ruled out
as being produced by SNRs (or pulsars).\\

Three variability indices for EGRET sources have been introduced
in the literature so far. The first of them, dubbed $V$, was
presented by MacLauglin et al. (1996), who computed it for the
sources contained in the Second EGRET Catalog. This method was
later used, also, for short timescale studies by Wallace et al.
(2000, 2002). The basic idea behind $V$ is to find $\chi ^2$ from
the measured fluxes, and to compute $V=-\log Q$, where $Q$ is the
probability of obtaining such a $\chi ^2$ if the source were
constant.  Several critiques have been mentioned concerning this
procedure, among them, that the scheme gets complicated when the
fluxes are just upper limit detections. It can be shown that
sources which have upper limits included in the analysis will have
a lower $V$ than what is implied by the data (Tompkins 1999). In
addition, a source can have a large value of $V$ because of
intrinsic reasons or because of small error bars in the flux
measurements. Similarly, a small value of $V$ can imply a constant
flux or big error bars. Each value of $V$ is obtained disregarding
those of a control population. Then, there could be pulsars with
very high values of $V$, or variable AGNs with very low ones.
Hence, the use of $V$ to classify the variability of $\gamma$-ray
sources seems not to produce confident results. Two other indices
have been computed for all $\gamma$-ray sources: the $I$ index and
the $\tau$ index (Torres et al. 2001a; Tompkins 1999,
respectively). See Torres et al. (2001c) for a comparison among
the results obtained with them. The idea behind the index $I$ is
to carry out a direct comparison of the flux variation of any
given source with that shown by pulsars, which is considered as
instrumental. It basically establishes how variable a source is
with respect to the pulsar population. Contrary to Tompkins'
index, the $I$-scheme uses only the publicly available data of the
3EG Catalog, and is defined as follows. Firstly, a mean weighted
value for the EGRET flux is computed as \be \left< F \right> =
\left[ \sum_{i=1}^{N_{{\rm vp}}} \frac{F(i)}{\epsilon(i)^2}
\right]\times \left[ \sum_{i=1}^{N_{{\rm vp}}} \frac
1{\epsilon(i)^{2}} \right]^{-1}.\ee Here $N_{{\rm vp}}$ is the
number of single viewing periods for each $\gamma$-ray source,
$F(i)$ is the observed flux in the $i^{{\rm th}}$-period, whereas
$\epsilon(i)$ is the corresponding error in the observed flux.
These data are taken directly from the 3EG Catalog. A fluctuation
index, $\mu$, is defined as (e.g. Romero et al. 1994): $ \mu
=100\times \sigma_{{\rm sd}}\times \left< F \right>^{-1} ,$ where
$\sigma_{{\rm sd}}$ is the standard deviation of the flux
measurements. This fluctuation index is also computed for the
confirmed $\gamma$-ray pulsars in the 3EG Catalog, and then the
averaged statistical index of variability, $I$, is introduced by
$I=\mu_{{\rm source}}/<\mu>_{{\rm pulsars}}$. We refer the reader
to the work by Torres et al. (2001a,c) for
details on the error estimates.\\

Since the $I$-scheme is a relative classification, a result like
$I=3.0$ says that the flux evolution is three times  more variable
than (equivalently, 4$\sigma$ above) the mean flux evolution for
pulsars. Torres et al. (2001c) mentioned that in order to get more
reliable results under the $I$-scheme at low-latitudes it seems
safer to consider a restrictive criterion. For instance, a source
will be considered very variable only when $I-\delta I > [I_p] + 3
\sigma_{p,I}$. Here $\delta I \sim 0.5 I$, $[I_p]$ is the mean
value of $I$ for pulsars, $\sigma_{p,I}$ is their standard
deviation, and $[I_p]+ 3 \sigma_{p,I}=2.5$. Rephrasing the
previous constraints just in terms of $I$, a source will be very
variable if $I>5$. This represents a deviation of $8\sigma$ from
the mean $I$-value for pulsars. With this criterion, 3EG
J1837-0423, with $I=5.41$, is a very variable source. We have 7
non-variable sources, those having $I < 1.7$ in Table 1. The rest
of the sources listed there are dubious under this classification
scheme. \\

Tompkins (1999) used the 145 marginal sources
that were detected but not included in the final official 3EG list,
and, simultaneously, all the detections within 25 deg of the source of
interest. The maximum likelihood set of source fluxes was then
re-computed. From these fluxes, a new statistic measuring the
variability was defined as $\tau=\sigma / \mu$, where $\sigma$ is
the standard deviation of the fluxes and $\mu$ their average
value. The final result of Tompkins' analysis is a table listing
the name of the EGRET source and three values for $\tau$: a mean,
a lower, and an upper limit (68\% error bars).\\

The mean value of $\tau$ for pulsars -again an assumed
non-variable population- is very low, 0.09, but the mean of the
upper limits is $\sim 0.2$. So pulsars are consistent with having
values of $\tau$ up to 0.2. The deviation for the mean value of
pulsars is 0.08. A source will be likely variable under the $\tau$
scheme when the lower limit is at least 0.6, $3\sigma$ above the
mean value of the $\tau$ upper limit for pulsars. A source will be
considered non-variable when the upper limit for $\tau$ is below
that threshold. Sources not fulfilling either classification will
be considered as dubious. This is also
consistent with the fact that the mean $\tau$ value for the known
population of AGNs is 0.9. Using this scheme, too,
3EG J1837-0423 is a variable source. In addition, 3EG
J0631+0642 and 3EG J0634+0521 are also variable under $\tau$. However,
Tompkins (1999) adds a word of caution for these two sources: the
fitted flux for them is zero. Such sources are most likely
variable, but unknown instrument systematics or numerical problems
within the $\tau$ scheme could conceivable change these results.
Also important is to note that many sources have a dubious
classification: within the 68\% error bars on $\tau$, they can be
as variable as an AGN, or as non-variable as a pulsar. This is,
unfortunately, a common situation for many of the sources. For
those ones, in particular, the $I$-index scheme can provide some
additional information.\\

Based on the variability of the $\gamma$-ray flux, then,  it is
highly unlikely that
3EG J1837-0423 is caused by SNR G27.8+0.6, or by the pulsars
PSR J1836-0436 (from Princeton Catalog) and PSR J1838-0453 (from
Parkes Catalog). In addition, the spectral index for
this 3EG source is very steep, possibly arguing against the pulsar
hypothesis (Fierro et al. 1993), although it should be noted that
Halpern et al. (2001)
have recently presented a strong argument for the association of PSR J2229+6114
  with 3EG J2227+6122, which a relatively soft index of
2.24$\pm$ 0.14.\\

Out of 19 sources in Table 1, 12 have the same
classification under the two schemes; in Table 2, 6 out of 7 have the
same classification. This confirms, for these
sources, that the schemes are statistically correlated
and that it is safe to consider both indices to smooth out any
particular problem with singular sources, as
apparently is the case for 3EG J0631+0642 and 3EG J0634+0521.\\

\begin{figure}[t]
\begin{center}
\end{center}
\vspace{-1cm} \caption{Comparison of the variability indices of
the 3EG sources superposed with SNRs and identified 3EG AGNs.}\label{com-I}
\end{figure}

Figure \ref{com-I} presents a comparison between the
variability indices for those 3EG sources superposed to SNRs and
the set of `A' AGNs identified in the same catalog. The
distributions are quite different, showing a more
non-variable population in the case of the sources herein investigated.
The most variable sources in the set are
identified for reference.

\section{ Observations and data analysis}

\subsection{CO data}

CO is a polar molecule with strong dipole rotational emission in
the mm waveband, and is considered a reliable tracer of molecular
hydrogen,
$H_{2}$, which, though much more abundant, has only a weak
quadrupole signature. Throughout this paper, molecular gas masses
are derived from observations of the J=1-0 rotational transition
of CO at 115 GHz, assuming a proportionality between
velocity-integrated CO intensity, $W_{\rm co}$, and molecular
hydrogen column density N(H2). Specifically, we adopt $$N({\rm
H2})/W_{\rm co} = 1.8 \times 10^{20} \;{\rm cm}^{-2} {\rm K}^{-1}
{\rm km}^{-1} {\rm s}^{-1},$$  the value derived by Dame,
Hartmann, \& Thaddeus (2001) from an intercomparison of
large-scale far-infrared, 21 cm, and CO surveys. To estimate total
nucleon densities in molecular clouds, we account for elements
heavier than hydrogen by assuming a mean molecular weight per $H_{2}$
molecule of 2.76.\\

All CO(1-0) data presented here are from the whole-Galaxy survey
of Dame, Hartmann, \& Thaddeus (2001). This survey is a composite
of 37 separate CO surveys carried out over the past 20 years with
two nearly identical 1.2 meter telescopes, one located at the
Harvard-Smithsonian Center for Astrophysics in Cambridge,
Massachusetts, and the other at the Cerro Tololo Interamerican
Observatory in Chile. The angular resolution of the composite
survey is $\sim 8.5$ arcmin and the velocity resolution $\sim 1$
km s$^{-1}$. Many of the 37 separate surveys have been published
previously and, where appropriate, the earlier papers are cited
instead of, or in addition to, Dame, Hartmann, \& Thaddeus
(2001).\\

In a few cases, observations of the CO(J=2-1) transition at the
same angular resolution, obtained with the Tokyo-NRO 60 cm Survey
Telescope (Sakamoto et al. 1995), are used to search for
enhancements of the CO(J=2-1)/CO(J=1-0) ratio as an indicator of
SNR-molecular cloud interaction (e.g., Seta et al. 1998).

\subsection{ Radio continuum data and diffuse background filtering}

In general, the diffuse radio emission of the Galaxy hampers the
detection of weak and extended sources with low surface
brightness, like SNRs. This large-scale diffuse emission can be
removed through different techniques for astronomical data
analysis. These techniques can range from detailed modeling of the
non-thermal emission in the Galaxy to model-independent filtering
algorithms. In this work, many of the radio continuum maps have
been cleaned of background diffuse contamination
using the method originally introduced by Sofue \& Reich (1979).
Basically, the technique consists of convolving the continuum map
with a Gaussian filtering beam, producing a new map with a
different brightness temperature $T_0^1$. A new temperature
distribution is computed as $T_0^{\prime 1}=T-\Delta T^1$, for
$\Delta T^1>0$ and $T_0^{\prime 1}=T$ for  $\Delta T^1<0$. In
these expressions, $T$ is the temperature distribution of the
original map, and $\Delta T^1=T-T_0^1$ are the residuals between
the original and the convolved maps. The procedure is repeated, now
convolving the $T_0^{\prime 1}$ map in order to obtain $T_0^2$,
$\Delta T^2$ and finally $T_0^{\prime 2}$. After $n$ iterations,
the difference $|T_0^{n}- T_0^{ n-1}|$ becomes smaller than the
rms noise, and a map of residuals $\Delta T^n=T-T_0^n$ is obtained
where all diffuse emission with size scales larger than the
original filtering beam has been removed. The result is completely
independent of the original mechanism that produced the
large-scale emission. In the following sections we apply this
technique to Effelsberg 100m-telescope data and to the MOST
Galactic plane survey in order to get images of the SNRs as clean
as possible at low galactic latitudes.

\section{ Case by case analysis}

\subsection{ $\gamma$-ray source 3EG J0542+2610 -- SNR G180.0-1.7}

The possible counterparts of
this $\gamma$-ray source were explored in detail in a recent paper
(Romero et al. 2001). No known
radio pulsar coincides with this source (see Table 3).
Additionally, a radio-quiet Geminga-like pulsar origin is disfavored
a priori because of the high variability and the steep spectral
index that this source presents ($\Gamma=2.67\pm0.22$, Hartman et al. 1999).
We have searched the EGRET location error box for other
compact radio sources. Although 29 point-like radio sources were
detected, none of them is strong enough to be considered a likely
counterpart (Romero et al. 2001). The strongest of the sources
detected have a radio flux one order of magnitude less than those
presented by known $\gamma$-ray blazars detected by EGRET.
Moreover, the absence of an X-ray counterpart to this source
suggest that it is not an accreting source, like a microquasar. \\

Some of us suggested that the only object within the 95\% error box
capable of producing the required $\gamma$-ray flux is the X-ray
transient A0535+26. This Be/accreting pulsar, not detected at all
in the radio band, can produce variable hadronic $\gamma$-ray
emission through the mechanism originally proposed by Cheng \&
Ruderman (1989, 1991). See Romero et al. (2001) for further
details.\\

On the basis of results discussed in that paper we
conclude that 3EG J0542+2610 and G180.0-1.7 are most likely
unrelated. An interesting comparison between this case and one of
the EGRET sources coincident with the Monoceros Loop is made
below.

\subsection{ $\gamma$-ray source 3EG J0617+2238 -- SNR G189.1+3.0 (IC443)}

\begin{figure}[t]
\begin{center}
\caption{CO distribution around the remnant IC443 (G189.1+3.0).
The 3EG $\gamma$-ray source J0617+2238 is superposed. Note the
positional coincidence of the contours of the latter with part of
the most dense regions of the CO distribution. The optical
boundary of the SNR is superposed as a black contour. The optical
emission seems to fade in regions where CO emission increases,
this indicates that the molecular material is likely located on the
forground side of the remnant, absorbing the
optical radiation. Optical contours are from Lasker et al. (1990).
\label{ic433}}\end{center}
\end{figure}

A detailed description of SNR IC433 was given by Chevalier
(1999). We have recently reviewed the spatially-resolved multiwavelength
spectrum of IC443 and argued that the morphology and spectrum of the
$\gamma$-ray
emission make it a likely hadronic cosmic-ray accelerator (Butt et al. 2002b.\\

Seta et al.(1998) have provided an analysis of the CO
environment of this remnant. They concluded, as was previously
reported by Scoville et al. (1977) and Cornett et al. (1977), that
IC433 is interacting with several ambient molecular clouds with a total
mass of about 10$^4 M_\odot$. They also analyzed the ratio
$R$=CO(J=2-1)/(J=1-0) in the environs of IC443 and concluded that
parts of the clouds
presented an abnormally high value, consistent with shock interaction.
The detected value of $R$ exceeds 3 (the average
galactic value is $\sim 0.6$) in some regions. Interestingly, the peak
of the CO(J=2-1)/(J=1-0) ratio is coincident with the location of the newly
discovered pulsar wind nebula by Olbert et al. (2001), which may indicate
an alternate way of exciting molecular gas. Dickman et al. (1992) has estimated
that the total perturbed molecular gas has a mass of 500--2000
$M_\odot$.\\

\begin{figure}[t]
\begin{center}
\caption{Hard energy band (3--10 keV) of the IC443 nebula. Most of the
thermal emission associated with IC443 is not present in this band.
The image shows several point sources, besides the plerion nebula itself.
The nebula can
be represented by an ellipse of 8 arcmin $\times$ 5 arcmin. The 95\%
confidence EGRET error circle for 3EG J0617+2238 is also shown. From
Bocchino \& Bykov (2001).
\label{bocchino}}\end{center}
\end{figure}

In recent years, X-ray
observations of IC443 have been carried out. IC443 was a target for
X-ray observations
with  HEAO 1 (Petre et al. 1988), Ginga (Wang et al. 1992), ROSAT
(Asaoka \& Aschenbach 1994), ASCA (Keohane et al. 1997), and more
recently, with Chandra and Beppo-SAX; we discuss the latter
in more detail below. IC443 was believed to be mostly
thermal in the X-ray band (Petre et al. 1988, Asaoka \& Aschenbach
1994), although it has been discovered to emit hard X-ray emission
(Wang et al. 1992). Keohane et al. (1997) later found that the
hard  X-ray emission was localized and non-thermal. They concluded
that most of the 2--10~keV
photons came from an isolated emitting feature and from the South
East elongated ridge of hard emission. Even more recently, Preite
Mart\'{\i}nez et al. (1999) and Bocchino \& Bykov (2000) reported
a hard component detected with the Phoswich Detector System (PDS)
on BeppoSAX and two compact X-ray sources corresponding to the
ASCA sources detected with the BeppoSAX Medium-Energy Concentrator
Spectrometer (MECS) (1SAX J0617.1+2221 and 1SAX J0618.0+2227).
1SAX J0617.1+2221 has also been observed with the Chandra satellite
by Olbert et al. 2001, who also obtained VLA observations at 1.46,
4.86 and 8.46 GHz and a polarization measurement. The hard radio
spectral index, the amount of polarization, and the overall X-ray
and radio morphology led them to suggest that the source is a
plerion nebula containing a point source whose
characteristic cometary shape is due to supersonic motion of the
neutron star. Bocchino \& Bykov (2001) have, in addition, recently
observed IC443 with XMM-Newton Observatory (see Figure
\ref{bocchino}). They resolve the structure of the nebula into a
compact core with a hard spectrum of photon index $\gamma =
1.63^{+0.11}_{-0.10}$ in the 2--10 keV energy range, and found
that the nebula also has an extended ($\sim$ 8 arcmin $\times 5$
arcmin) X-ray halo, much larger than the radio emission extension.
The photon index softens with distance from the centroid, a
behavior also found in other X-ray plerions such as 3C58
and G21.5-0.9. Bocchino \& Bykov (2001) also looked for periodic
signals from the NS but found none with 99\% confidence level in
the $10^{-4}-6.5$ Hz range. Assuming that $L_X/\dot E \sim 0.002$,
the spin-down luminosity of the central object results in $\dot E=
1.3 \times 10^{36} $ erg s$^{-1}$. If the power-law spectrum of the
nebula core region were extrapolated up to the GeV regime, it would
provide a flux of 2.0 (0.4--15.2)$\times 10^{-7}$ ph cm$^{-2}$
s$^{-1}$, which is consistent with the EGRET flux from the IC443
region (see Table 1). However, the fact that the nebula lies outside
the 95\% confidence circle of the source argues against an association.
Further, the EGRET source luminosity would require a substantial
fraction of the estimated spin-down power.\\

The H$_2$ mass near SNR IC443 that we report here, $1.1 \times
10^4 M_\odot$, is the total molecular mass in the IC443 velocity
range, $v=-40$ to +20 km s$^{-1}$, within a rectangle enclosing
the main clump coincident with the SNR: $l$=188.75 to 189.5,
$b$=2.5 to 3.375, see Figure \ref{ic433}. The average density of
that region is about 840 nucleons cm$^{-3}$. Assuming that the
energy of the explosion was $E_{51}=0.27$, and an unshocked
density of $0.21$ cm$^{-3}$ (from a 2D dynamic model by Hnatyk and
Petruk 1998), we found that the hadronic flux would be
$4.7\;\times 10^{-6}$ photons cm$^{-2}$ s$^{-1}$. Indeed, just
10\% of the ambient mass is necessary to produce
the observed flux of 3EG J0617+2238.\\

The energy of the explosion and the unshocked density yield a CR
enhancement factor, $k_{\rm s}\sim 600$ within the SNR, which appears to be
unusually high. However, computation of the energy transformed
into cosmic rays by the direct product $k_{\rm s} \epsilon_{\rm
CR} (4/3 \pi R^3)$ gives $0.4 E_{51}$, a value compatible with
that obtained for $\theta$ (the efficiency of SN energy conversion to CRs) in
the Morfill et al. (1984) prescription
discussed above. Alternatively, by direct use of Eq.
(\ref{hadro}), and using the mass considered above and the
observed $\gamma$-ray flux, we can obtain an estimate of the
enhancement factor $k_{\rm s}$ of 66. The
difference between the CR enhancement factor of the cloud ($k_{\rm
s}=66$) and the SNR ($k_{\rm s}=600$) could be explained in a variety of
ways: most simply, that the CR enhancement predicted
by the Morfill et al. (1984) prescription for SNRs
overestimates the value within the adjacent cloud.
Particularly if the cloud abuts the remnant, its enhancement will
naturally be smaller
than that calculated for the SNR interior. As Figure \ref{ic433} shows, the
optical
emission seems to fade in regions where CO emission increases,
which perhaps indicates that the molecular material is absorbing
the optical radiation, abutting the remnant on the near side.
The report by Cornett et al. (1977) also argues that the molecular mass is
located between us and the SNR.
In any case, we
remark that the previous estimations of the CR enhancement factors
assume that the
explosion proceeds in an homogeneous medium,
something which we know is not true in this case (Chevalier 1999).\\

We note that an electronic bremsstrahlung hypothesis for the origin
of the GeV flux
(eg. Bykov et al. 2000) is difficult to reconcile with the fact that the radio
synchrotron emission is concentrated towards the rims of the remnant, whereas
the GeV source is centrally located (Figure \ref{ic433}).\\

It is clear that IC443 will continue to be a primary target for
future satellites missions and telescopes. A better localization
of the EGRET sources, by AGILE, or GLAST, as well as the already
approved INTEGRAL
observations could help much in determining the ultimate nature of
3EG J0617+2238. The reader is referred to Butt et al. (2002b) for further
analysis of the likely hadronic origin of the $\gamma$-ray emission.


\subsection{ $\gamma$-ray source 3EG J0631+0642 and 3EG J0634+0521 --
SNR G205.5+0.5 (Monoceros nebula) \label{MON}}

\begin{figure}[t]
\begin{center}
\caption{Left: CO contours (white) on a background image from the Digital Sky
Survey. The map covers a somewhat larger region than the left panel
Figure, and shows the HII region and young cluster NGC 2244
producing a hole in the cloud. Most of the Monoceros Loop is also
seen faintly. EGRET sources contours are marked in black.
Right: CO emission plus contours (black) of 1.4 GHz emission
to mark the Rosette Nebula. The peak of the CO emission if near
3EG J0634+0521. The position of the X-ray source SAX J0635+0533 is
marked with a star. White lines are the confidence levels of the EGRET sources.
  \label{205-2}}\end{center}
\end{figure}

The large SNR G205.5+0.5 (Monoceros Loop nebula, 220 arcmin in
size) has been thoroughly studied in the past. Various papers have
proposed that the Monoceros Loop SNR is interacting with the
Rosette Nebula (e.g., Odegard 1986). A recent study of the
stars in NGC 2244, the cluster within the Rosette, finds a
distance of 1.39$\pm$0.1 kpc (Hensberge et al. 2000); we assume
this distance in the following computations. In the left panel of Figure
\ref{205-2}, we show an overlay of the CO contours on an image
from the Digital Sky Survey (Lasker et al. 1990). It shows very
nicely the HII region and
young cluster apparently carving a hole in the cloud. Most of the
Monoceros Loop is also seen very faintly. This justifies the distance adopted,
under the assumption that the Nebula and the Monoceros Loop are
equally distant from Earth. Bloemen et al. (1997) presented COMPTEL
observations of the
Monoceros region, and found  excessive 3-7 MeV emission which they
attributed to nuclear deexcitation lines at 4.44 and 6.13 MeV from
accelerated 12 C and 16 O nuclei.\\

Monoceros was already suggested by Esposito et al. (1996) as a
source of $\gamma$-rays, and it was also mentioned by Sturner and
Dermer (1995) and Sturner, Dermer and Mattox (1996) as a possible
case of $\gamma$-rays production by hadronic interactions. The
age of the remnant is not well determined, 3-20 10$^4$ yrs. A
study of Einstein IPC data (Leahy et al. 1986) shows diffuse X-ray
thermal emission in a region corresponding to the detection of
optical filaments (Odegard 1986). This is only possible if the gas
is sufficiently hot, and thus if the remnant is sufficiently
young. This would be in contradiction with the age one obtains
from the homogeneous Sedov solutions, which would give an age in
excess of 100,000 yrs. One direct interpretation of this discrepancy
(Leahy et al. 1986) is that the expansion of the remnant proceeds
in a non-homogeneous multi-component medium, where the homogeneous
Sedov solutions are not valid.\\

A large region covering the spatial extent of both 3EG sources was
studied by Jaffe et al. (1997): ($198 < l < 214 $, and $-6<b<8 $).
They presented an image
reconstruction of the region around the Rosette Nebula and
Monoceros using high-energy ($>$100 MeV) $\gamma$-ray data from
EGRET. The resulting image showed a 7$\sigma$ extended feature in
excess of the expected diffuse emission located at the
point-source position listed in the EGRET catalog (2EG at that time).
These authors proposed that this excess could be
evidence of an interaction between the Monoceros remnant and the
Rossette nebula. They concluded that if the $\gamma$-ray emission
arises solely in the interaction between the two nebulae then the
cosmic-ray enhancement would be around $k_{\rm s}$=300. This value
appears to be excessively high in this case, should the enhancement be the same
for all the SNR region. The energy in cosmic-rays, computed using
$k_{\rm s} \epsilon_{\rm CR} (4/3 \pi R^3)$ together with the size
of the Monoceros remnant ($\sim 60$ pc), imply an energy of the
explosion about one order of magnitude larger than the assumed
$E_{51}\sim 1$. This may indicate that the hadronic origin of
the $\gamma$-rays in the interaction of the Monoceros SNR and the
Rossete Nebula (i.e. 3EG 3EG J0634+0521) cannot be responsible for
the entire observed flux.\\

Indeed, within the 95\% contour of 3EG J0634+0521 there exists an
X-ray source SAX J0635+0533, and a Be-star/neutron-star X-ray binary
pulsar, probably with a relatively short orbital period (Kaaret et
al. 1999, Cusumano et al. 2000, Nicastro et al. 2000). The hard
X-ray source SAX J0635+0533 shows pulsations at a period of 33.8
ms but no radio flux was detected at the Be-star position (see
below). SAX J0635+0533 might be, as in the case of A0535+26
mentioned above, a source of $\gamma$-rays through hadronic
processes. Kaaret et al. (1999) suggested that SAX J0635+0533 and
3EG J0634+0521 are related. One fact favoring this
physical association is that the SAX satellite has not detected
extended emission in the region of 3EG J0634+0521, as would be
the case if the bulk of the radiation were produced in a SNR shock.
Additionally, the probability for chance positional coincidence
between a Be/X-ray binary and an EGRET
source is less than 4\% (Kaaret et al. 1999).\\

The situation, however, is far from resolved. The EGRET source,
for instance, is non-variable as it would be expected for a binary
with eccentric orbit. Recent results reported by Kaaret et al.
(2000), comparing observations obtained with BeppoSAX and RXTE
separated in time by 2 years, showed that the period derivative of
the pulsar has a lower bound equal to 3.8 $\times 10^{-13}$. This
value is 30 times larger than values found from accreting neutron
stars (Bildsten et al. 1997), and it implies a mass accretion rate
of 6 $\times 10^{-7} M_\odot$ yr$^{-1}$ (Kaaret et al. 2000,
Bildsten et al. 1997), which far exceeds the expected mass
capture rate of a neutron star $\sim 10^{-11} M_\odot$ yr$^{-1}$,
and is even slightly larger than the average mass loss rate of
Be-stars (Kaaret et al. 2000). Apparently, this would indicate
that the X-ray luminosity does not originate in the
accretion disk, and argues in favor of SAX J0635+0533 being a
rotation powered pulsar. In this case, the value of $\dot P$ would imply a
characteristic age of only 1400 years and a high spin-down
luminosity of 5 $\times$ $10^{38}$ erg s$^{-1}$, out of which less than
0.05\% could make a noticeable contribution to the observed
$\gamma$-ray flux (assuming a distance of 4 kpc, Kaaret et al.
1999). Additionally, arguing against an accretion origin of the
radiation, the derived X-ray luminosity (7.7 $\times$ 10$^{34}\; (d_{\rm
kpc}/4$ kpc)$^2$ erg s$^{-1}$) and magnetic field strength ($\sim
10^9$ G) are too low in comparison to other Be-X-ray binaries such
as A0535+26 (Cusumano et al. 2000).\\

Very recently, Monoceros  was the target of the HEGRA \v{C}erenkov
telescopes (see Lucarelli et al. 2001). HEGRA observed the
Monoceros-Rossete region for about 120 hours, with an energy
threshold of 500 GeV and an angular resolution of 0.1 deg, and
mapped a 2 $\times$ 2 deg$^2$ region centered in the source SAX
J0635+0533. The EGRET source 3EG J0634+0521 is also within the
field of view. Although the flux and spectrum have not yet been
officially reported, HEGRA found a tentative excess of counts in
four different pixels (0.2 $\times$ 0.2 deg$^2$) within the 3EG
contours; interestingly, none of them coinciding with SAX
J0635+0533 (Lucarelli et al. 2001). It is possible that TeV emission
coming from the binary is being re-absorbed in its neighborhood, as in the
case studied by Romero et al. (2001).\\


What HEGRA observations seem to imply is that the marginally significant TeV
radiation has an extended origin, different from that
producing GeV photons in the binary system SAX J0635+0533, but this
remains to be confirmed.\\

In the right panel of Figure \ref{205-2} we show the CO emission plus
contours of 1.4 GHz continuum emission marking the Rosette.
Using the standard CO-to-$H_{2}$ mass conversion and a mean molecular
weight per $H_{2}$ molecular of 2.76, we calculate  a total $H_{2}$ mass for
the associated cloud (in the region $l=205$ to 209, $b=-3$ to
$-1$, and velocity range $ v = -5$ to 30 km s$^{-1}$) of $1.2 \;
\times 10^5 \;M_\odot$. We have calculated the molecular masses in
small rectangles enclosing the two 3EG sources near the Rosette.
Using a distance of 1.39 kpc we get the following results: In the
case of 3EG J0634+0521 the region considered is $l$ = 205.5 to
206.875 and $b$ = $-2.125$ to $-0.375$, and the mass is
2.0 $\times 10^4 \; M_\odot$. For 3EG J0631+0643, in the region
$l$ = 204.25 to 205.5 and $b = -2.125$ to $-0.75$, the mass is 4.7
$\times 10^4 \; M_\odot$. In both cases, the velocity range
considered is $v = -5$ to 30 km s$^{-1}$ and the formal error on
these masses, based on the instrumental noise, is $\sim 0.2 \times
10^3 \; M_\odot$.\\

Because of uncertainties on the nature of this SNR (for instance,
the controversy on the SNR age) Morfill et al.'s method is
unreliable for estimating the SNR GeV-flux. However, using
directly Eq. (\ref{hadro}) and the observed flux (Table
\ref{green}), we can estimate the value of $k_{\rm s}$ needed to
generate the observed flux, resulting in $k_{\rm s} \sim 6.5$ for 3EG
J0634+0521.
Because of the high molecular density, just a modest enhancement of the
cosmic-ray density can explain
a substantial part of the detected $\gamma$-ray flux. We suggest,
then, that 3EG J0634+0521 might be a composite source: SAX J0635+0533
might be responsible for part of the GeV flux,
as well as the bulk of the emission at X-ray energies. The
interacting SNR and Rossete Nebula might also contribute to the
flux in the GeV range, and would provide the bulk of the possibly
detected TeV emission from the region. One direct way to test this
scenario would be through an analysis of the spectrum from GeV
to TeV. In the case of a composite source, there should be a break
in the spectrum between GeV and TeV energies, the latter
corresponding only to the accelerated particles in the SNR
remnant.\\

In the case of 3EG J0631+0643, a CR enhancement value of just
$k_{\rm s} \sim 3$ can explain the observed GeV flux. New
high-sensitivity radio measurements of the region would be of
great value in determining the relative importance of hadronic and
leptonic $\gamma$-ray emission.

\subsection{ $\gamma$-ray source 3EG J1013-5915 -- SNR G284.3-1.8 (MSH 10-53)}

\begin{figure}
\begin{center}
\caption{(a) Longitude-velocity map of CO integrated over 1 degree
of Galactic latitude roughly centered on the SNR G284.3-1.8 (MSH
10-53), $b$ = -2.5 to
-1.5. The longitude of the SNR is indicated by the dotted vertical
line. (b)  Spatial map of CO integrated over the velocity range
-22 to 3 km s$^{-1}$. The plus sign marks the center position of
the SNR G284.3-1.8, whose size is 25 arcmin.  The dotted circle is
the 95\% confidence radius about the position of 3EG J1013-5915
(Hartman et al. 1999). Note that the longitude range (x axis) of
both maps is the same.
  \label{g284}}\end{center}
\end{figure}

\mbox{} For this 3EG source, a natural candidate to generate a significant
part of the $\gamma$-ray emission seems to be the recently
discovered pulsar PSR J1013-5915; see Table 3 (Camilo et al.
2001). This pulsar has a characteristic age of $\tau = 21$\,kyr
and a spin-down luminosity of $\dot E = 2.6\times 10^{36}$\,erg
s$^{-1}$. If only the pulsar is considered, the efficiency
required for converting spin-down luminosity into $\gamma$-rays
is $\eta \sim 0.5$\% (Camilo et al.  2001), well within the
range of efficiencies for previously confirmed $\gamma$-ray pulsars
detected by EGRET. \\

Another Parkes' pulsar, PSR~J1013$-$5934, is also coincident with
3EG~1013$-$5915, but it can be ruled out on the basis of energetic
arguments. So also the Princeton pulsar PSR
J1012-5857 (Table 3), which requires an efficiency of about 100\%
in the generation of the $\gamma$-ray emission from the spin-down
losses. It is also interesting to note that 3EG~J1013$-$5915 has
a photon spectral index softer than typical for pulsars: $\Gamma=
2.32\pm0.13$.\\

The remnant is located in the near side of the Carina spiral arm, in a region
with a high density of molecular clouds. Ruiz \& May
(1986) found filamentary optical emission associated with the
remnant. These authors also found clear evidence of at least
three small CO clouds interacting with G284.3-1.8. Other small
clouds could have been disrupted by the supernova blast wave and
are now forming the shell. Their CO(J=1-0) mm line data shows
sudden changes with position in radial velocity, and the presence
of broad asymmetric lines with peak-shoulder profiles, both of
which indicate a shock wave disruption of the CO clouds. Since
Ruiz \& May (1986) gave only an upper limit for the mass of
the shell, we have re-analyzed the gas content for this region.
As the longitude-velocity map in Figure \ref{g284}a shows, nearly
all CO emission in the general direction of the 3EG source and the
SNR G284.3-1.8 lies in the velocity range $-22$ to 3 km s$^{-1}$.
A CO map integrated over this range is shown in Figure
\ref{g284}b. The mean velocity of the emission is $\sim -9$ km
s$^{-1}$, which is consistent with the terminal velocity in this
direction, suggesting a distance of approximately 2.1 kpc.
However, since radial velocity changes very slowly with distance
in this direction, the uncertainty on the kinematic distance is
large, approximately $\pm 1$ kpc. We adopt a distance of 2.9 kpc,
the value inferred by Ruiz \& May (1986) based on optical
observations of the SNR filaments, the $\Sigma-D$ distance, as
well as the CO kinematics.\\

There is no CO detected at any velocity toward the nominal center
of the 3EG source ($l=283.93$, $b=-2.34$), although the total
molecular mass within the 95\% confidence radius of the 3EG source
(dotted circle in Figure \ref{g284}b) is $5.9 \times 10^4
M_\odot$. Most of this mass does not coincide with the SNR, which
is completely included within the 3EG source. Since this emission
does not form a single well-defined cloud, it's possible that it
arises from gas spread over quite a large distance along the line
of sight, perhaps $1-2$ kpc. If so, the gas density near the SNR
could be quite low. Our study therefore reinforces the idea that
it is most likely the pulsar, and not the hadronic or
bremsstrahlung emission from the SNR neighborhood, that is
responsible for 3EG J1013-5915.


\subsection{ $\gamma$-ray source 3EG J1102-6103 -- SNR G290.1-0.8
(MSH 11-61A)/289.7-0.3}

\begin{figure}[t]
\begin{center}
\caption{CO integrated over the velocity range of the far Carina
arm, 0 to 45 km s$^{-1}$. Contours: 843 MHz continuum from the
MOST Galactic plane survey (Green 1997); the survey has been
smoothed to a resolution of 3 arcmin to highlight extended
sources. The contour interval is 0.04 Jy/beam, starting at 0.04
Jy/beam. The dotted circle is the 95\% confidence radius about the
position of 3EG 1102-6103 (Hartman et al. 1999). G290.1-0.8 is the
SNR MSH 11-61A (Kirshner \& Winkler 1979). Both HII regions are in
the catalog of Georgelin \& Georgelin (1970) and have
recombination line velocities in rough agreement with that of the
complex. The component clouds A and B are discussed in the text.
  \label{1102}}\end{center}
\end{figure}

Sturner \& Dermer (1995) proposed that this $\gamma$-ray source,
in its 2EG J1103-6106 incarnation, may have been related to
SNR G291.0-0.1. However, the more precise localization in the 3EG
catalog shifted the source's position such that it is no
longer superposed with that SNR.\\

Zhang \& Cheng (1998) argued against a newly discovered young
radio pulsar, PSR J1105-6107 (Kaspi et al. 1997), as the source of
the observed high-energy $\gamma$-ray emission. Its age ($\sim 6.3
\times 10^4$ yr) also seems high for a Vela-like pulsar. In
addition, the photon spectral index is very soft,
$\Gamma=2.47\pm0.21$, though this in itself does not disqualify
a possible pulsar origin, as seen in the case of PSR J2229+6114/3EG J2227+6122
(Halpern et al. 2001b).\\

The line of sight to 3EG 1102-6103 intersects both the near and
far sides of the Carina spiral arm, at velocities near $-20$ km
s$^{-1}$ and $+20$ km s$^{-1}$ respectively. There is a distinct
gap in the near side of the Carina arm in the direction of the 3EG
source, with almost no molecular gas within $\sim 1$ deg of the
source direction (see, e.g., Figure 2 of Dame, Hartmann, \&
Thaddeus 2001). On the other hand, as Figure \ref{1102} shows,
there is a very massive molecular complex in the far Carina Arm
overlapping the direction of the 3EG source; this complex is No. 13
in the Carina Arm cloud catalog of Grabelsky et al. (1988). There
is little doubt that the two component clouds labeled A and B in
Figure \ref{1102} are part of the same complex, since they have
approximately the same velocity of 22 km s$^{-1}$, and are
connected smoothly by weaker emission, also at the same velocity.
Also, the HII regions are evidence of abundant on-going star
formation in this molecular complex which additionally supports the
association of the SNR. Assuming a flat rotation curve
beyond the solar circle, the kinematic distance of the complex is
8.0 kpc and its total molecular mass is 2.1 $\times 10^6
M_\odot$.\\

It is worth noting that the composite CO line profile of cloud B
is very broad and complex, suggesting possible interaction with
SNR G290.1-0.8. In the case of cloud A, its radius ($\sim $48 pc)
and composite linewidth (17 km s$^{-1}$ FWHM) are roughly
consistent with the radius-linewidth relation found for large
molecular complexes by Dame et al. (1986). For cloud B, however,
its linewidth ($\sim $27 km s$^{-1}$) is about a factor of 3 too
large compared to its radius ($\sim $28 pc). We can also see that
the coinciding SNR G289.7-0.3 is far from Cloud B, in a region of
low molecular density. It is extremely unlikely that this SNR is
related with the 3EG source in question. The only remaining
candidate is, then G290.1-0.8. The total molecular mass within the
95\% confidence radius of the 3EG source (dotted circle in Figure
\ref{1102}) is 7.7 $\times 10^5 M_\odot$ and most of it is
localized in Cloud B (4.5 $\times 10^5 M_\odot$). \\

Assuming typical values for the energy of the explosion
($E_{51}$=1) and the unshocked ambient density ($n=0.1 $cm$^{-3}$)
we obtain a CR enhancement factor of $\sim250$. Assuming that the
same CR enhancement is applicable to the cloud overpredicts the
EGRET flux by about a factor of 10. Thus, it is likely that the
average CR enhancement factor within the cloud is ten times lower
than within the SNR, a reasonable result. It is possible, then,
that 3EG J1102-6103 and SNR G290.1-0.8 are indeed
related. Note that Bremsstrahlung, which we have neglected here,
will contribute still more to the predicted
flux from SNR-cloud interactions. If the outlined
scenario is correct, GLAST and AGILE ought to observe a
strong, compact $\gamma$-ray source coincident with the position of Cloud
B.\\

An alternative, and promising, hypothesis for explaining the
high-energy emission is stellar winds collisions, as developed by
Eichler \& Usov (1993) and Benaglia \& Romero (2003). Recently,
Contreras et al. (1997) have provided convincing evidence for
non-thermal radio emission from the colliding winds region in the
stellar system Cygnus OB2 No 5.\footnote{The HEGRA Cherenkov
telescope array group recently reported a steady and extended
unidentified TeV gamma-ray source lying at the outskirts of Cygnus
OB2, the most massive stellar association known in the Galaxy,
estimated to contain ~2600 OB type members alone (Aharonian et al.
2002). Butt et al. (2003) reported on near-simultaneous follow-up
observations of the extended TeV source region with the CHANDRA
X-ray Observatory and the Very Large Array (VLA) radio telescope.
The broadband spectrum of the TeV source region favors a
predominantly nucleonic – rather than electronic – origin of the
high-energy flux, possibly in a way similar to that proposed by
Romero and Torres (2003) in the case of NGC 253.}

The position of their radio-imaged shocked region is consistent
with the inferred location of the contact discontinuity of the
wind-wind interaction of the constituent stars. Benaglia et al.
(2001) have argued that the source 3EG J2033+4118 could be mainly
due to inverse Compton scattering of the stellar photons by the
locally accelerated electrons. Similarly, in the present case, 3EG
J1102-6103 might be the result of $\gamma$-ray production by the
stellar winds of the early-type stars WR37, WR38, WR38B and WR39,
all located within the 95\% confidence contour of the $\gamma$-ray
source and at 2 kpc from the Sun. In particular, WR39 presents an
unusually strong wind with a terminal velocity of about 3600 km
s$^{-1}$ (Romero et al. 1999a and references therein). Non-thermal
radio emission at the mJy level has been recently detected at
$\sim $3 arcsec from the optical position of the star by Chapman
et al. (1999). This emission is a clear indication of the
existence of a population of relativistic electrons in the region.
Chapman et al. (1999) have suggested that particle acceleration
could be occurring at the region where the wind of WR39 collides
with the wind of the neighboring Wolf-Rayet star WR38B. This
hypothesis is supported by the fact that synchrotron radiation is
located between both stars (2 arcsec from WR38B). The relativistic
electrons should interact with UV photons from the star, producing
IC $\gamma$-rays that could explain part of the emission of 3EG
J1102-6103.\\

Fortunately, the peak of the spectral energy distribution should
be in the IBIS energy range, the Imager on-Board INTEGRAL. Using
Benaglia et al.'s (2001) model with a spectral index of $-2$, we
get an integrated flux for the energy interval 100 keV - 200 keV
of $1.2 \times 10^{-4}$ ph s$^{-1}$ cm$^{-2}$. For the entire IBIS
energy range (20 keV-10 MeV) the expected value is $1.2 \times
10^{-3}$ ph s$^{-1}$ cm$^{-2}$. In addition to this wind-wind
contribution, single stars might also be sources of $\gamma$-rays
in the IBIS energy band through the IC emission of electrons
locally accelerated in shocks at the base of the winds. These
shocks are produced by line-driven instabilities (see Chen \&
White 1991 for a discussion of $\gamma$-ray emission from single
stars, and
Benaglia et al. 2001 for a particular application).\\

Consequently, further study of the possible association of 3EG J1102-6103
with the SNR G290.1-0.8 (MSH 11-61A) is of the utmost importance, since
there are at least two scenarios (aside from the pulsar possibility) that
might well contribute to the observed $\gamma$-ray flux.


\subsection{ $\gamma$-ray source 3EG J1410-6147 -- SNR G312.4-0.4}

Although there is no correlated Princeton pulsar within the
contours of 3EG J1410-6147, there are two Parkes pulsars near
  the $\gamma$-ray source (see Table 6 above). Both
of them require an unreasonable efficiency, at their
dispersion-measure distances, to explain the observed $\gamma$-ray
flux. Both pulsars are located (at least in projection) well
within the boundaries of the incomplete shell of SNR~G312.4$-$0.4
(Caswell \& Barnes 1985), to which Yadigaroglu \& Romani (1997)
estimate a $\Sigma-D$ distance of 1.9\,kpc, whereas Case \&
Bhattacharya (1999) find 3.1$\pm$1.0 kpc. At either of these
distances, the required efficiencies would be substantially
smaller. The photon spectral index of the $\gamma$-ray source,
$\Gamma=2.12\pm0.14$, seems to be in the range of other pulsar
cases. More recently, Doherty et al. (2002) have provided
important new HI absorption measurements toward SNR G312.4-0.4
which indicate that it may be
much further away, ~8.1 kpc (see below).\\

\begin{figure}[t]
\begin{center}
\caption{CO integrated over the velocity range of the Centaurus
arm, -75 to -35 km s$^{-1}$. Contours: 843 MHz continuum from the
MOST Galactic plane survey (Green et al. 1999); the survey has
been smoothed to a resolution of 3' to highlight extended sources.
The contour interval is 0.01 Jy/beam, starting at 0.01 Jy/beam.
The dotted circle is the 95\% confidence radius about the position
of 3EG J1410-6147 (Hartman et al. 1999). Many of the other radio
sources in the map are unrelated HII regions discussed by Caswell
\& Barnes (1985).
  \label{1410}}\end{center}
\end{figure}

Case \& Bhattacharya (1999) have made an in depth study of the
possible association between the remnant and the $\gamma$-ray
source and concluded that the $\gamma$-ray data alone cannot at
present provide conclusive evidence to decide whether the
$\gamma$-ray emission from 2EG J1412-6211 is due to a pulsar or
SNR-molecular cloud interaction, or both. They suggested that CO
observations of the environment surrounding G312.4-0.4 would help
in determining whether a molecular cloud of sufficient mass is
present in the right location to produce the observed $\gamma$-ray
intensity. Such observations are presented here.\\

The CO emission toward 3EG 1410-6147 is extremely bright and
complex, arising mainly from the tangent region of the Centaurus
spiral arm at $v < -30$ km s$^{-1}$. The line of sight also
intersects the near side of the Carina arm at less negative
velocities and the far side of Carina at positive velocities. The
cloud in the general direction of the 3EG
source is also by far the most massive and dense. This cloud, with
a velocity of $-49$ km s$^{-1}$, is labeled A in Figure
\ref{1410}. The near and far kinematic distances of the cloud are
3.3 kpc and 8.1 kpc respectively (Clemens 1985); we will adopt the
far kinematic distance since it agrees with the HI absorption
measurements of Doherty et al. (2002) towards SNR G312.4-0.4.\\

The mass of Cloud A is quite uncertain owing to the fact that
molecular gas in  both the near and far sides of the Centaurus arm
probably contribute to the CO emission at the cloud velocity; the
high-velocity limit of the cloud is also uncertain owing to
blending with emission at higher (less negative) velocities. If we
assume that all of the emission in the velocity range of Figure
\ref{1410} ($-75$ to $-35$ km s$^{-1}$) arises from cloud A at 8.1
kpc, the total molecular mass within the 95\% confidence radius of
the 3EG source (dotted circle in Figure 15) is 3.3 $\times 10^5
M_\odot$; given the uncertainties just discussed, the actual mass
might be as much as a factor of 2 lower.\\

Even at such a distance, the large quantity of molecular material is sufficient
  to explain a
significant part of the $\gamma$-ray flux observed. As in the case
of 3EG 1102-6103, with the usual assumptions, a CR enhancement
factor of $\sim $100 is necessary to explain the bulk of the
$\gamma$-ray emission from 3EG J1410-6147. It is most likely,
however, that this EGRET source is a composite. New X-ray
observations could be used to study the pulsars and evaluate their
$\gamma$-ray emissivities. An extrapolation of the GeV spectrum to
the TeV regime, if there is no break, would give a flux of 1
$\times 10^{-10}$ erg cm$^{-2}$ s$^{-1}$, which is above the HESS sensitivity
in the range 500 GeV--10 TeV. Even with a substantial break in the
spectrum the 3EG source should be observable by HESS (see the study
on SNR TeV observability below, Section \ref{TEV_S}).

\subsection{ $\gamma$-ray source 3EG J1639-4702 -- SNR
G337.8-0.1/338.1+0.4/338.3+0.0}

\begin{figure}[t]
\begin{center}
\caption{Relative positions of 3EG J1639-4702 (which occupies all
the box) and the SNRs G333.8-0.1, G338.1+0.4, and G338.3+0.0,
which are very small in comparison. The superposed grey-scaled
levels show the radio emission at 843 MHz of the three SNRs as
reported in the MOST Catalog prepared by  Whiteoak and Green
(1996). \label{338}}\end{center}
\end{figure}
\begin{figure}[t]
\begin{center}
\caption{CO integrated over the velocity range -65 to -45 km
s$^{-1}$. Contours: 843 MHz continuum from the MOST Galactic plane
survey (Green et al. 1999); the contour interval is 0.2 Jy/beam,
starting at 0.1 Jy/beam. The dotted circle is the 95\% confidence
radius about the position of 3EG J1639-4702 (Hartman et al.
1999).\label{338b}}\end{center}
\end{figure}

Figure \ref{338} shows the relative positions of these SNRs within
the large location contours of 3EG J1639-4702. One Princeton
pulsar is within the contour of 3EG 1639-4702, but it can be ruled
out as a possible counterpart because of the required energetics.
In addition, three Parkes pulsars coincide with the same 3EG
source (see Table 3). Two of them can be immediately discarded
based on the same grounds: the required efficiencies are
unphysically high. However, PSR J1637-4642 seems to be a plausible
candidate. Only a 12\% efficiency would be needed to convert this
pulsar into a plausible counterpart for the origin of the
$\gamma$-ray emission. Although the spectral index, $\Gamma=2.50\pm0.18$,
seems quite soft in comparison with detected EGRET pulsars, the work of
Halpern et al. (2001) suggests that a soft spectral index does not
automatically rule out a pulsar origin of the $\gamma$-rays: they present
a strong case for PSR J2229+6114 being responsible for 3EG J2227+6122, even
though it has a high-energy spectral index of 2.24$\pm$ 0.14.  \\

Based on HI absorption seen all the way up to the terminal
velocity, Caswell et al. (1975) placed the SNR G337.8-0.1 beyond
the tangent point at 7.9 kpc. Koralesky et al. (1998) detected
maser emission in the SNR at $-45$ km s$^{-1}$, implying a far
kinematic distance of 12.4 kpc. As Figure \ref{338b} shows, there
is a very massive giant molecular cloud adjacent to the SNR in
direction and close to the associated maser in velocity ($-56$ km
s$^{-1}$). The far kinematic distance for this giant molecular
cloud  is favored by (1) its likely association with both the
  far-side maser just mentioned and a group of far-side HII regions
(group 5 in Georgelin \& Georgelin 1976); (2) its location very
close to the Galactic plane; and (3) the radius linewidth relation for
giant molecular clouds (Dame et al. 1986). The mean velocity of
the complex is $-56$ km s$^{-1}$, implying a far kinematic
distance of 11.8 kpc.\\

The giant molecular cloud  has recently been discussed by Corbel
et al. (1999) because an adjacent cloud (just outside the velocity
integration range of Figure \ref{338b}) apparently harbors the
soft $\gamma$-ray repeater SGR 1627-41. Corbel et al. suggest that
collision or tidal interaction between these two giant molecular
clouds may have set off the burst of star formation evident in
both.\\

  Taking the total CO luminosity of the giant molecular
cloud to be that in the range $l = 337.625$ to $338.25$, $b =
-0.25$ to $0.25$, and $v = -65$ to $-45$ km s$^{-1}$, the total
molecular mass is $5 \times 10^6 M_\odot$; this mass may be
overestimated by 10-20\% owing to the inclusion of emission from
gas at the same velocity at the near kinematic distance. Even with
this correction, this giant molecular cloud ranks among the few
most massive GMCs in the Galaxy (see, e.g., Dame et al. 1986); its
composite CO linewidth of $\sim 20$ km s$^{-1}$ is correspondingly
very large. Adopting a mean radius of 0.31 deg, or 65 pc at 11.8
kpc, the mean nucleon density of the cloud  is 176 cm$^{-3}$. The
total mass within the 95\% confidence radius of the 3EG source is
$7.6 \times 10^6 M_\odot$; this mass too may be overestimated by
10-20\% owing to inclusion of near-side emission.\\

The enhancement factor obtained from Eq. (\ref{kk}) is large for
typical parameter values. However, since the SNRs seems to be
immersed in the molecular cloud, the Morfill et al. (1984)
prescription may be an oversimplification in this case.
This case is similar to that of 3EG J1903+0550 in that we have a
distant SNR that we would not expect to be able to detect with EGRET
in the neighborhood of a very large molecular cloud.  AGILE observations,
in advance of GLAST, would greatly elucidate the origin for this 3EG
source, since even a factor of 2 improvement in resolution would be
enough to favor or reject the SNR connection.


\subsection{ $\gamma$-ray source 3EG J1714-3857 -- SNR
G348.5+0.0/348.5+0.1/347.3-0.5 \label{G347}}

\begin{figure}[t]
\begin{center}
\caption{Intensity map of the CO(J=1-0) transitions in the region
around RX J1713.7-3946, from Butt et al. (2002a). Two massive
clouds, called Cloud A and Cloud B, are indicated. The X-ray
contours of the SNR (Slane et al. 1999) are superposed in black, as well as
the location confidence contours of the GeV $\gamma$-ray source
3EG J1714-3857 (coincident with Cloud A) and the significance
contours of the TeV detection of the remnant by Enomoto et al.
(2002), mostly coincident with the X-ray radiation.
\label{mapa-butt}}\end{center}
\end{figure}

\begin{figure}[t]
\begin{center}
\caption{The radio synchrotron spectrum which would be expected
from the region of the shocked molecular material located towards
the NE of the remnant RX J1713.7-3946 (Cloud A), under the
assumption that the observed GeV flux were due to
electron/positron Bremsstrahlung. Since this spectrum violates the
upper limit (dark) derived from the non-detection of the cloud in
the radio band by a factor of $\sim 20$ at 843 MHz (Slane et al.,
1999), we can rule out a predominantly leptonic origin of the GeV
luminosity. Furthermore, if the GeV flux of 3EG J1714-3857 were of
electronic origin, the cloud region would outshine even the
radio-brightest NW rim of the remnant, which is found to be
emitting only 4$\pm$1 Jy at 1.36 GHz (Ellison et al. 2001), as
shown by the light data point. An assumed low frequency turnover
at $\sim$100 MHz in the radio spectrum is shown by the red dotted
line.\label{spec-point}}\end{center}
\end{figure}

The supernova remnant RX J1713.7-3946 is probably the most
convincing case for a hadronic cosmic-ray accelerator detected so
far in the Galaxy. Butt et al. (2001) noted the positional
coincidence of the nearby $\gamma$-ray source 3EG J1714-3857 with
a very massive ($\sim 3\; \times 10^5$ $M\odot$) and dense ($\sim
500$ nucleons cm$^{-3}$) molecular cloud that is clearly
interacting with the SNR RX J1713.7-3946 (G347.3-0.5) (Slane et
al., 1999; Butt et al., 2001). Figure \ref{mapa-butt} shows the
CO(J=1-0) line intensity distribution in the vicinity of
the SNR. The remnant is a
strong X-ray source (Slane et al. 1999) whose ROSAT contours are
indicated in the figure. Two massive clouds, which we called Cloud
A and Cloud B, can be seen. The first one is coincident with the
$\gamma$-ray source 3EG J1714-3857, whose location confidence
contours are also superposed in the figure. The X-ray
emission is produced by TeV-range electrons radiating by the
synchrotron mechanism in the local magnetic field. These same
electrons were suggested to be responsible, through IC
up-scattering of cosmic microwave background photons, for the TeV
$\gamma$-ray emission detected from the NW rim of the remnant by
Muraishi et al. (2000) and Butt et al (2002a).\\

We have previously provided measures of line intensity ratios
$R=$CO(J=2-1)/(J=1-0) for the entire region  demonstrating
that the SNR is likely interacting with Cloud A (Butt et al. 2001).
For this cloud, $R \sim
2.4$, more than 3.5$\sigma$ above the average Galactic value.\\

Upper limits to the continuum radio emission of Cloud A and the use of
Eq. (\ref{ratio}) above, allow us to rule out a
Bremsstrahlung origin of the GeV radiation from 3EG J1714-3857
(see Figure \ref{spec-point}): the electron flux needed to explain the GeV
source in terms of bremsstrahlung emission overpredicts the radio
synchrotron emission for any reasonable molecular cloud magnetic
fields (Cructher, 1988,1994,1999). The GeV $\gamma$-rays seem to be
the result of $\pi^{0}$-decays produced when the population of
cosmic-rays accelerated at the remnant shock are injected into the
dense medium of Cloud A. We estimate a cosmic-ray enhancement
factor in the range $24<k_{\rm s}<36$ given the parameters of the
SNR (Slane et al. 1999).\\

In addition, the $\gamma$-ray spectrum of 3EG J1714-3857 (Hartman et
al. 1999) is
consistent with a narrow spectral bump at $\sim70$ MeV that could
correspond to the signature of the pion-decays resulting from an enhanced
population of low
energy (E$\sim$1 GeV) protons. This apparent peak, although highly
suggestive, is not statistically significant and improved
observations are needed to confirm its existence. \\

Uchiyama et al. (2002a) have reported the discovery of extended
($10'\times 15'$) and hard (spectral shape described by a flat
power-law photon index $\Gamma=1.0^{+0.4}_{-0.3}$) X-ray emission
from the position of Cloud A, using ASCA data. This emission is
interpreted as bremsstrahlung from a Coulomb-loss-flattened
distribution of nonthermal low-energy protons in the cloud or mildly
relativistic electrons
(see
also Uchiyama et al. 2002b) . Uchiyama et al. (2002a) estimate that
the energy content in subrelativistic protons within the cloud far
exceeds that in the relativistic protons, say by a factor $\sim
80$. The explanation could be that the bulk of the more energetic
particles have already diffused from the cloud whereas the
sub-relativistic population is
captured there. Alternatively, energetic secondary leptons may also be
producing low-level non-thermal X-ray and radio emission in the clouds.
\\

Regarding the highest energy particles produced in the SNR, a
CANGAROO re-observation of the NW-rim of RX J1713.7-3946 with a
new 10-m reflector (CANGAROO II) has allowed a determination of
the TeV $\gamma$-ray spectrum, which can be fitted with a
power-law of photon index $\Gamma\sim -2.8$ (Enomoto et al. 2002).
Such a steep spectrum is hard to explain by IC emission and
Enomoto et al. (2002) have claimed that the TeV emission is also
of hadronic origin. If this were the case, however, the GeV
gamma-ray flux would be much higher than observed, as recently
noted independently by Reimer \& Pohl (2002) and by Butt et al.
(2002a). (It is conceivable that protons with a hard index of
$\sim$-1.8 may alleviate the discrepancy but then it is more
difficult to explain how such a population of protons could
produce the measured -2.8 index TeV $\gamma$-rays) \footnote{One
possible solution would be to introduce the effects of difussion,
if the $\gamma$-rays originate in one of the nearby clouds and
{\em not} in the SNR itself as suggested by CANGAROO team. See
also Uchiyama et al. 2002c.}).
\\

The SNR RX J1713.7-3946 is perhaps the best natural
laboratory available today for studying the acceleration and
diffusion of cosmic-rays. The unique combination of a relatively close SNR
and a group of well defined molecular clouds in its
surroundings, none of them in front of the remnant itself,
makes this source a priority target for the
forthcoming generation of high-energy instruments such as HESS,
AGILE, INTEGRAL, and, GLAST, as well as for infrared, radio, mm
and sub-mm observatories.

\subsection{ $\gamma$-ray source 3EG J1734-3232 -- SNR G355.6+0.0}
\begin{figure}
\begin{center}
\caption{Upper panel: MOST image of the SNR G355.6+0.0 at 0.843
GHz (Gray 1994). The grey-scale representation ranges from $0.4
\times 10^{-2}$ to $10 \times 10^{-2}$ Jy beam$^{-1}$. Radio
contours are shown in steps of 1 Jy beam$^{-1}$, the resolution is
43 arcsec. Part of the $\gamma$-ray probability contours of 3EG J1734-3232 are
superposed. Lower panel: Detailed radio image of the SNR at the
same frequency. Radio contours are shown in steps of $1 \times
10^{-2}$ Jy beam$^{-1}$, starting from $0.5 \times 10^{-2}$ Jy
beam$^{-1}$. \label{g355}}\end{center}
\end{figure}

The shell type supernova remnant G355.6+0.0 (Figure \ref{g355})
was first identified in the MOST Galactic Center survey (Gray
1994) as a compact ($\sim$0.1 deg $\times \sim$0.1 deg) radio
source with both thermal and non-thermal emission. The  western
limb shows possible indications of an interaction   with the
ambient diffuse thermal gas present at that location. X-ray
emission from this SNR has also recently been reported by the ASCA
X-ray   satellite under the designation
AX J173518-3237 (Sugizaki et al., 2001).\\

The physical relation of even part of the $\gamma$-ray flux of
3EG J1734-3232 with SNR G355.6+0.0, however, is unclear because of
the lack of information about both the SNR and its environs. The lack
of a $\gamma$-ray spectral index
for 3EG J1734-3232 (Hartman et al. 1999, see Table 1) further
complicates any attempt to connect the SNR with the
$\gamma$-ray emission. The $\gamma$-ray error box is also
coincident with a very young open cluster, NGC 6383,
$(l,b)=(355.66,0.05)$, which is centered around the bright
spectroscopic binary HD 159176 (O7V+O7V) (eg. van den Ancker et
al., 2000).  Together with NGC 6530 and NGC 6531, NGC 6383 belongs
to the Sgr OB1 association. The nearby radio source G355.3+0.1,
also within the $\gamma$-ray error box, is most likely an HII
region at a distance of $\sim$10 kpc (Crovisier et al. 1973).\\

Although the Third EGRET catalog lists GeV 1732-3130 (Lamb \& Macomb 1997)
as an alternate name for this source, the large
positional offsets indicate that these two may be
separate $\gamma$-ray sources (Roberts et al. 2001a).
Interestingly, ``bridging"   these two sources is a ``possibly
variable" COS-B source, 2CG 356+00, located at
$(l,b)=(356.5,+0.3)$ (Swanenburg et al. 1981), which may be
related to one or both of them.\\

The report of a bright, transient hard X-ray source, KS/GRS
1730-312 $(l,b=356.6,+1.06)$ (Vargas et al. 1996), within the 95\%
error ellipse of GeV 1732-3130 may also explain part
of the detected $\gamma$-ray emission from this region. It is
possible that in the quiet state KS/GRS 1730-312  was actually seen
in ASCA data, as ``src 1" in Roberts et al. (2001). A mild
indication of variability for 3EG J1734-3232 ($I=2.9,
\tau={0.00}_{0.00}^{0.24}$) appears within the $I$-scheme, and
would support the flaring hard X-ray source being associated with
it, although this is not conclusive with the data now at hand. The
transient source could also be separately connected to GeV
1732-3130/2CG 356+00. 2CG 356+00 is listed as ``possibly variable",
supporting such an argument (Swanenburg et al. 1981).\\

It is clear that no conclusive determination can be made regarding
possible association among SNR G355.6+0.0, GeV 1732-3130, 2CG 356+00,
and 3EG J1734-3232 until the sizes of the
$\gamma$-ray error boxes are significantly reduced by future
GeV telescopes. It is almost certain that many GeV sources, especially
those located towards the inner Galaxy,
will eventually be resolved into several separate sources.


\subsection{ Near the Galactic Center:
$\gamma$-ray source 3EG J1744-3011 -- SNR G359.0-0.9/359.1-0.5 and
$\gamma$-ray source 3EG J1746-2851 -- SNR G0.0+0.0/0.3+0.0}


Since the 3EG sources J1744-3011 and J1746-2851 lie near the very
confused Galactic Center region, an analysis of the sort we present
for other SNR-EGRET source pairs is not possible in this case.
Galactic Center region sources must
to be considered as possibly extended (most likely composite) and
confused, embedded in high and structured background, and the analysis
procedures used for other sources in the EGRET catalogs may not
apply (Mayer-Hasselwander et al. 1998).\\

A detailed discussion of the Galactic Center region is beyond the
scope of this report; for this the reader is referred to
Markoff et al. (1999), Yusef-Zadeh et al. (2000), Melia \& Falcke (2001),
as well as the book edited by Falke et al.(1999).
Very recently, Goldwurm (2001) has presented a review of the
high-energy emission detected from the direction of the Galactic
center; the reader is referred to that paper for details on different
X-rays observations of the Galactic Center region.\\


A complete discussion of EGRET observations of the
Galactic Center region was presented by Mayer-Hasselwander et al.
(1998). They found 5488$\pm$516 counts above 30 MeV, representing
a high significance excess. The region analyzed was larger than
the position of the central EGRET sources by several degrees (see
Figure 1 in Mayer-Hasselwander et al. 1998). They also re-analyzed
COS-B data, showing that the COS-B observations of the region were
in agreement, despite previous claims, with the more recent EGRET
data. Several objects in the region are potential counterparts
for the $\gamma$-ray
radiation detected, including, for example, GRO 1744-28, 2S1743-2941,
E1740-2942, PSR 1742-30, GRS 1736-297, GRS 1739-278, and GX 359+02.
However, not all of these coincide with the positions of the
3EG sources we are considering. The $\gamma$-ray fluxes in
different energy bands give only the hint of variability, which
is consistent with 3EG Catalog estimates for $I$ and
$\tau$ (Table 1). The photon spectrum of the diffuse
emission of the Galactic Center region shows a clear break at
energies about 1 GeV, with a significant steepening thereafter (it
shifts from $-1.3$ to $-3.1$). The hard spectrum at energies above
100 MeV has to be compared with the already hard values obtained
with standard EGRET analysis techniques, quoted in Table 1 for the
sources of interest. These hard values argue against $\gamma$-rays
being produced in diffusive processes, i.e., by ambient
matter--cosmic-ray interactions. However, this is yet to be
confirmed.\\

Yusef-Zadeh et al. (2002) have suggested that the central
$\gamma$-ray source may be due to the interaction of the
G0.13-0.13 molecular cloud with the diffuse and filamentary X-ray
features discovered using Chandra, all lying within the 95\%
confidence location contours of 3EG J1746-2851. The hard spectrum
of the EGRET source ($-1.7$) seems to match the cloud spectrum at
about 10 keV when extended down to X-ray energies. Electron
Bremsstrahlung and inverse Compton may be responsible for the GeV
emission. Pohl (1997) raised the possibility that the radio arc at
the Galactic Center could be the counterpart of the high-energy
$\gamma$-ray source. Existing radio data on the arc support the
view that its synchrotron emission originates from cooling,
initially monoenergetic electrons that diffuse and convect from
their sources to the outer extensions of the arc. If the source of
high-energy electrons coincides with the Sickle region
(G0.18-0.04), as indicated by the radio data, then the ambient
far-infrared  photons could be subject to inverse Compton
interaction by high-energy electrons. Pohl (1997) showed that the
predicted {$\gamma$}-ray emission depends mainly on the magnetic
field strength in the arc and that both the flux and the spectrum
of the central source could be explained by such a
process.\\

On the other hand, the starving state of accretion flow around the
supermassive black hole make it a dubious counterpart for the
high-energy radiation. Note, however, that from an statistical
point of view the probability of such a good agreement in the
positions of the Galactic Center and the 3EG source is about
$10^{-4}$ (Mayer-Hasselwander et al. 1998). However, early
scenarios were presented in which the $\gamma$-ray emission is
produced by the
wind accretion from the nearby IRS16 cluster (Melia 1992,
Mastichiadis \& Ozernoy 1994). More recently,
Markoff et al. (1997) were able to reproduce the observed spectrum
with a combination of synchrotron radiation and pion decay. If the
$\gamma$-ray flux is directly related to the dissipation of
gravitational energy, i.e. if it is produced by relativistic
particles energized by a shock within the infalling plasma, Sgr
A$^*$ could still be the source of the $\gamma$-rays observed.
However, in a refined analysis, Markoff et al. (1999), using data
from the 3EG catalog and an improved physical treatment, concluded that
this was not the case. \\

Forthcoming satellites, particularly INTEGRAL, will scrutinize the
Galactic center region and hopefully resolve the nature of this
interesting region.

\subsection{ $\gamma$-ray source 3EG J1800-2338 -- SNR G6.4-0.1 (W28)}

\begin{figure}[t]
\begin{center}
\caption{Relative positions of 3EG J1800-2338 and SNR W28.
Contours in step of 5 mK, starting from 15 mK, from the 2695-MHz
map obtained with the Effelsberg 100-m single dish telescope
(F\"urst et al. 1990).
  \label{6}}\end{center}
\end{figure}

This EGRET-SNR positional coincidence  was originally proposed by
Sturner et al. (1996) and Esposito et al. (1996). However, the
previous 2EG J1801-2321 source does not coincide with 3EG
J1800-2338, since the new position has shifted position by about
half a degree. W28 was also presented as a possible candidate for
an association with a COS B source by Pollock (1985), after
noticing that SNR W28 appears to be interacting with molecular
clouds. Dubner et al. (2000) have made a recent study of the
remnant using the VLA. They concluded that the remnant is indeed
interacting with molecular clouds in the vicinity, and observed
maser emission, as earlier reported by Claussen et al. (1997,
1999). Arikawa et al. (1999) observed W28 and mapped the CO(J=3-2)
and CO(J=1-0) rotational transition lines toward the remnant, and
also concluded that the remnant is interacting with the clouds.
The mass of the clouds was found to be $2 \times
10^3 M_\odot$. This mass, however, is with respect to the
2EG source position. A new molecular mass estimate with respect
to 3EG J1800-2338 is given below.\\

\begin{figure}
\begin{center}
\caption{(a) CO integrated over the velocity range 0 to 28 km
s$^{-1}$ corressponding to the distance of the SNR W28. Contours:
4850 MHz continuum from the survey of Condon
et al. (1991); the lowest contour is at 0.32 Jy/beam and the
contours are logarithmically spaced by a factor 2. The three
labeled sources are HII regions: M20 the Trifid Nebula and M8 the
Lagoon Nebula. The dotted circle is the 95\% confidence radius
about the position of 3EG J1800-2338 (Hartman et al. 1999). (b) CO
intensity integrated over the latitude range of the molecular gas
associated with W28, b = -0.5 to -0.125 deg.
  \label{w28b}}\end{center}
\end{figure}

The H-alpha filaments in W28 have a mean velocity of 18$\pm$5 km
s$^{-1}$ (Lozinskaya 1974).  HI absorption measurements by
Radhakrishman et al. (1972) are consistent with this velocity,
since absorption features at 7.3 and 17.6 km s$^{-1}$ are seen
against the SNR continuum. As Figure \ref{w28b} (lower panel)
shows, there is a molecular cloud at about the same velocity
($\sim$19 km s$^{-1}$) which is most likely the birth place of the
supernova progenitor. Directly toward the radio-bright rim of W28
( $l \sim$6.6 deg,  $b \sim$ -0.3 deg), which has been proposed by
Wootten (1981) and Arikawa et al. (1999) as the site of SNR-giant
molecular cloud interaction, a jet-like CO feature is seen in
Figure \ref{w28b}, extending to a smaller cloud at $\sim$7 km
s$^{-1}$. Arikawa et al. (1999) proposed the 7 km s$^{-1}$
component as the systemic velocity of W28, whereas Claussen et al.
(1997) suggested that the entire 7 km s$^{-1}$ cloud has been
accelerated by the SNR from the main cloud $\sim$10 km s$^{-1}$
higher in velocity. The alignment of the 7 km s$^{-1}$ cloud with
the bright interacting rim of W28 and the jet-like feature linking
it to the larger cloud at higher velocity supports the Claussen et
al. proposal.\\

The low longitude of W28 (6.5 deg) makes both the kinematic
distance and the mass of the associated molecular cloud difficult
to measure. Assuming a systemic velocity of 19 km s$^{-1}$, the
near kinematic distance is 3.7$\pm$1.5 kpc. If all emission in the
velocity range $0-28$ km s$^{-1}$ is associated with W28 (see
Figure \ref{w28b}b), the total molecular mass within the 95\%
confidence radius of the 3EG source is $3.9 \times 10^5 M_\odot$.
Owing to the severe velocity crowding at this Galactic longitude,
this mass estimate may be overestimated by as much as
50\%.\\

Even reduced by 50\%, the ambient molecular material can still account
for the observed EGRET flux.
Ignoring bremsstrahlung, a CR enhancement factor of about $k_{\rm
s} \sim 20$, similar to that found in other cases, is necessary
to explain the emission by hadronic interactions. This result is
stable against reasonable variations in the input parameters and
is compatible with consistency tests. Note that the distance used here
is near the upper limit of those shown in Table \ref{snr}; a smaller
distance would make a physical association even more likely. Vel\'azquez
et al. (2002), in a recent analysis of large-scale neutral hydrogen around
W28, adopt a distance of $\sim$1.9 kpc. They concluded that the SN
energy was  $\sim$ $1.6 \times 10^{50}$ erg about $3.3 \times
10^4$ years ago. \\


An intriguing possibility in this case is that there are actually
two (or more) $\gamma$-ray sources, each associated with a different
molecular cloud in Figure \ref{w28b} and/or with
the pulsar PSR B1758-23 (which, coincides with the GeV source
as reported in the Roberts et al. (2001) catalog). Investigation
of this possibility will require the next generation of GeV
and TeV-telescopes, with their improved sensitivity and
angular resolution. \\

W28 is one of the few remnants observed with \v{C}erenkov
telescopes. Rowell et al. (2000) observed it with CANGAROO 3.8-m
telescope and were able to set an upper limit on the flux (for
photons with $E>1.5$ TeV) of a diffuse source encompassing the
clouds discovered by Arikawa et al. (1999) and part of the 3EG
source: 6.64 $\times 10^{-12}$ ph cm$^{-2}$ s$^{-1}$. A simple
extrapolation of the 3EG flux, with the same spectral index, up to
TeV energies yields a value higher than this upper limit by more
than an order of magnitude (see Figure 5 of Rowell et al. 2000).
This implies the existence of a break in the spectrum in the
GeV-TeV region. But even considering such a break (see Section
\ref{TEV_S} and Table \ref{tev-snr}) the 3EG J1800-2338 region
could be visible to an observatory such as HESS in a matter of
hours.\\

The possibility of a leptonic origin for the $\gamma$-ray source
cannot be ruled out in this case. The SNR is ranked sixth when
ordered by radio flux among all entries in Green's (2000) Catalog
(Table \ref{snr}). With standard values of magnetic fields appropriate to
molecular clouds (eg. Crutcher 1988, 1994, 1999), the
radio flux that would be generated by the same electronic
population producing the GeV emission would not overpredict the currently
detected radio flux. A model such as the one
presented by Bykov et al. (2000) could explain the EGRET source
without invoking the dominance of hadronic interactions.

\subsection{ $\gamma$-ray source 3EG J1824-1514 -- SNR G16.8-1.1}

\begin{figure}[t]
\begin{center}
\caption{High-resolution radio map of the nearby star LS 5039
obtained with the VLBA and the VLA in phased array mode at 6 cm
wavelength. The presence of radio jets in this high- mass x-ray
binary is the main evidence supporting its microquasar nature. The
contours shown correspond to 6, 8, 10, 12, 14, 16, 18, 20, 25, 30,
40, and 50 times 0.085 mJy beam. From Paredes et al. (2000).
  \label{paredes}}\end{center}
\end{figure}

Paredes et al. (2000) proposed that the massive star LS 5039 is
part of a newly discovered microquasar system, and that it can be
identified with the $\gamma$-ray source 3EG J1824-1514. The
detection of radio jets and strong variability at different
wavelengths (including $\gamma$-rays) supports their
claim. Gamma-ray emission from microquasars with high-mass
companions has been recently discussed by Kaufman-Bernad\'o et al.
(2002), who show that GeV $\gamma$-rays can result from the
up-scattering of UV stellar photons by the relativistic jet (see
also Georganopoulos et al. 2002). Variability is naturally
produced by the changing viewing angle as the jet precesses due to
tidal forces from the accretion disk.\\

The supernova remnant G16.8-01.1, also within the 95 \% confidence
contour of the $\gamma$-ray source, has been recently studied by
Rib\' o et al. (2002). Although the radio structure of the remnant is not
well-resolved because of contamination from the partially
superposed HII region RCW 164 (Rodgers et al. 1960), but its size
is $\sim30$ arcmin and its total flux at 5 GHz is $\sim 1$ Jy
(Rib\'o et al. 2002). The distance to the source is not known, but
a lower limit of $\sim 2$ kpc has been established by Rib\'o et
al. (2002) through H166$\alpha$ line observations of the
foreground HII region. HI observations by the same authors
indicate that the ambient density around the remnant is $\sim 5$
cm$^{-3}$, with no evidence of interacting clouds. For a SN energy
release $E_{51}\sim 1$ and a distance $d\sim 3$ kpc, we found that
the expected pion-decay $\gamma$-ray flux from this SNR should be
$F(E>100\;{\rm MeV})\sim 10^{-7}$ ph cm$^{-2}$ s$^{-1}$
($\theta\sim0.5$). This is about one third of the observed flux
from 3EG J1824-1514. It is possible that bremsstrahlung from SNR
shell-cloud interactions could also generate a fraction of the
observed $\gamma$-rays.\\

In addition, there is one Princeton pulsar superposed on this 3EG
source, but it is not energetic enough to be a plausible
alternative to the microquasar found by Paredes et al. (see
Section \ref{difff}). The apparent (though marginal) variability
of the $\gamma$-ray source, in addition, argues weakly against a
pulsar origin of the GeV flux. A variability analysis (Torres et
al. 2001) suggests that 3EG J1824-1514 is marginally variable, but
this is not confirmed by Tompkins's $\tau$ index, which is
compatible with a steady source (although the difference between
the upper and the lower limit on Tompkins' $\tau$ is large). These
results are not conclusive, but it seems unlikely
that most of the $\gamma$-ray flux could come from either the SNR or
pulsar alone, especially when
the uncertainties in distance are taken into account.\\

The source 3EG J1824-1514 remains the best candidate for a
$\gamma$-ray emitting microquasar. Future tests of this interesting
source will surely be carried out with AGILE and GLAST.


\subsection{ $\gamma$-ray source 3EG J1837-0423 -- SNR G27.8+0.6}

\begin{figure}[t]
\begin{center}
\caption{Flux history of the 3EG J1837-0423. Only for one viewing
period 423.0, the source was undoubtedly detected, all others
being only upper limits. The x-axis in the figure does not
represent a linear scale of time. Rather, each point represent the
measurement for a different single viewing period given in Hartman
et al. (1999).
  \label{1837}}\end{center}
\end{figure}

This source is variable between EGRET observations, which argues
against a SNR or pulsar origin (see above). Indeed, 3EG J1837-0423
was detected only once, in viewing period 423.0 with 5.8 $\sigma$
significance level. In all other single viewing periods in which
it was observed, only an upper limit to the flux could be
established (see Figure \ref{1837}). This behavior is
compatible with objects presenting flares, such as AGNs. A
microlensing model might be a plausible alternative given the
absence of a strong radio emitter in the 3EG field (Torres et al.
2002, 2003). Another possibility is a non-pulsating black
hole of the sort discussed by B. Punsly (1998a,b and Punsly
et. al. 2000). In any case, the variability clearly indicated by
both the $I$ and $\tau$ indices make 3EG J1837-0423 incompatible
with a SNR or pulsar origin.


\subsection{ $\gamma$-ray source 3EG J1856+0114 -- SNR G34.7-0.4 (W44)}

\begin{figure}[t]
\begin{center}
\caption{The colors are CO(1-0) integrated over the velocity range 30 to 65
km/s. The solid contours are 4.85 GHz continuum from the survey of
Condon et al. (1994). The dashed contours are the 50\%, 68\%, 95\%, and
99\% confidence contours for 3EG J1856+0114.  The white "+" marks the
position with the highest ratio CO(2-1)/CO(1-0) as determined by Seta
et al. (1998).
\label{w44}}\end{center}
\end{figure}

The possible association of the $\gamma$-ray source 3EG J1856+0114
and SNR G34.7-0.4 (W44) has already been proposed by Esposito et
al (1996) and by Dermer et al. (1997). A comprehensive review of
the morphological properties of W44 at different frequencies is
given by De Jager \& Mastichiadis (1996). The radio emission of
the SNR is shell-like, but
the X-ray emission is centrally peaked (Rho et al. 1994). A pulsar
is found near the center of the 3EG source, PSR B1853+01 (quoted
as PSR J1856+0113 in Table \ref{pulsar}). Frail et al. (1996)
discovered a corresponding radio pulsar wind nebula, with a
tail-shape pointing back to the center of W44. In addition, they
found that the transverse velocity of the pulsar is compatible with
the expansion speed of the radio shell. Thus, it is probable that
this pulsar is the compact remnant of the supernova.\\

Recently, two new studies concluded that the remnant is in
interaction with molecular clouds: a new radio and optical study
by Giacani et al. (1997) and a complete CO study by Seta et al.
(1998). In Figure \ref{w44}, a map of  CO(1-0) integrated over
the velocity range 30 to 65 km/s is shown.
There are six giant molecular clouds, with masses between 0.3 and
$3 \times 10^5 M_\odot$, in the vicinity of W44. Three of them, with
a total mass of $4.1 \times 10^5 M_\odot$ (Seta et al. 1998) are
apparently interacting with the remnant.\\

The total molecular mass in the vicinity of the 3EG source in W44,
specifically in the region $l =  34.5$ to 34.875, $b = -0.75$ to $-0.375$,
and  $v = 30 - 65$ km s$^{-1}$, is $6.2 \times 10^4\;
M_\odot$. Based on the Clemens (1985) rotation curve, the main clump
near the 3EG source has a kinematic distance of 2.8 kpc. The molecular
mass is obtained based on the usual assumption that the H2 column
density is proportional to the CO velocity-integrated intensity.
The mass in the velocity-perturbed wings can be a factor $\sim$100
less than the total mass of the cloud. With these values for the
masses of passive targets and an CR enhancement factor of $\sim$40, the
entire GeV flux could, in principle, be explained by hadronic
interactions, although an additional Bremsstrahlung
component will be present also in the SNR shell-cloud
interactions. The enhancement factor of $\sim$40 is obtained
assuming $E_{51}=0.67$ and a pre-shock ambient density
$n_0=1$, but the conclusions are robust for reasonable
variations in these parameters.\\

We can compute mean densities for the well-defined
clumps (lighter colors in Figure \ref{w44}) by assuming
they are roughly spherical. However, the 3EG source contours enclose
less intense emission that is spread throughout the whole W44
complex, which is about $\sim 1$ deg across, (49 pc at the assumed
distance of 2.8 kpc). Thus it is reasonable to assume that the
molecular gas enclosed by the 3EG contours is spread over about
49 pc along the line of sight as well. The mean density would then be
$n \sim 6.2 \times 10^4
M_\odot/ [ \pi (9 {\rm pc} )^2 \times 49 {\rm pc} ] = 188$
H/cm$^{-3}$; here, 9 pc is the 95\% location contour of the 3EG
source. This density is 6 times larger than that used by
de Jager \& Mastichiadis' (1997) in computing the hadronic emission.\\

De Jager \& Mastichiadis (1997) developed a specific model
for this EGRET source (at the time, 2EG J1857+0118) and proposed
that the $\gamma$-ray radiation could be accounted for by
relativistic Bremsstrahlung and inverse Compton scattering. One of
their main motivations was to explain the hard spectral index of the
corresponding $\gamma$-ray source, $\sim -1.80$, as well as the
hard index of the radio source $\sim -0.3$, both difficult to
reconcile with the standard Fermi first order acceleration
process. The model of de Jager \& Mastichiadis (1997) propose the
pulsar PSR B1853+01 as the source of the gamma-rays; indeed, the
required efficiency to produce all the $\gamma$-ray radiation is
13\%, which appears marginally plausible, given the uncertainties.
However, the luminosity of the pulsar wind nebula (PWN) is
negligible in comparison with the total X-ray luminosity of W44
(Harrus, Hughes, \& Helfand 1996). The bulk of the X-ray emission
from W44 is thermal (Jones, Smith, \& Angellini 1993, Rho et al.
1994). But even though the X-ray luminosity of
the PWN is negligible, the pulsar could have injected a
significant amount of electrons in an earlier stage.
Mastichiadis' paper 
Good fits of the 2EG spectrum were obtained for a range of
particle density, whereas the magnetic field was required
to be $\sim 10 \mu$ G and the synchrotron cutoff frequency, $\nu_b
\sim 10^{12.5}$ Hz. A reasonable field strength could then explain
the $\gamma$-ray spectrum of W44 as originating in leptonic
processes. \\

W44 is then a complicated case. While the $\gamma$-ray source may
be due to SNR shock-cloud interactions (hadronic and leptonic), a
significant part of the $\gamma$-ray flux observed -- perhaps
even all of it -- could be due to a pulsar.
Application of Eq. (\ref{ratio}) does not allow us to discard, as
in the case of SNR G347.3-0.5, a leptonic origin for the
high-energy radiation. Here, then, future satellites and
telescopes will play an essential role in disentangling the
different possibilities. W44 was not observed to be emitting at
TeV energies, where an upper limit has been imposed by the Whipple
observatory at ($F(E>250 {\rm GeV})= 8.5 \times 10^{-11}$
cm$^{-2}$ s$^{-1}$, Leslard et al. 1995), implying that there is
a possible break in the spectrum from GeV to TeV. This can be
tested by forthcoming higher sensitivity TeV-telescopes.


\subsection{ $\gamma$-ray source 3EG J1903+0550 -- SNR G39.2-0.3}

\begin{figure}[t]
\begin{center}
\caption{Left: Relative positions of 3EG J1903+0550 and the SNR
G39.2-0.3. Radio contours in step of 2 mK, starting from 5 mK,
from the 2695-MHz map obtained with the Effelsberg 100-m single
dish telescope (F\"urst et al. 1990). Right: CO integrated over
the velocity range of the far Sagittarius arm, 48 to 70 km
s$^{-1}$. Contours: 4.85 GHz continuum from the survey of Condon
et al. (1994). The angular resolution is 7 arcmin; the contour
interval is 0.5 Jy/beam, starting at 0.3 Jy/beam. The dotted
circle is the 95\% radius about the central position of 3EG
J1903+0550 (Hartman et al. 1999). \label{39}}\end{center}
\end{figure}

Figure \ref{39} shows the relative positions of 3EG J1903+0550 and
the SNR G39.2-0.3. The 3EG source is also coincident with two
Princeton pulsars (Table 3). One of them can be readily
discarded as the origin of the $\gamma$-ray emission because of
the unrealistically high value required for the efficiency. The other
pulsar, PSR J1902+0615, lacks a measurement of the period derivative and
so we cannot judge the likelihood of this particular association.
The SNR G39.2-0.3 has been searched for OH maser emission by
Koralesky et al. (1998), but none was detected, although, of
course, this does not imply a lack of possible interaction. The
SNR is more than 8 kpc away, which seems to be -a priori- a
problem for generating the requisite flux via shock interactions.
We find, however, that the large distance may be, at least
partially, compensated for by the large amount of
molecular material in the neighborhood.\\

Caswell et al. (1975) detected HI absorption all the way up to the
terminal velocity towards the SNR G39.2-0.3, with almost
continuous strong absorption between 60 km s$^{-1}$ and the
terminal velocity. They therefore concluded that the remnant was
certainly beyond the tangent point, and most likely at the far
distance corresponding to 60 km s$^{-1}$, $\sim$9.6 kpc. Such a
large distance is consistent with the high foreground hydrogen
column inferred both by Becker \& Helfand (1987) based on 21 cm
absorption measurements with the VLA, and by Harrus \& Slane
(1999) based on ASCA observations. A distance of 9.6 kpc would
place the SNR in the far Sagittarius arm, where as Figure \ref{39}
shows, the remnant is nearly coincident with a massive molecular
complex. The complex is (40,59) in the catalog of Dame
et al. (1986), who assigned the far kinematic distance based on 2
associated HII regions.  The mass of this complex is estimated to
be $2.1 \times 10^6 M_\odot$. The mass within the 95\% confidence
radius of the 3EG source (dotted circle in Figure \ref{39}) is
even higher, $3.4 \times 10^6 M_\odot$, because the radius also
includes part of another molecular complex at higher longitude. If
part of the mass contained in the molecular complex could serve as
target material for the relativistic particles accelerated in the
SNR shock, this 3EG detection could plausibly be produced by a
combination of Bremsstrahlung and pion decay. With the mass
quoted, a CR enhancement factor of less than 10 is all that is
needed to produce the bulk of the observed $\gamma$-ray emission.
However, it is clear that not all of the molecular mass can
be illuminated by the SNR shock front. The SNR itself is less
than 8 arcmin in size, while the 3EG source is $\sim $1 deg in size.
Only 0.1\% of the molecular material need to serve as a target for
the particles accelerated in G39.2-0.3 in order to produce the 3EG source.
In this case, however, as in the case of 3EG J1639-4702, the
enhancement factor (Eq.\ref{kk}) is very large ($\sim 1000$), as a
result of the use of the Sedov solutions with typical values for the
energy of the explosion and ambient unshocked density.\\

There is an additional
nearby SNR, G40.5-0.5 (Downes et al. 1980), which is also associated with the
same cloud complex as G39.2-0.3 and at a similar distance. Though the center of
G40.5-0.5 does not coincide with the 3EG source
herein analyzed, it is suggesstive that the location of the 3EG source lies
between the two SNRs, and could thus have a composite origin. (SNR G40.5-0.5
  does not appear
in Table \ref{green} due to the high ellipticity of the confidence level
contours of 3EG J1903+0550.)
A large region, $-10 < b < 5$, $38<l<43$,
comprising this latter SNR as well as the 3EG source
was subject of a search for gamma-ray emission using the HEGRA system
of imaging
atmospheric telescopes (Aharonian et al. 2001). No evidence for emission
from point sources was detected, and upper limits
imposed were typically below 0.1 Crabs for the flux above 1 TeV.

\subsection{ $\gamma$-ray source 3EG J2016+3657 -- SNR G74.9+1.2 (CTB 87)}

\begin{figure}[t]
\begin{center}
\caption{CTB 87 region: The dashed contours are the usual EGRET
50\%, 68\%, 95\%, and 99\% confidence levels. The solid contours are
taken from the
4.85 GHz  continuum survey of Condon et al. (1994). The color is CO
integrated over the range v=-65 to -50 km/s. The positions of the
blazar G74.87+1.22
and of the WR-star WR138 are marked.
  \label{ctb87}}\end{center}
\end{figure}

Although it has been suggested that SNR G74.9+1.2 (CTB 87) may be interacting
  with ambient clouds (Huang et
al. 1983, Huang and Thaddeus 1986), the coincident 3EG J2016+3657
source has been proposed as a counterpart of the blazar-like radio
source G74.87+1.22 (B2013+370) (Halpern et al. 2001a, Mukherjee et
al. 2000). B2013+370 is a compact, flat spectrum, 2 Jy radio
source at 1 GHz. Its multiwavelength properties were compiled by
Mukherjee et al. (2000), and since they resemble other blazars
detected by EGRET, make B2013+370 an interesting possible counterpart
for this 3EG source.\\

Optical photometry of B2013+370 shows that it is variable,
providing additional evidence of its blazar nature (Halpern et al.
2001a). Additionally, the same authors presented a complete set of
classifications for the 14 brightest ROSAT X-ray sources in the
error circle of the 3EG source, of which B2013+370 remained the
most likely source of the $\gamma$-rays, should these come from a
point-like source. The Crab-like supernova remnant CTB 87 is
located at more than 10 kpc (Green 2000), seemingly disfavoring
its shell interactions as the cause of the EGRET source.
There are also WR stars in the field (Romero et al.
1999), which might produce $\gamma$-ray emission. This
possibility remains to be analyzed. INTEGRAL observations
would help in determining if there is $\gamma$-ray emission coming
from the stars.\\

Figure \ref{ctb87} shows a CO map for the CTB 87 region.  One
clearly defined molecular cloud appears in the map. The mean
velocity of the molecular cloud is -57 km/s. Assuming a flat
rotation curve beyond the solar circle, the cloud's kinematic
distance is 10.4 kpc. The total molecular mass within the 95\%
confidence radius of the 3EG source is $1.7 \times 10^5 M_\odot$.
With such a high value for the molecular mass, there is still
a chance that the hadronic or leptonic $\gamma$-ray emission may be
contributing to
3EG J2016+3657. \\

As in previous cases, only a precise determination of
the $\gamma$-ray source position will disentangle the origin of
this $\gamma$-ray source. Contrary to
other EGRET-SNR pairs, though, this one has the particularity of
enclosing a good candidate for an extra-galactic origin of
the radiation.


\subsection{ $\gamma$-ray source 3EG J2020+4017 --
SNR G78.2+2.1 ($\gamma$-Cygni Nebula, W66) \label{sw66}}

\begin{figure}[t]
\begin{center}
\caption{The filtered radio emission at 2.7 GHz of the SNR
G78.2+2.1 is shown in black. Radio contours, from the 2695-MHz map
obtained with the Effelsberg 100-m single dish telescope (F\"urst
et al. 1990), are labeled in steps of 1 K in brightness
temperature, starting at 3.0 K. The superposed white levels
represent the 99\%, 95\%, 68\%, and 50\% statistical probability
that a $\gamma$-ray source lies within each contour according to
the EGRET catalog (Hartman et al. 1999). In the background, a
CO(J=1-0) map of the region, integrated over the range $v=-20$ to
20 km s$^{-1}$, which includes all the emission in this direction
except for a small amount in the range $-50$ to $-40$ km s$^{-1}$
which is probably associated with the Perseus Arm is shown.
  \label{78}}\end{center}
\end{figure}

The SNR G78.2+2.1 lies in a very complex region of the sky, where
more than forty HII regions and a large number of shell structures
exist. Recently, Lozinskaya et al. (2000) have made an in depth
analysis of the SNR, which included new optical observations and
re-analysis of archival X-ray data. We refer the reader to their
paper for appropriate details on the SNR structure and other
features, other than those commented here, related with the
possible association with the 3EG source. The main result of
Lozinskaya et al. is that X-ray observations lead to a
self-consistent model of a young SNR at an early stage of
adiabatic expansion into a medium of relatively low density
($t=(5-6)\times 10^3$ yr, $n_0=0.14-0.3$ cm$^{-3}$).\\

The possible association between 3EG J2020+4017 (with the largest
signal-to-background ratio of any of the sources in the third
EGRET catalog that are positionally coincident with shell-type
supernova remnants) and G78.2+2.1 was previously suggested by
Pollock (1985), Sturner and Dermer (1995) and Esposito et al.
(1996), each using the $\gamma$-ray catalog available at
the time. Brazier et al. (1996)
studied this source (2EG 2020+4626 then) and reported the
discovery of a point like X-ray source, RX J2020.2+4026, which lies
close to the center of the remnant. If one is to assume that this
source and the 3EG detection are related, the ratio of the gamma
to X-ray fluxes is about 6000, similar to what is detected for the
radio quiet Geminga-pulsar. This prompted Brazier et al. to
suggest that 3EG J2020+4017 can indeed be a new Geminga-like radio
quiet pulsar, something which would be in tune with the hard
spectra, the low variability index, and the absence of `em'
classification in the 3EG Catalog (see Table \ref{green}). This
possible association remains, then, very suggestive. However,
Brazier et al. have reported no evidence for pulsations in the
range $13.3>f$(Hz)$>2$, $\dot F < 10^{-11}$ s$^{-2}$. Also, no
radio pulsar is known to be superposed to the 3EG contours (see
Table \ref{pulsar}). The absence of pulsar $\gamma$-ray radiation
can well be due to a statistical limitation of the EGRET data.
Analysis of GLAST (LAT) performance shows that periodicities
should be detectable if present in any of the known low-latitude
EGRET sources (Carrami\~nana 2001, Thompson et al. 2001).\\

In particular, Brazier et al. considered that the absence of em
(i.e. extended) classification in the 3EG Catalog was definitive
in disregarding the SNR as a possible site of cosmic ray
acceleration: the SNR is about 1 deg in size, and it should be
visible as extended by EGRET if the ambient matter were uniformly
distributed. This, however, may not be the case if the shock
accelerated particles interact and emit $\gamma$-rays most
intensely at a localized nearby concentration of molecular
material.\\

Yamamoto et al. (1999) have studied the possible SNR-cloud
interaction here and observed a very high
CO(J=2-1)/CO(J=1-0) intensity ratio ($\sim 1.5$), very suggestive
of an interacting cloud, is observed at $(l,b)$=(78,2.3).
Interestingly, this position exactly coincides with that of the
3EG source (see Figure \ref{78}).
After re-analyzing the same set of data used by Yamamoto et al.
(1999), we agree with their result, but have also found another
high ratio CO(J=2-1)/CO(J=1-0) at an adjacent position
($l=77.875$, $b=2.25$), and yet another moderately high ratio
($\sim 1$) at another adjacent position: $(l,b)=(78.00,2.25)$.
These high
ratios coincide nicely with the $\gamma$-ray source and with a
fairly well-defined CO cloud, and make a good case for the
interaction between the SNR and the cloud.\\


Assuming a distance of 1.7 kpc (Lozinskaya et al. 2000),
the molecular mass of this cloud is 4700 $M_\odot$
(calculated over the range $l = 77.75-78.125$ deg,
$b =  2.00 - 2.375$ deg, and $v=-10$ to 0 km s$^{-1}$).
This mass is rather uncertain since W66
lies in the so-called Cygnus X region of the Galaxy where the CO
emission is very complex and strong, probably because we are
viewing the Local spiral arm tangentially. It is hard to
judge what amount of CO might be associated with the SNR and what
amount is seen in projection. The Local arm emission is mainly in
the range $-20$ to 20 km s$^{-1}$, and the Perseus Arm is seen at
more negative velocities. Kinematic distances, particularly at low
velocity, are very unreliable because our line of sight is almost
tangent to the solar circle. \\

Using the data obtained for the mass of the cloud in the vicinity
of the 3EG source and apparently in interaction with the SNR, as
well as the mean value for the unshocked density ($n_0=0.22$
cm$^{-3}$) and distance (from Lozinskaya et al. 2000), it is
possible to explain the $\gamma$-rays by hadronic processes. The
use of Eq. (\ref{ratio}) with plausible values of the magnetic
fields in molecular clouds (Crutcher 1988, 1994, 1999) rules out
leptonic processes as the source of most of the radiation. \\

We cannot discard, however, a composite origin for the
$\gamma$-rays, with part of the radiation coming from the putative
$\gamma$-ray pulsar proposed by Brazier et al. (1996) and part
from cloud interactions with the nucleonic component of
freshly accelerated CRs.

\subsection{ An example beyond $|b|>10$:
$\gamma$-ray source 3EG J0010+7309 -- SNR G119.5+10.2  (CTA 1)}

The SNR-EGRET connection that we have explored so far was based
on the sample shown in Table 1,  constructed using
3EG sources within $10^{\rm o}$ of the Galactic plane. Although most SNRs
fall into this latitude range, there is the possibility of finding a
few nearby SNRs related with $\gamma$-ray sources at higher latitudes.
An example of such is briefly described in this section.\\

CTA 1 is a shell-type SNR, but its shell is incomplete and
broken-out towards the NW. This breakout phenomenon may be
caused by more rapid expansion of the blast wave shock into a
lower density region toward the NW.
HI observations supported this interpretation (Pineault et al.
1993, 1997). Since CTA 1 is located at a relatively high latitude (10
deg) and is nearby (1.4$\pm$0.3 kpc, Pineault et al. 1993), it has a
large angular size (90 arcmin), little foreground or background confusion,
and it can be observed at exceptionally high linear resolution.
The age of the SNR was estimated to be 10$^4$ yr by Pineault et al.
(1993), but it could be younger by a factor of 2 (Slane et al.
1997). CTA 1 was subject of intense observational campaigns in the
past years. There have been both ROSAT and ASCA X-ray observations (Seward et
al. 1995, Slane et al. 1997), as well as optical, infrared and
radio (see Pineault et al. 1997 and Brazier et al. 1998
for a review).\\

The ROSAT observation confirmed that CTA 1 belongs to the class of
composite SNRs, which show a shell-type morphology in the radio
band and are centre-filled in X-rays. Five point sources were
detected with ROSAT, one of which was found to coincide with the
EGRET source (at the time of the analysis, 2EG J0008+7307).  ASCA
data later revealed that this source, named RX J0007.0+7302,
has a non-thermal spectrum, suggesting that it is the pulsar left
from the supernova explosion (Slane et al. 1997). Optical
observations were carried out by Brazier et al. (1998), with a
2.12-m telescope, but no object was found within the positional
error box of the X-ray source. This allowed an upper limit to
be set on the optical magnitude of any counterpart to the putative
pulsar. \\

  The 3EG J0010+7309 is a non-variable
source under the $I$ and $\tau$ schemes, and has a hard spectral
index of 1.85$\pm$0.10, compatible with those of the Vela pulsar.
This source was also detected in the Second EGRET Catalog, but
with a shifted position which made it coincide with a nearby AGN
-- which was at the time proposed as a possible counterpart (Nolan
et al. 1996). This AGN is not coincident with the 3EG source and
can not be considered a plausible counterpart any longer. Based on
positional coincidence, on the hard spectral index, and on
physical similarities between the Vela pulsar and RX J0007.0+7302,
Brazier et al. (1998) proposed that the 3EG source and this X-ray
source were related. For an assumed 1 sr beaming, the observed
100-2000 MeV flux corresponds to a luminosity of $4\times 10^{33}$
erg s$^{-1}$, compatible with other $\gamma$-ray
pulsar detections (see Table \ref{eg-p}).\\

Although a thorough CO investigation has not yet been done for
this SNR, current data from HI observations do not support the
presence of very dense and massive molecular clouds in its
neighborhood (Pineault et al. 1997). It seems CTA 1 and 3EG
J0010+7309 might be related only through the compact object left
by the latter (RX J0007.0+7302). A better localization of the
$\gamma$-ray source with AGILE and GLAST will certainly test this
suggestion.

\section{ SNRs discovered by their likely
associated high-energy radiation}

\begin{figure} [t]
\begin{center}
\end{center}
\caption{The Capricornus SNR uncovered by the likely associated
high-energy emission. Left: Background filtered radio emission at
408 MHz (radio data from Haslam et al. 1982) of the region
surrounding three 3EG sources, whose contours are marked. Middle:
Radio map at 2.3 GHz (data from Jonas et al. 1998). The use of
both maps shows that the SNR is a non-thermal radio source. Right:
A map of the integrated column density of HI (velocity interval -3
to +5 km s$^{-1}$). Label units are 10$^{19}$ atoms cm$^{-2}$. Use
of this map shows that small enhancement factors could produce all
three EGRET sources. From Combi et al. (2001).}\label{capri}
\end{figure}

The SNR catalog compiled by Green (2000) is by no means complete.
A large number of low surface brightness remnants remain to be
discovered, hidden in the diffuse non-thermal continuum emission
produced by the diffuse component of the cosmic-ray electrons in
the Galaxy. In recent years, the application of filtering
techniques in radio continuum data has lead to the detection of
several new SNR (e.g. Duncan et al. 1995, 1997, Jonas 1999, Combi et al. 1999,
Klothes et al. 2001). The use of unidentified non-variable
$\gamma$-ray sources as tracers of interacting remnants can result
in new SNRs being discovered by their (plausibly associated)
high-energy emission.\\

As it was emphasized in the previous sections, not all SNR generate
observable $\gamma$-rays (say, at EGRET sensitivity) at typical Galactic
distances. A second ingredient is necessary: a target, i.e., a
dense medium such as a molecular cloud. If a SN explodes in a cloudy
medium then more than a single cloud could be illuminated by $p-p$
collisions and thus multiple $\gamma$-ray sources can emerge.
Clusters of steady unidentified $\gamma$-ray sources could trace
these situations and lead to the discovery of new, very extended
SNRs of low radio surface brightness.\\

This approach of considering EGRET sources as tracers of SNRs has
been applied by Combi et al. (1998, 2001), leading
to the discovery of two new SNRs. The basic technique consists of
making HI line observations in the direction of clearly
non-variable $\gamma$-ray sources. If some well-defined but small
clouds ($M\sim10^3-10^4$ $M_{\odot}$) are found within the 95\%
contours of the 3EG sources at velocities that correspond,
according to the galactic rotation curve, to distances of less
than 1 kpc, then large-scale (i.e. several degrees) radio
continuum observations are carried out. These observations aim at
detecting, through image filtering techniques, large SNRs of low
surface brightness. Observations at more than a single frequency
are necessary in order to determine if the radio spectral index
is non-thermal, as expected from such remnants.\\

The SNR candidates discovered by this procedure are G327-12.0,
located in the ARA region (Combi et al 1998) and G06.5-12.0, located in
Capricornus (Combi et al 2001): see Figure \ref{capri}. The first
one appears to be responsible for the $\gamma$-ray source 3EG
J1659-6251, whereas the second could be related to a cluster of
three 3EG sources: 3EG J1834-2803, 3EG J1847-3219, and
3EG1850-2652. HI clouds have been found at the positions of all
these sources except for 3EG J1847-3219. Both remnants are
thought to be so close that cosmic-ray enhancement factors
in the range $5<k_{\rm s}<45$ are sufficient to explain the
$\gamma$-sources through $\pi^{0}$-decays alone. Their low radio
surface-brightnesses rule out electronic Bremsstrahlung as the source
of the GeV flux. \\

We expect that the use of filtering techniques in
interferometric radio observations could lead to the discovery of
more such remnants in the near future (see, eg., Klothes et al.
2001), and with the improved capabilities of the next generation satellites,
$\gamma$-ray emission from far more distant interacting SNRs could
be detected.

\section{ SNRs and their neighborhoods as TeV sources}

\begin{table}
\caption{TeV Observations of SNRs.  Plerions are shown in the
first panel and shell-type SNRs in the second. The third panel
shows results for two 3EG sources coincident with SNRs for which
there have also been TeV observations. Partially adapted from
Fegan (2001) and Mori (2001).}
\begin{center}
{\small
\begin{tabular}{lll}
\hline
Object & Exposure time & Flux/Upper Limit or spectrum \\
Name & (hours) & $10^{-11} cm^{-2}s^{-1}$ or $10^{-11}$ cm$^{-2}$
s$^{-1}$ TeV$^{-1}$ \\\hline
{\bf  All TeV Observatories} & & \\
Crab Nebula & $\rightarrow\infty$ & 7.0 ($>$ 400GeV) \\
{\bf  CANGAROO} & & \\
PSR 1706-44 & 60 & 0.15 ($>$1TeV) \\
Vela Pulsar & 116 & 0.26 (E/2 TeV)$^{-2.4}$ TeV$^{-1}$  \\
{\bf  Durham} & & \\
PSR 1706-44 & 10 & 1.2 ($>$300GeV) \\
Vela Pulsar & 8.75 & $<$5.0 ($>$300GeV) \\
\hline
{\bf CANGAROO} & & \\
RXJ 1713.7-3946 & 66 & 0.53 ($\ge$1.8 TeV) \\
RXJ 1713.7-3946$^f$ & 32 & 0.53 $1.63 \pm 0.15 \pm 0.32)
E^{-2.84\pm0.15\pm0.20}$ \\
SN1006  & 34 & 0.46 ($\ge$1.7 TeV) \\
W28 & 58 & $<$0.88 ($>$ 5 TeV){\it $^a$}\\
{\bf HEGRA} & & \\
Cas A & 232 & 0.058 ($>$ 1 TeV){\it $^b$} \\
$\gamma$-Cygni & 47 & $<$1.1 ($>$500GeV){\it $^c$} \\

Monoceros  & 120 & ?$^e$\\
{\bf Durham} & & \\
SN1006 & 41 & $<$1.7 ($>$300GeV) \\
{\bf Whipple} & & \\
Monoceros & 13.1 & $<$4.8 ($>$500GeV) \\
Cas A & 6.9 & $<$0.66 ($>$500GeV)\\
W44 & 6 & $<$3.0 ($>$300GeV) \\
W51 & 7.8 & $<$3.6 ($>$300GeV) \\
$\gamma$-Cygni & 9.3 & $<$2.2 ($>$300GeV) \\
W63 & 2.3 & $<$6.4 ($>$300GeV) \\
Tycho & 14.5 & $<$0.8 ($>$300GeV)\\
{\bf CAT} & & \\
CasA & 24.4 & $<$0.74 ($>$400GeV) \\\hline
J2016+3657 & 287 & 5.8{\it $^d$} \\
J2020+4017 & 513 & 0.990{\it $^d$} \\
\hline
\end{tabular}
}
\end{center}
{\small $^{a}$A different definition of Energy Threshold is used.
$^{b}$Evidence for emission at the 4.9$\sigma$ level
(P\"{u}hlhofer et al. 2001). $^{c}$Limits converted from Crab
units  using flux of Hillas et al. (1998). $^{d}$Integral Flux
Above 400 GeV. $^e$Not yet reported. $^f$Observations made with
CANGAROO-II (Enomoto et al. 2002).  } \label{snr-tev-o}
\end{table}

As can be seen in Table \ref{snr-tev-o}, several observations of
SNRs such as W44, W51, $\gamma$-Cygni, W63 and Tycho's SNR, selected
because of their possible association with molecular clouds and/or
EGRET sources, have only produced TeV emission upper limits
(Buckley et al. 1998). This, as discussed below, could indicate
spectral cutoffs or breaks in the GeV-TeV energy range. For
instance, the required differential source spectrum would have to
steepen to $\sim$ E$^{-2.5}$ for $\gamma$-Cygni in order to escape
detection at TeV energies (Fegan 2001).\\

Very recently (Aharonian et al. 2002a),
the HEGRA system of imaging atmospheric Cherenkov telescopes reported
a survey of one quarter of the Galactic plane ($-2 < l < 85 $).
TeV gamma-ray emission from point sources and moderately extended
sources (diameter less than 0.8 deg), including
86 known pulsars (PSR), 63 known supernova remnants (SNR) and 9 GeV
sources, were searched
with negative results. Upper limits range from 0.15 Crab units up to
several Crab units, depending on the observation time and zenith
angles covered:
no TeV source was detected above 4.5$\sigma$ in a total observation
time of 115 h. At the same time, a search for point sources of radiation above 15 TeV
has been conducted with HEGRA AIROBICC array (Aharonian et al. 2002b), but only
flux upper limits of around 1.3 times the flux of Crab nebula were obtained
for candidate sources (including SNRs), depending, again, on the observation
time and zenith angles covered.\\

Positive detections, however, already exist. SN1006 (Tanimori et
al. 1998) and RXJ1713.7-3946 (Muraishi et al. 2000, Enomoto et al.
2002) were detected by the CANGAROO  telescopes. Observations of
SN1006 in 1996 and 1997 show a significant excess from the NW rim
of the SNR. The excess is consistent with the location of
non-thermal X-rays detected by ASCA (Koyama et al. 1995). Theoretical modeling
for SN 1006 has also been performed (e.g. Berezhko et al. 2003).\\

Cassiopeia A  was recently announced as TeV source by the HEGRA
collaboration (Aharonian et al. 2001c). 232 hours of observations
yield an excess at the 4.9$\sigma$ level, and a flux of
$\mathrm{F}=5.8\pm1.2_{{\rm stat}}\pm2_{{\rm syst}}\times10^{-13}
\mathrm{cm}^{-2}\mathrm{s}^{-1}$ at (E$>$ 1 TeV). The origin of
the $\gamma$-rays has been recently discussed by Berezhko et al. (2003).
Cas A has already been
associated with a bright source of hard X-rays which indicates a
population of non-thermal electrons with energies up to 100 TeV
(Allen et al. 1999b). 
Nevertheless, Cas A is not
a 3EG source, but this could be due only to the small EGRET sensitivity.   \\

There have also been observations of unidentified $\gamma$-ray
sources whose identifications are, perhaps, more
tentative. These observations were made with the Whipple 10-m
telescope (Buckley et al. 1997) Two of them were related with the
3EG sources in our sample, and details are provided in Table
\ref{snr-tev-o}.\\

The Crab Nebula has been detected  by several Cherenkov
observatories and it is often used to provide a check on the
calibration of new instruments (e.g. Aharonian et al. 2000).
STACEE and CELESTE, on which we
comment below, have published significant detections of the Crab
Nebula in an energy range lower than that obtained by others
ground-based instruments, $E>190\pm60$ GeV (Oser et al. 2000) and
$E>50\mathrm{GeV}$ (De Naurois et al. 2001), respectively. The Crab's
energy spectrum between 300GeV and 50TeV has been well
established (see below), and it steepens with energy
(Hillas et al. 1998, Aharonian et al. 2000). No pulsation has yet
been seen (Gillanders et al. 1997, Burdett et
al. 1999, Aharonian et al. 1999).\\

The CANGAROO team has detected the pulsar PSR 1706-44 in 60 hours
of observations (Kifune et al. 1995). They have also detected the
Vela pulsar at the $6\sigma$ level, based on $\sim120$ hours of
observation. Observations by the Durham group (Chadwick et al.
1997) also confirmed these detections. In the case of Vela, the
VHE signal, which is offset from the location of the pulsar by
$0.14^\circ$, is thought to originate from a synchrotron nebula,
powered by a population of relativistic electrons which were
created in the supernova explosion and which have survived since
then due to the low magnetic field in the nebula.

\section{ Concluding remarks}

The coming years will be exciting times again for $\gamma$-ray
astronomy -- after the unfortunate forced demise of the Compton
Observatory. INTEGRAL, AGILE, MAGIC, and the new stereo-IACTs are
or should be on-line soon; and in the longer term, GLAST will
surely bring about another revolution, answering existing
questions, such as the one reviewed here, and posing further
challenges. Ultimately, it will take a sensitive and high
spatial-resolution MeV-GeV detector such as GLAST working in close
consort with the ground-based TeV and radio telescopes to address
the Galactic cosmic ray origin problem: the single most definitive
test that SNRs (or any other putative sources) are accelerating CR
nuclei would be the statistically significant detection of the
signature neutral pion $\gamma$-ray hump centered at 67.5 MeV (in
$\log E_\gamma$), as has been seen already by EGRET in the diffuse
$\gamma$-ray background (Hunter et al. 1997). Though such a
detection, by itself, only proves the existence of lower energy
($\sim$1 GeV/n) nuclei, the detection allows to normalize the
hadronic vs. electronic contributions in a model--independent
fashion at those `lower' energies. Thus, such a detection together
with the extension of the spectrum into the TeV regime, and a
multiwavelength spectrum inconsistent with an electronic origin,
ought to be sufficient evidence to conclude that nuclei are being
accelerated by SNRs. It remains to be seen if this will cleanly be
borne out by the data. \\

In any case, the SNR-EGRET connection looks stronger than ever.
Several cases for a physical association, based on the analysis of
gamma-ray data and the molecular environment of the SNR, has been
shown promising and a definite conclusion about the origin of
cosmic rays in SNR shocks seems to be reachable in a nearby
future. Very recently, Erlykin and Wolfendale (2003) have shown,
in addition, that  for some nearby SNRs, and even when there is no
gamma-ray signature, the idea of cosmic ray production in the
shocks cannot be discarded since there could be a possible
evacuation of ambient gas by the stellar wind of the progenitor
star, or by the explosion of a nearby earlier supernova. The case
by case analysis shows, moreover, that it is at least plausible
that EGRET has detected distant (more than 6 kpc) SNRs. There are
5 coinciding pairs of 3EG sources and SNRs for which the latter
apparently lie at such high values of distance (disregarding those
related with SNRs spatially close to the galactic center). For all
these cases, we have uncovered the existence of nearby, large, in
some cases giant, molecular clouds that could enhance the GeV
signal through pion decay. It is possible that the physical
relationship between the 3EG source and the coinciding SNR could
provide for these pairs a substantial part of the GeV emission
observed. This does not preclude, however, composite origins for
the total amount of the radiation detected. Some of these cases
present other plausible scenarios. AGILE observations, in advance
of GLAST, would greatly elucidate the origin for these 3EG
sources, since even a factor of 2 improvement in spatial
resolution would
be enough to reject the SNR connection.\\

Kilometer-scale neutrino telescopes have also proposed as viable
detectors of hadronic cosmic ray sources (Halzen \& Hooper 2002,
Anchordoqui et al. 2003), and will be a welcomed addition to the
arsenal of space- and ground-based detectors that ought to be
lined up by the time the large-scale neutrino telescopes are
functional. Surely, with all this instrumentation focused on the
problem we will finally be able to test Shklovskii's suspicion
that ``it is possible that ionized interstellar atoms are
accelerated in the moving magnetic fields connected with an
expanding [SNR] nebula." (Shklovskii, 1953).

\section{ Appendix:
Reviewing the prospects for the forthcoming GeV satellites}

\begin{table}[t]
   \caption{Instrumental parameters of some of the forthcoming
satellite missions
   in the MeV-GeV range. EGRET data (first panel) is shown for
comparison. The second panel
   corresponds to AGILE-GRID and the third to GLAST-LAT.}\vspace{0.2cm}
   \begin{center}
   {\small
\begin{tabular}{llllllll}\hline
  Energy  & Energy & Effective & Field & Angular
  & Minimum & Source &Bremsstrahlung
Life\\
   range &  resolution & area & of view & resolution
&flux & location &
\\
   &$\Delta E/E$ & cm$^2$&  sr&  &ph cm$^{-2}$ s$^{-1}$ & arcmin& yr\\
  \hline
&\\
  20 MeV &$\sim 0.1$ &1500 & 0.5 & 100 MeV: 5.8$^{\rm o}$ &
$\sim 10^{-7} $ & $\sim 30$ & 91-96\\
  30 GeV& & & & 1 GeV: 1.7$^{\rm o}$\\ &\\ \hline
&\\
  30 MeV  & $\sim 1$ &540 & 3.0 & 1 GeV:
$0.6^{\rm o}$ & $\geq 6 \times 10^{-8}$ &5-20 & 03-06\\
  50 GeV \\&\\ \hline
&\\
  20 MeV  & $\sim 0.1 $ & 8000 & 2.5 & 100 MeV: $\sim
3.5^{\rm o}$ & $\sim4\times
10^{-9}$ & $<1 $ & 06-11 \\
  300GeV & &  &  &  10 GeV: $\sim0.1^{\rm o}$ & \\ &\\ \hline
\end{tabular}
}
\end{center}\label{glast}
\end{table}

\subsection{ INTEGRAL}

The International Gamma-ray Astrophysics Laboratory (INTEGRAL) is
now in orbit. It has main scientific instruments, named SPI
(Spectrometer of INTEGRAL) and IBIS (Imager onBoard the Integral
Satellite). SPI will work in the range 20 KeV -- 8 MeV and will
perform high resolution spectroscopy; it can measure $\gamma$-ray
line profiles with an accuracy of 2 keV. The sensitivity of IBIS
lies in the range 10 keV -- 10 MeV, and it will focus on achieving
good angular resolution (12 arcmin). INTEGRAL will also have three
monitors, the twin JEM-X's (3 keV -- 35 keV) and a OMC in the
optical band (500-600 nm). The JEM-X's will have an angular
resolution of 3 arcmin and a 4.8$^{\rm o}$ fields of view. The OMC
will have a pixel resolution of 16.6 arcsec and a 5$^{\rm o}
\times 5^{\rm o}$ field of view
(Sch\"onfelder 2001).\\

INTEGRAL could identify compact objects with good spatial
resolution that are likely counterparts of EGRET sources, and will
help confirm or reject interpretations based, for instance, on
microquasars and $\gamma$-ray blazars. The INTEGRAL galactic plane
exposure will help to clarify the confused region near Vela.
Another 1 Ms exposure, part of the satellite Open Program, will
focus on the Carina region. That part of the sky is populated
with five EGRET sources, two of which were analyzed above in the
case-by-case study: 3EG J1102-6103 and 3EG J1013-5915. As an
example of its potential, we will briefly discuss what
INTEGRAL can do in establishing the nature of these sources. \\

The former case, 3EG J1102-6103,  was briefly mentioned in the
corresponding section above. Although a hadronic origin for the
$\gamma$-ray emission is possible for this source, it could also
be the result of inverse Compton up-scattering of UV photons by
electrons accelerated in the winds of one or several Wolf-Rayet
stars (particularly in the region where the winds of
W39 and WR38B collide). If the latter explanation is correct,
INTEGRAL should see a source where the winds collide.\\

In the latter case, the source 3EG J1013-5915 is likely mostly
produced by  PSR J1013-5915 (Camilo et al. 2001). The JEM-X
monitor, in particular, could then probe the putative non-thermal
emission from this region, especially the tail of the synchrotron
spectrum and, in this way, help to determine the high-energy
cutoff of the electron population in the source. \\

In addition, there have been suggestions that INTEGRAL could
detect the 0.511 MeV line from giant molecular clouds
(see e.g. Guessoum et al. 2001). Giant molecular clouds are
typically surrounded by HII regions, ionized in an uncertain
fraction. If cosmic-rays are diffusing into the cloud, the cores,
too, may be ionized. If a sufficient density of cosmic-rays is
present, they could excite CNO nuclei as well as lead to the
production of nuclear gamma-rays lines and positron-production,
the latter being emitted by radioactive nuclei. The fluxes and
line separations predicted are on the verge of detectability by
INTEGRAL (Guessoum et al. 2001), but it will be an interesting
arena to explore with forthcoming observations, particularly for
those giant molecular clouds closer to Earth (regrettably not the ones we
find superposed with the SNRs in Table 1).\\

Thus, although it is expected that no direct observations with
INTEGRAL can prove that cosmic-rays are accelerated in SNR shocks
(the energy range of the relevant phenomena, around 100 MeV and
beyond, is out of the INTEGRAL energy band), they can be used to
explore alternative explanations for the unidentified EGRET
sources, and so are important to get an overall picture of the
SNR-EGRET source connection. Nuclear gamma-ray line signatures in
the 0.1-10 MeV band will certainly also provide important hints
regarding the CR acceleration processes in the Galaxy.

\subsection{ AGILE }

AGILE (Astro-rivelatore Gamma a Immagini Leggero), is expected to
be launched in 2003 (for a recent review see Tavani et al. 2001).
AGILE will have a very large field of view, covering at one time
approximately 1/5 of the sky at energies between 30 MeV and 50 GeV.
The angular resolution will be a factor of two better than that of
EGRET, specifically due to the GRID instrument, whose
parameters are given in Table \ref{glast}. However, the
sensitivity for point-like sources will remain comparable to that
of EGRET. AGILE will also have detection and imaging capabilities
in the hard X-ray range provided by the Super-AGILE detector. The
main goals of Super-AGILE are the simultaneous $\gamma$-ray and
hard X-ray detection of astrophysical sources (which was never
achieved by previous $\gamma$-ray instruments), improved source
positioning (1-3 arcmins, depending on intensity), fast burst
alert, and on-board triggering capability.\\

AGILE is thus very well suited for studying compact
objects, particularily those presenting $\gamma$-ray variability or
pulsed emission. AGILE will search for pulsed $\gamma$-ray
emission from all recently discovered Parkes pulsars coincident with
EGRET sources (D'amico et al. 2001, Torres et al. 2001d, Camilo et
al. 2001), so establishing their contribution to the
EGRET $\gamma$-ray flux. Unpulsed $\gamma$-ray emission
from plerionic SNRs, and search for time variability in
pulsar-wind nebula interactions will also be possible targets for
AGILE. Finally, AGILE will be essential in assessing the possible
existence of new populations of variable $\gamma$-ray sources in
the galaxy, such as non-pulsating black holes, X-ray binaries, and
microquasars. In the case of 3EG J0542+2610, for instance, AGILE
could test the hypothesis that the $\gamma$-ray
emission is produced in A0535+26 (Romero et al. 2001). The key
prediction of this model is anti-correlation of the X-ray and
$\gamma$-ray emissions.

\subsection{ GLAST\label{gg}}

\begin{figure} [t]
\begin{center}
\end{center}
\caption{Models of the neutral pion decay, non-thermal
Bremsstrahlung (NB), inverse Compton (IC), and pulsar (PSR)
$\gamma$-ray spectra associated with W66. The sum of the three
cosmic-ray components ($\pi^0$ + NB + IC) of the shell and the sum
of all four components ($\pi^0$ + NB + IC + pulsar emission) are
shown. EGRET spectral data is also included. From Allen et al.
(1999).}\label{allen-spe}
\end{figure}

The Large Area Telescope (LAT) on the upcoming Gamma-ray Large
Area Space Telescope (GLAST) mission will be suitable for studying
the relationship between $\gamma$-ray sources and SNRs. The
instrument's parameters are given in Table \ref{glast} (Michelson
2001). GLAST, expected to be launched in 2006, will explore the
energy range from 30 MeV to greater than 100 GeV with 10\% energy
resolution between 100 MeV and 10 GeV.  GLAST uniquely combines high
angular resolution with superb sensitivity, and has a moderate
effective area. Sources below the EGRET threshold ($\sim 6 \times
10^{-8}$ photons cm$^{-2}$ s$^{-1}$) will be localized to arcmin
scales. This clearly will improve our understanding of the SNR shock
acceleration of cosmic-rays and hadronic $\gamma$-ray
production.\\

In what follows, we provide a brief example of GLAST capabilities
as applied to the SNR  $\gamma$-cygni (W66), courtesy of Seth
Digel and NASA (see also, Ormes et al. 2000 astro-ph/0003270 from
which this example is adapted).  Allen et al. (1999) studied what
information could be obtained from GLAST observations based on a 1-year
all-sky survey assuming that 60\% of the $\gamma$-ray flux
was produced by a pulsar at the location proposed by Brazier et
al. (1996). This pulsar was assumed to have a differential photon
index of 2.08, that of the coincident source, 3EG J2020+4017. The
remainder of the photon flux was assumed to come from
relativistically accelerated particles through both leptonic and
hadronic processes -- the large dominance of the latter in
agreement with what we found in the corresponding section above.
The position of the molecular cloud assumed in the simulations
corresponds well with the CO clouds actually found -- see Figure \ref{78}).\\

Allen et al. additionally assumed that the electron and proton
spectra of W66 have the shape specified by Bell (1978, see their
Eq. 5), with a common relativistic spectral index of
$\Gamma=2.08$. The normalization of the electron spectrum was
determined from the radio data, by assuming that the magnetic
field strength is 100 $\mu$G. The normalization of the proton
spectrum, instead, was determined by assuming that the total
number of non-thermal electrons is a factor of 1.2 larger than the
number of non-thermal protons.\\

\begin{figure} [t]
\begin{center}
\end{center}
\caption{Comparison between the observed EGRET data and a GLAST
simulation for the W66 SNR region. The large circle shows the
position and extent of the radio shell of the SNR.   GLAST will be
able to localize distinctively the position of the putative
$\gamma$-ray pulsar (marked here as the X-ray source) proposed by
Brazier et al. as well as the SNR shock interacting with the
molecular cloud, should both contribute as assumed to the
$\gamma$-ray flux observed. Courtesy of S. Digel and NASA; adapted
from Ormes et al. 2000 astro-ph/0003270.
}\label{simw66}
\end{figure}

The spectral results of the simulations are shown in Figure
\ref{allen-spe}. It shows the $\gamma$-ray spectra produced by the
putative pulsar, by the decay of neutral pions, as well as by the
leptonic processes: Bremsstrahlung radiation of the electrons, and
inverse Compton scattering of electrons on the cosmic microwave
background radiation. The latter three of these four spectra were
obtained using the $\gamma$-ray emissivity results of Gaisser,
Protheroe, \& Stanev (1998), and Baring et al. (1999). The
non-thermal Bremsstrahlung and neutral pion spectra were obtained
assuming that the average density of the material with which the
cosmic-rays interact was $n_0 = 190$ atoms cm$^{-3}$, which should
be compared with the actual density we found, of 188 atoms
cm$^{-3}$ (see Section \ref{sw66}. Perhaps most illustrative is
Figure \ref{simw66}, which demonstrates that the resolution
of GLAST will allow one to distinguish the contribution of the
pulsar from that of the interacting molecular cloud. Should a picture like
Figure \ref{simw66} be the result of an actual GLAST observation
the proposed composite origin for 3EG J2020+4017 would be proved.

\section{ Appendix: Future TeV telescopes and their look at SNRs
- Adapted from Petry (2001) \label{TEV_S}}

At higher and higher energies the sky appears darker and darker.
There are just 57 sources detected between 1 and 10 GeV (Lamb \&
Macomb 1997) and higher frequencies remain largely unobserved. In
order to reach the highest energy $\gamma$-rays, new ground-based
telescopes are being built and older ones upgraded. A very brief
description of them, and of their impact on the SNR-$\gamma$-ray
source association problem is given here.\\

One approach to constructing high-energy $\gamma$-ray detectors is
to use existing solar farms, which have fields of large heliostats
focusing sunlight on a central tower; such facilities lie unused
at night. The arrival of the \v{C}erenkov wavefront at groups of
heliostats is precisely measured, and this information is used to
differentiate $\gamma$-rays from cosmic-ray primaries (Ong 1998).
STACEE (Chantell et al. 1998), CELESTE (Smith et al. 1997),
Solar-2 (T\"{u}mer et al. 1999), and GRAAL
(Arqueros et al. 1999) are all examples of such facilities.\\

MILAGRO (Sinnis et al. 1995) and Tibet HD (Amenomori et al. 1999)
are examples of air showers detectors. These arrays can in
principle operate 24 hours a day and are expected to achieve a
larger energy range, comparable to the next
generation imagers discussed below.\\


Imaging Atmospheric \v{C}herenkov Telescopes (IACTs) are
ground-based $\gamma$-ray detectors using the atmosphere as a
tracker and calorimeter and having good all-around performance,
both in sensitivity and angular resolution, for sources above 100
GeV. They typically have point spread functions with $\theta_{68}
< 0.16^\circ$. The number of detected primary gamma-photons will
typically be $> 100$, and therefore source locations, computed as
$\theta_{68} / \sqrt{N}$, can reach arcmin accuracies.
Generically, the possibility of separating two nearby point
sources can be realized if their angular distance is $>
3\theta_{68}$. Some forthcoming IACT
telescopes in the TeV regime are detailed in Table \ref{tev}.\\

The successful introduction of stereo imaging technology in the
HEGRA instrument has paved the way for bigger and more complex
instruments. The unprecedented accuracy in air shower stereo
reconstruction of the HESS telescopes in Namibia will allow one to
obtain an angular resolution of $\sim 0.1^{\circ}$ with a source
location accuracy of only 10 arcseconds (Konopelko 2001). After 50
hours of observation of a point source, HESS will detect a flux of
$\sim 10^{-11}$ ph cm$^{-2}$ s$^{-1}$ at $E>100$ GeV. In the case
of extended sources of angular size $\Theta$, the flux detected
will be $\sim (\Theta/0.1\;{\rm deg})\times 10^{-11}$ ph cm$^{-2}$
s$^{-1}$ in the same energy range. It is clear that such a
powerful instrument will provide answers to many of the pending
questions mentioned in this review, at least for the case of
southern SNRs. \\

A natural next step in the development of stereo imaging arrays
for TeV astronomy is to place one of these systems at high
altitude, where it could detect much lower-energy gamma-rays. It
has been estimated that at an altitude of 5 km, a threshold of 5
GeV might be achieved (Aharonian et al. 2001b). These systems,
then, might replace orbital
observatories in the GeV band in the foreseeable future.\\

\begin{table}
\caption{Next-generation Imaging \v{C}herenkov Telescopes and
their main characteristics. The locations of the sites chosen for
these observatories and their estimated minimum energy thresholds
are given.  Roman numbers behind some of the project names denote
the project phases. The Type refers to the number of individual
telescopes and diameter of their mirror dish. The last column is
the predicted energy threshold for $\gamma$-ray photons at zenith
angle $\vartheta = 0^\circ$. (From Petry 2001).}
\begin{center}
{\small
\begin{tabular}{llllll}
\hline Project name & latitude & longitude & altitude & Type &
$E_{\mathrm{thr}}(0^\circ)$ \\
\hline CANGAROO III & 31$^\circ$ S & 137$^\circ$ E & 160 m & 4
$\times$ 10 m
& 80 GeV \\
HESS I & 23$^\circ$ S & 17$^\circ$ E & 1800 m & 4  $\times$ 13 m & 40 GeV \\
MAGIC I & 29$^\circ$ N &  17$^\circ$ W & 2200 m & 1  $\times$ 17 m & 30 GeV \\
VERITAS & 32$^\circ$ N & 111$^\circ$ W & 1300 m & 7  $\times$ 10 m & 60 GeV \\
\hline
\end{tabular}
}
\end{center}
  \label{tev} \end{table}

In an attempt to establish an agenda for the new TeV telescopes,
Petry (2001) and Petry \& Reimer (2001) have analyzed the
observational possibilities for each of the unidentified and
tentatively identified EGRET sources. Several problems have to be
taken into account. The first is that of the apparent spectral
steepening of several 3EG sources between the GeV and the TeV band
(see Reimer and Bertsch 2001). This could be why strong sources
such as $\gamma$-Cygni, IC443, W28, and CTA 1 were not observed in
the TeV band (eg. Buckley et al. 1998). In order to get an
estimate of the impact of this effect on any 3EG source, Petry
(2001) proposed to use the same spectral steepening as that of the
Crab Nebula. The differential spectral index $\Gamma_{0.1 {\rm
GeV}}$ of the Crab Nebula at 0.1 GeV is $\Gamma_{0.1} = 2.19 \pm
0.02$ (Hartman et al. 1999) while it was measured by the Whipple
telescope at 500 GeV to be $\Gamma_{500 {\rm GeV}} = 2.49 \pm 0.06
\pm 0.04$ (Hillas et al. 1998). The latter authors showed that the
steepening towards higher energies can be described by an increase
in spectral index of 0.15 per decade of energy. Petry (2001)
showed that this would then imply, if a similar situation is the
case for the unidentified EGRET sources, that
\begin{equation}
\begin{array}{rcl}
    F ( E > x [\mathrm{GeV}]) & = & F_0 \cdot 10^{-\alpha}
    \cdot 10^{-(\alpha+0.15)} \cdot \left(\frac{x}{10}\right)^{-(\alpha+0.30)}\\
                   & = & F_0 \cdot 10^{-(\alpha - 0.15)} \cdot
x^{-(\alpha+0.30)}\\
                 & & \mathrm{ for}\ \  10 < x < 100 \\
   \\
    F ( E > x [\mathrm{GeV}]) & = & F_0 \cdot
    10^{-\alpha} \cdot 10^{-(\alpha+0.15)} \cdot 10^{-(\alpha+0.30)}
\cdot \left(
    \frac{x}{100}\right)^{-(\alpha+0.45)}\\
                   & = & F_0 \cdot 10^{-(\alpha - 0.45)} \cdot
x^{-(\alpha+0.45)}\\
                 & & \mathrm{ for}\ \  100 < x < 1000 \\
\end{array}
\end{equation}
where $F_0$ is the integrated flux above 0.1 GeV and $\alpha$ is
the flux spectral index (obtained by subtracting 1.0 from the
differential spectral index $\Gamma$, quoted for instance in Table
1) of a given source taken from the Third EGRET Catalog. An
extrapolation without taking into account possible breaks in the
spectrum could give a large over-estimation of the high-energy
flux. Although not valid for all sources, the previous scheme can
be used to give an idea of the expected flux from interesting 3EG
detections (Petry 2001, Petry \& Reimer 2001). Nonetheless, not
only the spectral cutoffs, but other technical problems affect
actual observation. \footnote{\v{C}herenkov telescopes typically
cannot observe at zenith angles much larger than 70$^\circ$. The
zenith angle $\vartheta$ at the upper culmination of an
astronomical object depends on the the latitude $\phi$ of the
observatory and the declination DEC of the object according to $
\vartheta = | \phi - \mathrm{DEC} | $. Therefore, the condition $
| \phi - \mathrm{DEC} | \leq 70^\circ$ has to be imposed in the
selection of observable objects. The area perpendicular to the
optical axis illuminated by the \v{C}herenkov light at the
position of the telescope is proportional to the square of the
distance $d$ to the shower maximum. $d$ grows with $\vartheta$ as
$d \propto 1/\cos(\vartheta)$. The energy threshold
$E_{\mathrm{thr}}$ of a \v{C}herenkov telescope is equivalent to a
\v{C}herenkov photon density threshold $\rho_{\mathrm{thr}}$ at
the position of the telescope. $\rho_{\mathrm{thr}}$ is an
instrumental constant determined by the trigger condition of the
data acquisition system. These parameters depend on the zenith
angle $\vartheta$ and the primary photon energy, and impact
directly on the required flux sensitivity for an observation to be
plausible. Under reasonable assumptions, $\rho(E, \vartheta)
\propto E\cdot\cos^2(\vartheta)$ (Petry 2001), and then, to
satisfy the trigger condition $\rho(E,\vartheta) >
\rho_{\mathrm{thr}}$, $E$ has to increase with $\vartheta$ as $
E_{\mathrm{thr}}(\vartheta) =  E_{\mathrm{thr}}(0^\circ) \cdot
\cos^{-2}(\vartheta)$, with the values for
$E_{\mathrm{thr}}(0^\circ)$ shown in Table \ref{tev}. An increase
in effective collection area is accompanied by a proportional
increase in hadronic background rate, such that the gain in flux
sensitivity is therefore only the square-root of the gain in area.
If we define, $F_{5\sigma}( E_{\mathrm{thr}})$ as the integral
flux above the energy threshold $E_{\mathrm{thr}}$ which results
in a $5 \sigma$ detection after 50~h of observation time,
$
F_{5\sigma}( E_{\mathrm{thr}}(\vartheta), \vartheta ) =
      F_{5\sigma}( E_{\mathrm{thr}}(0^\circ), 0^\circ ) \cdot
\cos^{-1}(\vartheta).
$
The last main ingredient that has to be considered is the needed
observation time itself. It can be computed as (Petry 2001)
$
    T_{5\sigma}(E_{\mathrm{thr}})
    = \left( \frac{F(E_{\mathrm{thr}})}{F_{5\sigma}( E_{\mathrm{thr}})} \right)
    ^{-2} \cdot 50 \mathrm{h}.
$
Objects requiring much more than 50-100 hours will probably be
excluded from the first years of operation of the next generation
IACTs, since, typically, the maximum expected duty
cycle for these telescopes will be about 1000 hours yr$^{-1}$.}\\

Taking these considerations into account it is possible to
determine which of the 3EG sources analyzed here might be detected
by the next generation IACTs. 12 out of the 19 SNR-EGRET cases
listed in Table \ref{green} can be considered likely candidates,
from a practical point of view. The complete results for all 3EG
sources superposed to SNRs that might be detected in less than 50
hours is compiled in Table \ref{tev-snr}. Columns are as follows:
3EG name, radius of the 95\% confidence contour, spectral index at
0.1 GeV, all from Hartman et al. (1999), minimum zenith angle,
IACTs minimum energy threshold, expected flux at the minimum
energy threshold, the integral spectral index at the IACTs minimum
energy threshold for this source, the observation time to obtain a
detection with 5 $\sigma$ significance (in brackets are the
corresponding data if the spectrum at 0.1 GeV was steeper by one
standard deviation). A $^\ddag$-mark indicates that the detection
is photon flux limited: the observation time was increased such
that 100 photons are detected.\\

\begin{table}
{\small \caption{TeV observability of EGRET-SNR pairs, adapted
from Petry (2001).}
\begin{center}
\begin{tabular} {llllllll}
\hline object  & \multicolumn{1}{c}{$\theta_{95}$}
     & \multicolumn{1}{c}{$\alpha$}
     & $\vartheta_{\mathrm{min}}$
     & \multicolumn{1}{c}{$E_{\mathrm{thr}}$}
     & \multicolumn{1}{c}{$F(E_{\mathrm{thr}})$}
     & \multicolumn{1}{c}{$\alpha(E_{\mathrm{thr}})$}
     & \multicolumn{1}{c}{$T_{5\sigma}$}  \\
name    &\multicolumn{1}{c}{$[^\circ]$}
     &
     & \multicolumn{1}{c}{$[^\circ]$}
     & \multicolumn{1}{c}{$[$GeV$]$}
     & \multicolumn{1}{c}{$[$cm$^{-2}$s$^{-1}$$]$}
     &
     & \multicolumn{1}{c}{$[$h$]$}  \\
\hline \multicolumn{3}{c}{3EG} & \multicolumn{5}{|c}{ CANGAROO III
} \\\hline
  0617$+$2238  &   0.13 &  $1.01\pm0.06$ &   54 &    $228$ &    5.13$
\times 10^{-11}$ (3.22$ \times 10^{-11}$) & 1.46(1.52) &  $ 22$$(
55)$\\
  0631$+$0642 &   0.46 &  $1.06\pm0.15$ &   38 &    $128$ &    4.12$
\times 10^{-11}$ (1.41$ \times 10^{-11}$) & 1.51(1.66) &  $
19$$(161)$ \\
  0634$+$0521 &   0.67 &  $1.03\pm0.26$ &   36 &    $123$ &    5.39$
\times 10^{-11}$ (8.47$ \times 10^{-12}$) & 1.48(1.74) &  $
11$$(430)$ \\
  1410$-$6147  &   0.36 &  $1.12\pm0.14$ &   31 &    $108$ &    8.78$
\times 10^{-11}$ (3.30$ \times 10^{-11}$) & 1.57(1.71) &  $3.51$$(
25)$ \\
  1714$-$3857 &   0.51 &  $1.30\pm0.20$ &    8 &    $ 82$ &    2.70$
\times 10^{-11}$ (7.06$ \times 10^{-12}$) & 1.60(1.80) &  $
28$$(409)$ \\
  1744$-$3011 &   0.32 &  $1.17\pm0.08$ &    1 &    $ 80$ &    9.72$
\times 10^{-11}$ (5.70$ \times 10^{-11}$) & 1.47(1.55) &
$2.86^\ddag$$(6.17)$ \\
  1746$-$2851 &   0.13 &  $0.70\pm0.07$ &    2 &    $ 80$ &    4.22$
\times 10^{-9}$ (2.64$ \times 10^{-9}$) & 1.00(1.07) &
$0.07^\ddag$$(0.11^\ddag)$ \\
  1800$-$2338  &   0.32 &  $1.10\pm0.10$ &    7 &    $ 81$ &    1.46$
\times 10^{-10}$ (7.45$ \times 10^{-11}$) & 1.40(1.50) &
$1.91^\ddag$$(3.73^\ddag)$ \\
  1824$-$1514 &   0.52 &  $1.19\pm0.18$ &   16 &    $ 86$ &    4.18$
\times 10^{-11}$ (1.24$ \times 10^{-11}$) & 1.49(1.67) &  $
12$$(141)$ \\
  1826$-$1302 &   0.46 &  $1.00\pm0.11$ &   18 &    $ 88$ &    2.78$
\times 10^{-10}$ (1.32$ \times 10^{-10}$) & 1.30(1.41) &
$1.00^\ddag$$(2.11^\ddag)$ \\
  1837$-$0606 &   0.19 &  $0.82\pm0.14$ &   25 &    $ 97$ &    6.30$
\times 10^{-10}$ (2.40$ \times 10^{-10}$) & 1.12(1.26) &
$0.44^\ddag$$(1.16^\ddag)$ \\
  1856$+$0114  &   0.19 &  $0.93\pm0.10$ &   32 &    $112$ &    3.33$
\times 10^{-10}$ (1.65$ \times 10^{-10}$) & 1.38(1.48) &
$0.83^\ddag$$(1.68^\ddag)$ \\
\hline \hline
\multicolumn{3}{c}{3EG} & \multicolumn{5}{|c}{ HESS I } \\
\hline
  0617$+$2238  &   0.13 &  $1.01\pm0.06$ &   46 &    $ 82$ &    2.21$
\times 10^{-10}$ (1.48$ \times 10^{-10}$) & 1.31(1.37) &
$4.22$$(9.44)$ \\
  0631$+$0642 &   0.46 &  $1.06\pm0.15$ &   30 &    $ 53$ &    1.42$
\times 10^{-10}$ (5.53$ \times 10^{-11}$) & 1.36(1.51) &  $6.69$$(
44)$ \\
  0634$+$0521 &   0.67 &  $1.03\pm0.26$ &   28 &    $ 52$ &    1.77$
\times 10^{-10}$ (3.49$ \times 10^{-11}$) & 1.33(1.59) &
$4.17$$(107)$ \\
  1410$-$6147  &   0.36 &  $1.12\pm0.14$ &   39 &    $ 66$ &    1.80$
\times 10^{-10}$ (7.28$ \times 10^{-11}$) & 1.42(1.56) &  $5.11$$(
31)$ \\
  1714$-$3857 &   0.51 &  $1.30\pm0.20$ &   16 &    $ 43$ &    7.44$
\times 10^{-11}$ (2.21$ \times 10^{-11}$) & 1.60(1.80) &  $
20$$(224)$ \\
  1744$-$3011 &   0.32 &  $1.17\pm0.08$ &    7 &    $ 41$ &    2.63$
\times 10^{-10}$ (1.63$ \times 10^{-10}$) & 1.47(1.55) &
$1.48$$(3.88)$ \\
  1746$-$2851 &   0.13 &  $0.70\pm0.07$ &    6 &    $ 40$ &    8.36$
\times 10^{-9}$ (5.49$ \times 10^{-9}$) & 1.00(1.07) &
$0.03^\ddag$$(0.05^\ddag)$ \\
  1800$-$2338  &   0.32 &  $1.10\pm0.10$ &    1 &    $ 40$ &    3.93$
\times 10^{-10}$ (2.16$ \times 10^{-10}$) & 1.40(1.50) &
$0.71^\ddag$$(2.17)$ \\
  1824$-$1514 &   0.52 &  $1.19\pm0.18$ &    8 &    $ 41$ &    1.28$
\times 10^{-10}$ (4.34$ \times 10^{-11}$) & 1.49(1.67) &  $6.28$$(
55)$ \\
  1856$+$0114  &   0.19 &  $0.93\pm0.10$ &   24 &    $ 48$ &    9.55$
\times 10^{-10}$ (5.15$ \times 10^{-10}$) & 1.23(1.33) &
$0.29^\ddag$$(0.54^\ddag)$ \\
\hline \hline
\multicolumn{3}{c}{3EG} & \multicolumn{5}{|c}{ MAGIC I } \\
\hline
  0010$+$7309  &   0.24 &  $0.85\pm0.10$ &   44 &    $ 58$ &    7.87$
\times 10^{-10}$ (4.16$ \times 10^{-10}$) & 1.15(1.25) &
$0.35^\ddag$$(0.67^\ddag)$ \\
  0617$+$2238  &   0.13 &  $1.01\pm0.06$ &    6 &    $ 30$ &    8.11$
\times 10^{-10}$ (5.75$ \times 10^{-10}$) & 1.31(1.37) &
$0.34^\ddag$$(0.48^\ddag)$ \\
  0631$+$0642 &   0.46 &  $1.06\pm0.15$ &   22 &    $ 35$ &    2.49$
\times 10^{-10}$ (1.03$ \times 10^{-10}$) & 1.36(1.51) &
$1.12^\ddag$$(3.15)$ \\
  0634$+$0521 &   0.67 &  $1.03\pm0.26$ &   24 &    $ 36$ &    2.89$
\times 10^{-10}$ (6.26$ \times 10^{-11}$) & 1.33(1.59) &
$0.96^\ddag$$(8.75)$ \\
  1744$-$3011 &   0.32 &  $1.17\pm0.08$ &   59 &    $114$ &    5.64$
\times 10^{-11}$ (3.21$ \times 10^{-11}$) & 1.62(1.70) &  $
35$$(107)$ \\
  1746$-$2851 &   0.13 &  $0.70\pm0.07$ &   58 &    $106$ &    3.16$
\times 10^{-9}$ (1.94$ \times 10^{-9}$) & 1.15(1.22) &
$0.09^\ddag$$(0.14^\ddag)$ \\
  1800$-$2338  &   0.32 &  $1.10\pm0.10$ &   53 &    $ 82$ &    1.45$
\times 10^{-10}$ (7.42$ \times 10^{-11}$) & 1.40(1.50) &  $3.71$$(
14)$ \\
  1824$-$1514 &   0.52 &  $1.19\pm0.18$ &   44 &    $ 58$ &    7.48$
\times 10^{-11}$ (2.38$ \times 10^{-11}$) & 1.49(1.67) &  $ 10$$(
99)$ \\
  1837$-$0423 &   0.52 &  $1.71\pm0.44$ &   33 &    $ 43$ &    4.44$
\times 10^{-11}$ (3.08$ \times 10^{-12}$) & 2.01(2.45) &  $
21$$(>500)$ \\
  1856$+$0114  &   0.19 &  $0.93\pm0.10$ &   28 &    $ 38$ &    1.26$
\times 10^{-9}$ (6.97$ \times 10^{-10}$) & 1.23(1.33) &
$0.22^\ddag$$(0.40^\ddag)$ \\
  1903$+$0550  &   0.64 &  $1.38\pm0.17$ &   23 &    $ 35$ &    9.10$
\times 10^{-11}$ (3.35$ \times 10^{-11}$) & 1.68(1.85) &  $4.12$$(
30)$ \\
  2016$+$3657 &   0.55 &  $1.09\pm0.11$ &    8 &    $ 31$ &    3.43$
\times 10^{-10}$ (1.83$ \times 10^{-10}$) & 1.39(1.50) &
$0.81^\ddag$$(1.52^\ddag)$ \\
  2020$+$4017  &   0.16 &  $1.08\pm0.04$ &   11 &    $ 31$ &    1.26$
\times 10^{-9}$ (1.00$ \times 10^{-9}$) & 1.38(1.42) &
$0.22^\ddag$$(0.28^\ddag)$ \\
\hline
\multicolumn{3}{c}{3EG} & \multicolumn{5}{|c}{ VERITAS } \\
\hline
  0010$+$7309  &   0.24 &  $0.85\pm0.10$ &   41 &    $141$ &    2.70$
\times 10^{-10}$ (1.31$ \times 10^{-10}$) & 1.30(1.40) &
$1.03^\ddag$$(2.12^\ddag)$ \\
  0617$+$2238  &   0.13 &  $1.01\pm0.06$ &    9 &    $ 82$ &    2.20$
\times 10^{-10}$ (1.47$ \times 10^{-10}$) & 1.31(1.37) &
$1.26^\ddag$$(1.89^\ddag)$ \\
  0631$+$0642 &   0.46 &  $1.06\pm0.15$ &   25 &    $ 98$ &    6.16$
\times 10^{-11}$ (2.19$ \times 10^{-11}$) & 1.36(1.51) &
$4.51^\ddag$$( 13)$ \\
  0634$+$0521 &   0.67 &  $1.03\pm0.26$ &   27 &    $100$ &    7.34$
\times 10^{-11}$ (1.22$ \times 10^{-11}$) & 1.48(1.74) &
$3.78^\ddag$$( 42)$ \\
  1800$-$2338  &   0.32 &  $1.10\pm0.10$ &   56 &    $251$ &    2.61$
\times 10^{-11}$ (1.19$ \times 10^{-11}$) & 1.55(1.65) &  $
23$$(110)$ \\
  1856$+$0114  &   0.19 &  $0.93\pm0.10$ &   31 &    $108$ &    3.48$
\times 10^{-10}$ (1.73$ \times 10^{-10}$) & 1.38(1.48) &
$0.80^\ddag$$(1.61^\ddag)$ \\
  1903$+$0550  &   0.64 &  $1.38\pm0.17$ &   26 &    $ 99$ &    1.62$
\times 10^{-11}$ (5.00$ \times 10^{-12}$) & 1.68(1.85) &  $
24$$(249)$ \\
  2016$+$3657 &   0.55 &  $1.09\pm0.11$ &    5 &    $ 81$ &    8.92$
\times 10^{-11}$ (4.27$ \times 10^{-11}$) & 1.39(1.50) &
$3.11^\ddag$$(6.50^\ddag)$ \\
  2020$+$4017  &   0.16 &  $1.08\pm0.04$ &    8 &    $ 82$ &    3.34$
\times 10^{-10}$ (2.55$ \times 10^{-10}$) & 1.38(1.42) &
$0.83^\ddag$$(1.09^\ddag)$ \\
  \hline
\end{tabular}
\end{center}
\label{tev-snr} }
\end{table}

Many of the SNR-EGRET pairs studied will be primary candidates for
one or several of the forthcoming IACTs. A combination of TeV and
GeV observations, together with an understanding of the molecular
material distribution of the region, will be crucial in
determining the nature of several of the unidentified EGRET
sources analyzed here.

\section*{Acknowledgments}

We thank Fumio Yamamoto, Masumichi Seta, Toshihiro Handa, and
Tetsuo Hasegawa for providing CO(2-1) survey data toward the SNRs
IC443, W44, and W66. We acknowledge F. Aharonian, M. Mori, F.
Bocchino, J. Paredes and S. Digel for their kind permission to
reproduce Figures \ref{dif} and \ref{difc}, \ref{snr-egret},
\ref{bocchino}, \ref{paredes}, and \ref{simw66}, respectively. We
thank D. Petry for his permission to adapt part of his work (Petry
2001) in Section \ref{TEV_S}. We further acknowledge Felix
Aharonian, Seth Digel, Dave Thompson, Don Ellison, Paula Benaglia,
Mischa Malkov, Olaf Reimer, and Isabelle Grenier for useful
comments. We remain grateful to the Referee, who provided a
detailed review which made us improve the manuscript. D.F.T. was
supported by Princeton University, CONICET, and Fundaci\'on
Antorchas during different stages of this research. Also, part of
his work was performed under the auspices of the U.S. Department
of Energy (NNSA) by University of California Lawrence Livermore National
Laboratory under contract No. W-7405-Eng-48. He thanks Princeton
University, SISSA, and the University of Barcelona for their kind
hospitality. G.E.R. and J.A.C. were supported by CONICET (under
grant PIP N$^o$ 0430/98), ANPCT (PICT 03-04881), as well as by
Fundaci\'on Antorchas. G.E.R. also thanks the Max Planck
Association for additional support at the MPIfK, Heidelberg, as
well as the University of Paris VII and the Service
d'Astrophysique, Saclay, for kind hospitality. Y.M.B. acknowledges
the support of the High Energy Astrophysics division at the CfA
and the {\em Chandra\/} project through NASA contract NAS8-39073.
This research would have been impossible without the effort of D.
A. Green at the Mullard Radio Astronomy Observatory, Camridge (UK)
in providing a web-based SNR catalog. NASA's HESEARC, SIMBAD,
Goddard's EGRET archive and MPE's ROSAT All-Sky Survey were also
invaluable to this study. The optical (and part of the radio) data
is from the Digitized Sky Survey, accessible through
http://skyview.gsfc.nasa.gov/ Parts of this work were based on
photographic data obtained using The UK Schmidt Telescope. The UK
Schmidt Telescope was operated by the Royal Observatory Edinburgh,
with funding from the UK Science and Engineering Research Council,
until 1988 June, and thereafter by the Anglo-Australian
Observatory.  Original plate material is copyright (c) the Royal
Observatory Edinburgh and the Anglo-Australian Observatory.  The
plates were processed into the present compressed digital form
with their permission.  The Digitized Sky Survey was produced at
the Space Telescope Science Institute under US Government grant
NAG W-2166.


\begin{thebibliography}{999}

\bibitem{} Aharonian F.A., Drury L.O'C., \& V\"olk H.J. 1994, A\&A 285, 645
\bibitem{} Aharonian F.A., \& Atoyan, A.M. 1996, A\&A 309, 91
\bibitem{} Aharonian F.A., et al. 1999, A\&A 346, 913
\bibitem{} Aharonian F.A. et al. 2000, ApJ 539, 317
\bibitem{} Aharonian F.A. 2001, Space Sci. Rev. 99, 187
\bibitem{} Aharonian F.A. et al. 2001b, A\&A 375, 1008
\bibitem{} Aharonian et al., 2001c, A\&A, 370, 112
\bibitem{} Aharonian F.A., Konopelko A.K., V\"olk H.J., \& Quintana H., 2001b,
Astrop. Phys. 15, 335
\bibitem{} Aharonian F.A., et al. 2002, A\&A 395, 803
\bibitem{} Aharonian F.A., et al. 2002b, A\&A 390, 39
\bibitem{} Aharonian F. et al. 2002, A\&A 393, L37


\bibitem{} Allen G.E., Digel S.W., \& Ormes J.F. 1999a, Proc. Int.
Cosmic Ray Conf., Utah, 5, 515
\bibitem{} Allen G.E., Gotthelf E.V., \& Petre R. 1999b, Proc. Int.
Cosmic Ray Conf., Utah,  5, 480

\bibitem{} Amenomori M., et al. 1999, ApJ 525, L93

\bibitem{}
Anchordoqui L.A., Torres D.F., McCauley T., Romero G.E. \&
Aharonian F.A.,
hep-ph/0211231, in press in ApJ.

\bibitem{} Anderson S.B.,  et al. 1996, ApJ  468, L55




\bibitem{} Arikawa Y., Tatematsu, K., Sekimoto Y., \&  Takahashi, T.
1999, PASJ 51, L7

\bibitem{} Arqueros F. et al. 1999, Proc. Int. Cosmic Ray Conf., Utah, 5, 211

\bibitem{}{Asaoka} I., {Aschenbach} B. 1994,  A\&A  284, 573

\bibitem{} Aschenbach B. 1998, Nature 396, 141

\bibitem{} Atoyan A.M., Aharonian F.A., \& Voelk H.J. 1995, Phys.
Rev. D52, 3265



\bibitem{} Bamba A., Yokogawa J., Sakano M., \& Koyama K. 2000, PASJ 52, 259

\bibitem{} Baring, M.G. et al. 1999, ApJ  513, 311

\bibitem{} Becker R.H., \& Helfand D.J. 1987, AJ 94, 1629

\bibitem{} Bednarek W. 1993, A\&A 278, 307

\bibitem{} Bell A.R. 1978, MNRAS 182, 147

\bibitem{} Benaglia P., Romero G.E., Stevens I. \& Torres D.F. 2001, A\&A
366, 605

\bibitem{} Benaglia P. \& Romero G.E. 2003, A\&A 399, 1121


\bibitem{}Berezhko E.G.;
 P\"uhlhofer G., \& V\"olk, H.J. 2002, A\&A 400, 971

\bibitem{} Berezhko E.G.;
 Ksenofontov  L.T., \&
 V\"olk, H.J. 2002, A\&A 395, 945


\bibitem{} Bignami G.F., \& Hermsen W. 1983, ARA\&A 21, 67
\bibitem{} Bignami G.F., Bennet K., Buccheri R., Caraveo P.,
  \& Hermsen W. 1981, A\&A 93, 71

\bibitem{} Bildsten L. 1997, ApJS 113, 367

\bibitem{} Bloemen H., et al. 1997, ApJ 475, L25

\bibitem{} Bocchino F., \& Bykov A.M. 2000, A\&A 362, L29
\bibitem{} Bocchino F., \& Bykov A.M. 2001, A\&A 376, 248

\bibitem{} Brazier K.T.S.,
Kanbach G., Carrami\~nana A., Guichard J., \& Merck M.
1996, MNRAS 281, 1033

\bibitem{} Brazier K.T.S., Reimer O.,  Kanbach G., \&
  Carrami\~nana A. 1998, MNRAS 295, 819


\bibitem{} Buckley J.H., et al. 1997, Proc. 25th Int. Cosmic Ray Conf., Durban,
3, 237
\bibitem{} Buckley J.H., et al. 1998, A\&A  329, 639

\bibitem{} Burdett A., et al. 1999, Proc. Int. Cosmic Ray Conf., Utah,  5, 448

\bibitem{} Butt Y., Torres D.F., Combi J.A., Dame T., \& Romero G.E.
2001, ApJ 562, L167
\bibitem{} Butt Y., Torres D.F., Romero G.E., Dame T., \& Combi J.A.
2002a, Nature
418, 499
\bibitem{} Butt Y., Torres D.F, Combi J.A., Dame T., \& Romero G.E. 2002b,
To appear in the proceedings of 22nd Moriond Astrophysics Meeting:
The Gamma Ray Universe, Les Arcs, Savoie, France, 9-16 Mar 2002.
astro-ph/0206132
\bibitem{} Butt Y. et al. 2003, astro-ph/0302342, submitted to
ApJ

\bibitem{} Bykov A.M., Chevalier R.A., Ellison D.C., \& Uvarov Y.A. 2000,
ApJ 538, 203

\bibitem{} Camilo F., et al. 2001, ApJ 557, L51

\bibitem{} Caraveo P.A., De~Luca A., Mignani R.P., \& Bignami G.F.
2001, ApJ 561,  930

\bibitem{} Carrami\~nana A. 2001, in Proc. Int. Workshop on The
Nature of Galactic Unidentified
Gamma-ray Sources, O. Carrami\~{n}ana, O. Reimer, D. Thomson Eds.,
Kluwer Academic Press, p.107

\bibitem{} Case G., \& Bhattacharya D. 1998, ApJ 504, 761
\bibitem{} Case G., \& Bhattacharya D. 1999, ApJ 521, 246

\bibitem{} Caswell J.L., \& Barnes P.J. 1985, MNRAS, 216, 753
\bibitem{} Caswell J.L., Murray J.D., Roger R.S., Cole D.J., \& Cooke
D.J. 1975, A\&A 45, 239


\bibitem{} Chadwick P.M. et al. 1997,  Proc. 25th Int. Cosmic Ray Conf.
Durban, 3, 189

\bibitem{} Chantell M.C., et al. 1998, Nucl. Instrum. Methods A408, 468

\bibitem{} Chapman J.M., Leitherer  C., Koribalski B., Bouter R., \&
Storey M.  1999, ApJ 518, 890

\bibitem{} Chen W. \& White R. 1991, ApJ 381, L63

\bibitem{} Cheng K.S., \&  Ruderman M. 1989, ApJ 337, L77
\bibitem{} Cheng K.S., \& Ruderman M. 1991, ApJ 373, 187

\bibitem{} Chevalier R.A. 1999, ApJ 511, 798

\bibitem{} Clark D.H., \& Caswell J.L. 1976, MNRAS 174, 267
\bibitem{} Clark G.W., Garmire G.P. \& Kraushaar W.L. 1968
ApJ 153, L203

\bibitem{} Claussen M.J., Frail D.A., Goss W.M., \& Gaume R.A. 1997,
ApJ  489, 143

\bibitem{} Clemens D.P. 1985, ApJ 295, 422

\bibitem{} Combi J.A., \&  Romero G.E. 1995, A\&A 303, 872
\bibitem{} Combi J.A., Romero G.E., \& Azac\'arate I. 1997,
Ap\&SS, 250, 1
\bibitem{} Combi J.A., Romero G.E., \& Benaglia P. 1998, A\&A 333,
L91
\bibitem{} Combi J.A., Romero G.E., \& Benaglia P. 1999, ApJ  519,
L177
\bibitem{} Combi J.A., Romero G.E., Benaglia P., \& Jonas J. 2001,
A\&A 366, 1047

\bibitem{} Condon, J.J., Broderick, J.J., \& Seielstad, G.A. 1991, AJ 102, 2041

\bibitem{} Contreras M.E., et al. 1997, ApJ 488,  L153

\bibitem{}  Corbel S., Chapuis C.,
Dame T.M., \& Durouchoux P. 1999, ApJ 526, L29

\bibitem{} Cornett R.H., Chin G., \& Knapp G.R.  1977, A\&A 54, 889

\bibitem{} Crovisier J., Fillit R., \& Kazes I. 1973, A\&A 27, 417

\bibitem{} Crutcher R.M. 1988, in ``Molecular Clouds, Milky-Way \&
External Galaxies",
R. Dickman, R. Snell, \& J. Young Eds., New York,  Springer, p.105

\bibitem{} Crutcher R.M. 1994,  ``Clouds, cores and low mass stars",
Astronomical Society of
the Pacific Conference Series, volume 65; Proceedings of the 4th Haystack
Observatory, edited by D. P. Clemens and R. Barvainis, p.87

\bibitem{} Crutcher, R.M. 1999, ApJ 520, 706

\bibitem{} Cusumano G., Maccarone M.C.,
  Nicastro L., Sacco B., \& Kaaret P. 2000, ApJ 528, L25


\bibitem{} Dame T.M., Elmegreen B.G., Cohen R.S., \& Thaddeus P. 1986,
ApJ  305, 892
\bibitem{} Dame T.M., Hartmann D., \& Thaddeus P. 2001, ApJ 547, 792

\bibitem{} D'Amico N., et al. 2001, ApJ  552, L45

\bibitem{} De Jager O.C., \&  Mastichiadis A. 1997, ApJ 482, 874

\bibitem{} De Naurois M., et al. 2001, Proc. Int. Symp. on High Energy
Gamma-Ray Astro. (Heidelberg), F. Aharonian and H.J. V\"olk
(Eds)., AIP, New York, p.540

\bibitem{} Dermer C.D. 1986, A\&A 157, 223
\bibitem{} Dermer C.D., et al. 1997, AJ 113, 1379

\bibitem{} Dermer C.D. 1997, in Proceedings of the
Fourth Compton Symposium, Editors Charles D. Dermer, Mark S. Strickman,
and James D. Kurfess, Williamsburg, VA April 1997: AIP Conference
Proceedings 410, p. 1275.


\bibitem{} Dickman R.L., Snell R.L., Ziurys L.M., \&
Huang Y.-L. 1992, ApJ 400, 203

\bibitem{} Doherty et al. 2002, MNRAS, submitted.

\bibitem{} Dorfi E.A. 1991, A\&A 251, 597
\bibitem{} Dorfi E.A. 2000, Ap\&SS. 272, 227

\bibitem{} Downes A.J.B., Pauls T, \& Salter C.J. 1980, A\&A 92, 47

\bibitem{} Drury L.O'C.,  Aharonian F., \&  V\"olk H.J. 1994, A\&A 287, 959
\bibitem{} Drury L.O'C. et al. 2001. Report of working group
number four at the ISSI workshop on Astrophysics of Galactic Cosmic
Rays: astro-ph/0106046

\bibitem{} Dubner G.M., Vel\'azquez P.F.,
  Goss W.M., \& Holdaway M. A.  2000, AJ 120, 1933

\bibitem{} Duncan A.R., Stewart R.T.,
  Haynes R.F., \& Jones K.L. 1995, MNRAS 277, 36

\bibitem{} Duncan A.R., Stewart R.T., Haynes R.F., \& Jones K.L. 1997, MNRAS
287, 722


\bibitem{} Eichler D., \& Usov V.V. 1993, ApJ 402, 271

\bibitem{} Ellison D.C., Slane P., \& Gaensler B.M.  2001, ApJ 563, 191

\bibitem{} Enomoto R., et al. 2002, Nature 416, 823

\bibitem{} Erlykin A.D., \& Wolfendale A.W. 2003, astro-ph/0301653, accepted by J.Phys.G: Nucl.Part.Phys

\bibitem{} Esposito J.A., Hunter S.D., Kanbach G., \& Sreekumar P. 1996,
ApJ 461, 820

\bibitem{} Falcke H., Cotera A.,
  Duschl W.J.,
  Melia F., \& Rieke M.J. 1999,
  ``The Central Parsecs of the Galaxy'', ASP Conference Series, Vol. 186.

  \bibitem{} Fegan S. 2001, Proc. Int. Workshop on The Nature of
Galactic Unidentified
Gamma-ray Sources, O. Carrami\~{n}ana, O. Reimer, D. Thomson Eds.,
Kluwer Academic Press, p.285

\bibitem{} Fesen R.A. 1984, ApJ 281, 658



\bibitem{} Fierro J.M., et al. 1993, ApJ  413, L27


\bibitem{} Frail D.A., Kulkarni S.R., \&  Vasisht G.  1993, Nature 365, 136

\bibitem{} Frail D.A., Giacani E.B., Goss W.M., \& Dubner G. 1996,
ApJ 464, L165

\bibitem{} F\"{u}rst E., Reich W., Reich P., \& Reif K. 1990, ApJS 85, 691


\bibitem{} Gaisser T.K., Protheroe R.J., \& Stanev T. 1998 ApJ  492, 219

\bibitem{} Gehrels N., Macomb D.J., Bertsch D.L., Thompson D.J.,
\& Hartman R.C. 2000, Nature 404, 363
\bibitem{} Gehrels N., \& Shrader C.R. 2001, in
2001, in Gamma 2001, edited by S. Ritz, N. Gehrels, \& C. R.
Schader, AIP Conference Proceedings, New York, p.3

\bibitem{} Georganopoulos M., Aharonian F. A., \& Kirk J. 2002,
A\&A 388, L25

\bibitem{} Georgelin Y.P. \& Georgelin Y.M. 1970, A\&A 7, 133
\bibitem{} Georgelin Y.P. \& Georgelin Y.M. 1976, A\&A 49, 57

\bibitem{} Giacani E.B., et al. 1997, AJ 113, 1379

\bibitem{} Gillanders G., et al. 1997, Proc. 25th Int. Cosmic Ray Conf.
Durban,  3, 185

\bibitem{} Ginzburg V.L., Syrovatskii S.I. 1964, ``The Origin of
Cosmic Rays'', Pergamon Press, London


\bibitem{} Goldwurm A. 2001,
In ``Exploring the gamma-ray universe'', Proceedings of the Fourth
INTEGRAL Workshop, Eds. B. Battrick, A. Gimenez, V. Reglero \& C.
Winkler. ESA SP-459, Noordwijk: ESA Publications Division, p.455

\bibitem{} Grabelsky D.A., Cohen R.S., Bronfman L., \& Thaddeus P.
1988, ApJ 331, 181

\bibitem{} Gray A.D. 1994, MNRAS 270, 847

\bibitem{} Grenn A.J.,
Frail D.A., Goss W.M.,\&
  Otrupcek R. 1997, AJ 114, 2058

\bibitem{} Green A.J., Cram L.E., Large M.I., \& Ye T. 1999, ApJS 122, 207.
\bibitem{} Green D.A. 1997, PASA 14, 73
\bibitem{} Green D.A. 2000, A Catalogue of Galactic Supernova Remnants,
Mullard Radio Astronomy Observatory, Cambridge, UK (available at
http://www.mrao.cam.ac.uk/surveys/snrs/)

\bibitem{}Grenier I.A. 1995, Advances in Space Research 15, 73

\bibitem{}Grenier I.A. 2000, A\&A 364, L93
\bibitem{}Grenier I.A. 2001, in The Nature of Unindentified
Galactic Gamma-Ray Sources, eds. A. Carraminana, O. Reimer \& D.
Thompson, Kluwer Academic Publishers, Dordrecht, p.51

\bibitem{} Guessoum N., Von Ballmoos P., Knodleseder J., \& Vedrenne G.,
2001, in Gamma 2001, edited by S. Ritz, N. Gehrels, \& C. R.
Schader, AIP Conference Proceedings, New York, p.16


\bibitem{} Halpern J.P., Eracleous M., Mukherjee R., \&
  Gotthelf E.V. 2001a, ApJ 551, 1016

  \bibitem{} Halpern J.P., et al. 2001, ApJ 552, L125

  \bibitem{} Halpern J.P., Gotthelf E.V., Mirabal N., \& Camilo F.
2002, ApJ 573, L41

\bibitem{} Halzen F. \& Hooper D. 2002, Rept. Prog. Phys. 65, 1025

\bibitem{} Hartman R.C., et al. 1999, ApJS 123, 79


\bibitem{} Harrus I.M., Hughes J.P., \& Helfand D.J. 1996, ApJ 464, L161
\bibitem{} Harrus I.M. \& Slane P.O. 1999, ApJ 516, 811

\bibitem{} Haslam C.G.T., Salter C.J., Stoffel H., \& Wilson W.E.
    1982, ApJS 47, 1

\bibitem{} Hensberge H., Pavlovski K., \&  Verschueren W. 2000, A\&A 258, 553

\bibitem{} Hermsen W., et al. 1981, in Proc. Int. Cosmic Ray Conf.
17th, Paris, 1, 320

\bibitem{} Hillas A.M., et al. 1998, ApJ 503, 774

\bibitem{} Hnatyk B., \& Petruk O. 1998,
Condensed Matter Physics 1,  655

\bibitem{} Huang Y.-L., Dame T.M., \& Thaddeus P. 1983, ApJ, 272, 609
\bibitem{} Huang Y.-L., \& Thaddeus P. 1986, ApJ, 309, 804



\bibitem{} Hunter, et al. 1997, ApJ 481, 205



\bibitem{} Jaffe T.R., Bhattacharya D.,
Dixon D.D., \& Zych A.D. 1997, ApJ  484, L129

\bibitem{} Jonas J.L. 1999, PhD Thesis, Rhodes University

\bibitem{} Jones L.R., Smith A., \& Angellini L. 1993, MNRAS 265, 631

\bibitem{} Jones T.W. 2001, in Proc. of the
7th Taipei Astrophysics Workshop on Cosmic Rays in the Universe,
ASP Conference Proceedings, Vol. 241. Edited by Chung-Ming Ko.
San Francisco: Astronomical Society of the Pacific, astro-ph/0012483


\bibitem{} Kaaret P., \& Cottam J. 1996, ApJ 492, L35

\bibitem{} Kaaret P., Piraino S., Halpern J., \& Eracleous M. 1999,
ApJ 523, 197
\bibitem{} Kaaret P., Cusumano G., \& Sacco B. 2000,
ApJ 542, L41

\bibitem{} Kaspi V.M., et al. 1997, ApJ 485, 820
\bibitem{} Kaspi V.M., et al.  2000, ApJ 528, 445

\bibitem{} Kassim N.E., \& Frail D.A.  1996, MNRAS 283, L51

\bibitem{}Kaufman-Bernad\'o M.M., Romero G.E., Mirabel I.F. 2002, A\&A 385,
L10

\bibitem{} Keohane J.W., Petre R., Gotthelf E.V., Ozaki M., \& Koyama K. 1997,
   ApJ 484, 350

\bibitem{} Kifune T., et al. 1995, ApJ 438, L91

\bibitem{} Kirk J.G., \& Dendy R.O. 2001, J. Phys. G27, 1589

\bibitem{} Kirshner R.P., \& Winkler P.F. 1979, ApJ  227, 853


\bibitem{} Klothes R., Landecker T.L., Foster T., \& Leahy D.A. 2001, A\&A 376,
641

\bibitem{} Kniffen D.A., et al. 1974,  Nature 25, 397

\bibitem{} Koralesky B., Frail D.A., Goss W.M., Claussen M.J.,
\& Green A.J. 1998,  AJ, 116, 1323

\bibitem{} Konopelko A.K. 2001, in: High Energy Gamma-Ray Astronomy, F.A.
Aharonian \& H.J. V\"olk (eds.),
AIP, Melville, p. 568





\bibitem{} Kraushaar W.L.,  et al.  1972, ApJ 177, 341




\bibitem{} Lamb R.C., \& Macomb D.J. 1997, ApJ 488, 872

\bibitem{} Lasker B.M., et al. 1990, AJ 99, 2019

\bibitem{} Leahy D.A., Naranan S., \& Singh K.P. 1986, MNRAS 220, L501

\bibitem{} Leslard  R.W. et al. 1995, Proc. Int. Cosmic Ray
Conference, Rome, 2, 475

\bibitem{} Longair M.S. 1994, ``High Energy Astrophysics, Vol.2:
Stars, the Galaxy
and the Interstellar Medium", Cambridge University Press, 2nd ed.

\bibitem{} Lozinskaya T.A. 1974, Soviet Astronomy 17, 603

\bibitem{} Lozinskaya T.A. 1992, ``Supernovae and stellar wind
in the interstellar medium", AIP, New York

\bibitem{} Lozinskaya T.A., Pravdikova V.V., \&
  Finoguenov A.V.  2000, AstL 26, 77

\bibitem{} Lucarelli F., Konopelko A., Rowell G, Fonseca V. \& the
HEGRA collaboration,
2001, in High-energy gamma-ray astronomy, edited by F. Aharonian
and H. V\"oelk, AIP Conference Proceedings, New York, p.779


\bibitem{} Manchester R.N., et al. 2001, MNRAS 328, 17

\bibitem{} Markiewicz W. J., Drury L. O'C., \& V\"olk H J. 1990, A\&A 236, 487

\bibitem{} Markoff S., Melia F., \& Sarcevic I.
1997, ApJ  489, L47

\bibitem{} Markoff S., Melia F., \& Sarcevic I.
1999, ApJ  522, 870

\bibitem{} Mastichiadis A. 1996, A\&A 305, L53

\bibitem{} Mastichiadis A., \& Ozernoy L.M. 1994, ApJ 426, 599


\bibitem{} Mayer-Hasselwander H.A., et al. 1998, A\&A 335, 161

\bibitem{} McLaughlin M.A., Mattox J.R., Cordes J.M., \& Thompson D.J. 1996,
ApJ 473, 763

\bibitem{} Melia F. 1992, ApJ 387, L25
\bibitem{} Melia F., \& Falcke H. 2001, ARA\&A 39, 309

\bibitem{} Merck M., et al. 1996, A\&AS 120, 465

\bibitem{} Michelson P.F. 2001, in Gamma 2001, edited by S. Ritz,
N. Gehrels, \& C. R. Schader, AIP Conference Proceedings, New
York, p.713

\bibitem{} Milne D. K. 1979, Aust. J. Phys., 32, 83

\bibitem{} Mirabal N., \& Halpern J.P. 2001, ApJ 547, L137
\bibitem{} Mirabal N., \& Halpern J.P., Eracleous M., \&
Becker R.H. 2000, ApJ 541, 180

Ray Conf., Paris, 1, 17

\bibitem{} Montmerle T. 1979, ApJ 231, 95
\bibitem{} Morfill G.E., Forman M., \& Bignami G. 1984, ApJ 284, 856
\bibitem{} Morfill G. E., \& Tenorio-Tagle G. 1983, Space Sci. Rev. 36, 93
\bibitem{} Mori M. 2001,  J. Phys. Soc. Japan Suppl. B70, 22
\bibitem{} Mukherjee R., Gotthelf E.V., Halpern J., \& Tavani M. 2000, ApJ
542, 740
\bibitem{} Muraishi H., et al. 2000, A\&A 354, 57L


\bibitem{} Naito T. \& Takahara F. 1994, J. Phys. G20, 477



\bibitem{} Nicastro L., Gaensler B.M., \&
  McLaughlin M.A. 2000, A\&A 362, L5

\bibitem{} Nolan P.L., et al. 1996, ApJ 459, 100
\bibitem{} Odegard N. 1986, ApJ 301, 813

\bibitem{}\"{O}gelman H., \& Finley J.P  1993, ApJ 413, L31

\bibitem{} Olbert C., Clearfield R.C., Williams N., Keohane J., \& Frail D.A.
   2001, ApJ 554, L205

\bibitem{} Oliver R.J., Masheder M.R.W., \& Thaddeus P. 1996,
A\&A 315, 578

\bibitem{} Ong R.A. 1998, Phys. Rep. 305, 93

\bibitem{} Oser S., et al. 2000, ApJ 547, 949


\bibitem{} Paredes J.M., Mart\'{\i} J., Rib\'o M., Massi M., 2000, Science
288, 2341

\bibitem{}
Petre R., Keohane J., Hwang U., Allen G., \& Gotthelf E. 1998 in
``The Hot Universe", Proceedings of IAU Symposium 188. Edited by
Katsuji Koyama, Shunji Kitamoto, \& Masayuki Itoh. Dordrecht:
Kluwer Academic Press, 1998., p.117

\bibitem{} Petry D., \& Reimer O. 2001, in Proceedings Gamma 2001 Workshop,
edited by S. Ritz, N. Gehrels, \& C. R. Schader, AIP Conference
Proceedings, New York, p.696

\bibitem{} Petry D. 2001, Proc. Int. Workshop on The Nature of
Galactic Unidentified
Gamma-ray Sources, O. Carrami\~{n}ana, O. Reimer, D. Thomson
Eds., Kluwer Academic Press, p.299

\bibitem{} Pineault S., et al. 1993, AJ 105, 1060
\bibitem{} Pineault S., et al. 1997, A\&A 324, 1152

\bibitem{} Plaga R. 2002, New Astron. 7, 317

\bibitem{} Pohl M. 1996, A\&A 307, 57

\bibitem{} Pohl M. 1997, A\&A 317, 441

\bibitem{} Pollock A.M.T. 1985, A\&A 150, 339

\bibitem{} Preite-Martinez A., Feroci M., Strom R.G., Mineo T. 2000,
in Proceedings of the Fifth Compton Symposium, American Institute
of Physics (AIP), edited by Mark L. McConnell and James M. Ryan,
AIP Conference Proceedings, Vol. 510., p.73


\bibitem{} Punsly B. 1998a, ApJ 498, 640
\bibitem{} Punsly B. 1998b, ApJ 498, 660
\bibitem{} Punsly B., Romero G.E., Torres D.F., \& Combi J.A 2000,
A\&A 364, 556



\bibitem{} Radhakrishman V., Goss W.M., Murray J.D., \& Brooks J.W.
1972, ApJS 24, 49

\bibitem{}Rib\' o M., et al. 2002, A\&A 384, 954


\bibitem{} Reich W., F\"urst E., \& Sofue Y. 1984, A\&A 133, L4

\bibitem{} Reimer O., \& Bertsch D.L. 2001, Proc. of Int. Cosmic Ray
Conf., Hamburg,
to appear.

\bibitem{} Reimer O., \& Pohl M. 2002, A\&A 390, L43

\bibitem{} Reynoso, E. \& Mangum J.G. 2000, ApJ 545, 874

\bibitem{} Reynolds S.P. 1996, ApJ 459, L13

\bibitem{} Reynolds S.P. 1998, ApJ 493, 375

\bibitem{} Rho J., \& Petre R. 1998, ApJ 503, L167

\bibitem{} Rho J., Petre R., Schlegel E.M., \& Hester J.J. 1994, ApJ 430, 757

1999, ApJ 515, 712
\bibitem{} Roberts M.S.E., Romani R.W., \& Kawai N.  2001, ApJS 133, 451

\bibitem{} Rodgers A.W., Campbell C.T., \& Whiteoak J.B. 1960, MNRAS 121, 103

\bibitem{} Romero G.E., Combi J.A., \& Colomb F.R. 1994, A\&A 288, 731
\bibitem{} Romero G.E. 1998, Rev.~Mex.~A\&A 34, 29
\bibitem{} Romero G.E., Benaglia P., \& Torres D.F. 1999a, A\&A 348,
868
\bibitem{} Romero G.E., Torres D.F., Andruchow I., Anchordoqui
L.A., \& Link B. 1999b, MNRAS 308, 799
\bibitem{} Romero G.E., Kaufman-Bernad\'o M., Combi J., \& Torres
D.F. 2001, A\&A 376, 599
\bibitem{}Romero, G. E. 2001, in The Nature of Unindentified
Galactic Gamma-Ray Sources, eds. A. Carraminana, O. Reimer \& D.
Thompson, Kluwer Academic Press, p.65

\bibitem{} Romero G.E., \& Torres D.F. 2003,
ApJ 586, L33

\bibitem{} Rowell G.P., et al.  2000, A\&A 359, 337
\bibitem{} Ruiz M.T., \& May J. 1986, ApJ  309, 667

\bibitem{} Sakamoto S., Hasegawa T., Hayashi M., Handa T., \& Oka T.
1995, ApJS, 100, 125
\bibitem{} Seta M., et al. 1998, ApJ 505, 286

\bibitem{} Seward F.D., Schmidt B., \& Slane P. 1995, ApJ 453, 284


\bibitem{} Sch\"onfelder V. 2001, in Gamma 2001, edited by S. Ritz,
N. Gehrels, \& C. R. Schader, AIP Conference Proceedings, New
York, p.809

\bibitem{} Scoville N.Z., Irvine W.M., Wannier P.G.,  \& Predmore C.R. 1977,
ApJ 216, 320

\bibitem{} Sedov L.I. 1959, ``Similarities and dimensional methods
in mechanics", Academic Press, New York


\bibitem{} Stecker F.W. 1977, ApJ 212, 60

\bibitem{} Shklovskii I. S. 1953, Dokl. Akad. Nauk SSSR 91,No. 3, 475

\bibitem{} Sigl G., Torres D.F, Anchordoqui L.A., \& Romero G.E.
2001, Phys. Rev. D63, 081302

\bibitem{} Sinnis G., et al. 1995, Nucl. Phys. B (Proc. Suppl.) 43, 141

\bibitem{} Slane P., et al. 1997, ApJ 485, 221
\bibitem{} Slane P., et al. 1999, ApJ 525, 357

\bibitem{} Smith D.A., et al. 1997, Nucl. Phys. (Proc. Suppl) 54, 362

\bibitem{} Sofue Y., \& Reich W. 1979, A\&AS 38, 251

\bibitem{} Sturner S.J., \& Dermer C.D. 1995, A\&A 293, L17
\bibitem{} Sturner S.J., Dermer C.D., \& Mattox J.R. 1996, A\&AS
120, 445
\bibitem{} Sturner S.J., Skibo J. G., Dermer C. D., \& Mattox J. R.
1997, ApJ 490,
617
\bibitem{} Sugizaki M., et al.  2001, ApJS 134, 77
\bibitem{} Swanenburg, et al. 1981, ApJ 243, L69
\bibitem{} Swanenburg, et al. 1978, Nature 275, 298

\bibitem{} Tanimori T., et al. 1998, ApJ 497, L25

\bibitem{} Tavani M., et al. 2001, in Gamma 2001, edited by S. Ritz,
N. Gehrels, \& C. R. Schader, AIP Conference Proceedings, New
York, p.729

\bibitem{} Taylor J.H., et al. 1993, ApJS 88, 529 (updated at
ftp://pulsar.princeton.edu)

\bibitem{} Thompson D.J., et al. 1975, ApJ 200, L79
\bibitem{} Thompson D.J., et al. 1995, ApJS 101, 259
\bibitem{} Thompson D.J., et al. 1996, ApJS 107, 227

\bibitem{} Thompson D.J., et al. 1999, ApJ  516, 297

\bibitem{} Thompson D.J. 2001,
Proc. Int. Symp. on High Energy
Gamma-Ray Astro. (Heidelberg), F. Aharonian and H.J. V\"olk
(Eds)., AIP, New York, p.103

\bibitem{} Thompson D.J., Digel S.W., Nolan P.L., \& Reimer O. 2001,
in Proc. Neutron
Stars in Supernova Remnants, ASP Conference Series, Vol. 9999, 2002,
P. O. Slane and B. M. Gaensler Eds., in press. astro-ph/0112518

\bibitem{} Tompkins W. 1999, Ph.D. Thesis, Stanford University.

\bibitem{} Torres D.F., et al.  2001a, A\&A 370, 468
\bibitem{} Torres D.F., Combi J.A., Romero G.E., \& Benaglia P.
2001b, Proc. Int. Workshop on The Nature of Galactic Unidentified
Gamma-ray Sources, O. Carrami\~{n}ana, O. Reimer, D. Thomson Eds.,
Kluwer Academic Press, p.97
\bibitem{} Torres D.F., Pessah M.E., \& Romero G.E. 2001c,
Astronomische Nachrichten, 322, 223
\bibitem{} Torres D.F., Butt Y.M. \& Camilo F. 2001d, ApJ 560, L155
\bibitem{} Torres D.F, Romero G.E., \&  Eiroa E.F. 2002,
ApJ 560, 600
\bibitem{} Torres D.F, Romero G.E.,  Eiroa E.F., Wambsganss J., \&
Pessah M.E. 2003a, MNRAS 339, 335

\bibitem{}
Torres D.F. \& Nuza S.E. 2003b,
ApJ  583, L25

\bibitem{}  T\"{u}mer, T., et al. (1999), Astropart. Phys.  11, 271


\bibitem{} Uchida K., Morris M., \& Yusef-Zadeh F. 1992, AJ 104, 1533

\bibitem{} Uchiyama Y., Takahashi T., \& Aharonian F.A. 2002a,
PASJ54, L73

\bibitem{} Uchiyama Y., Takahashi T., Aharonian F.A., \& Mattox J.R. 2002b,
ApJ 571, 866

\bibitem{} Uchiyama Y., Aharonian F.A.  \& Takahashi T. 2002c, to
appear in A\&A


\bibitem{} Usov V.V. 1994, ApJ 427, 394

\bibitem{} van den Ancker M.E., Th\'{e} P.S.,
  \& de Winter D. 2000, A\&A 362, 580

\bibitem{} Vargas M. et al. 1996, A\&A 313, 828

\bibitem{} Vel\'azquez P.F., Dubner G.M., Goss W.M., \& Green A.
2002. astro-ph/0207530

\bibitem{} V\"olk H.J. 2001, To appear in the proceedings of 21st
Moriond Astrophysics Meeting:
High energy astrophysical phenomena, Les Arcs, Savoie, France, 20-27 Jan 2001.
astro-ph/0105356

\bibitem{} V\"olk H.J. 2002, in Proc. Int. Cosmic Ray Conf., Hamburg,
to appear. astro-ph/0202421


\bibitem{} Wallace P.M., et al.
2000, ApJ 540, 184

\bibitem{} Wallace P.M., Halpern J.P.,
Magalhaes A.M., \& Thompson D.J. 2002, ApJ 569, 36

\bibitem{} Wang Z.R., Asaoka I., Hayakawa S. \& Koyama K.
1992, PASJ 44, 303



\bibitem{} Whiteoak J.B.Z. \& Green A.J 1996 A\&AS 118, 329



\bibitem{} Wootten, A. 1981, ApJ 245, 105

\bibitem{} Yadigaroglu I.-A.,  \& Romani R.W. 1997, ApJ 476, 356

\bibitem{} Yamamoto F., Hasegawa T., Morino J., Handa T.,
  Sawada T., \& Dame T. M 1999,  in Proc. of Star
  Formation 1999,
  T. Nakamoto Ed., Nobeyama Radio Observatory, p.110

\bibitem{} Yusef-Zadeh F., Melia F., \& Wardle M. 2000, Science 287, 85

\bibitem{} Yusef-Zadeh F., Law C., \& Wardle M. 2002, ApJ 568, L121

\bibitem{} Zhang L., \& Cheng K.S. 1998, A\&A 335, 234

\bibitem{} Zhang L., Zhang Y.J., \& Cheng K.S. 2000, A\&A 357, 957


\end{thebibliography}
\end{document}